\documentclass[11pt]{article}
\usepackage{amsfonts,amsmath,amssymb}
\usepackage{graphicx}
\usepackage{epic,eepic,color,graphicx}
\usepackage{textcomp}

\newfont{\twelvemsb}{msbm10 scaled\magstep1}
\newfont{\eightmsb}{msbm8}
\newfam\msbfam
\textfont\msbfam=\twelvemsb \scriptfont\msbfam=\eightmsb
\catcode`\@=11
\def\Bbb{\ifmmode\let\next\Bbb@\else
\def\next{\errmessage{Use \string\Bbb\space only in math mode}}\fi\next}
\def\Bbb@#1{{\fam\msbfam{{#1}}}}

\newcommand{\be}{\begin{equation}}
\newcommand{\ee}{\end{equation}}
\newcommand{\ba}{\begin{eqnarray}}
\newcommand{\ea}{\end{eqnarray}}

\begin{document}
\sloppy
\renewcommand{\thefootnote}{\fnsymbol{footnote}}
\newpage
\setcounter{page}{1} \vspace{0.7cm}

\vspace*{1cm}
\begin{center}
{\bf Regular and Floquet bases for gauge and gravity theories: a non perturbative approach}\\
\vspace{1.8cm} {\large Davide Fioravanti $^a$ and Marco Rossi $^b$
\footnote{E-mail: fioravanti@bo.infn.it, rossi@cs.infn.it}}\\
\vspace{.5cm} $^a${\em Sezione INFN di Bologna, Dipartimento di Fisica e Astronomia,
Universit\`a di Bologna\\
Via Irnerio 46, 40126 Bologna, Italy} \\
\vspace{.3cm} $^b${\em Dipartimento di Fisica dell'Universit\`a della Calabria and
INFN, Gruppo collegato di Cosenza\\
Arcavacata di Rende, 87036 Cosenza, Italy} \\
\end{center}
\renewcommand{\thefootnote}{\arabic{footnote}}
\setcounter{footnote}{0}
\begin{abstract}
This paper presents physical interpretations of the change of basis in Heun-type equations (Heun's and their confluences and reductions) from solutions decaying at two singular points to Floquet solutions. A crucial element in these connection relations is the phase acquired by Floquet solutions when transitioning between (ir)regular singularities. We exploit a novel `kink method' to compute the quantum momentum of the solutions as convergent series, explicitly at all orders. Consequently, by integrating the Floquet solutions term by term, the acquired phase can be explicitly derived as a series as well. Given that Heun equations and their confluences describe ${\cal N}=2$ supersymmetric Yang-Mills (SYM) theories in the Nekrasov-Shatashvili (NS) background (corresponding to the quantization of Seiberg-Witten differentials) as well as black hole perturbation theory, tests and predictions can be established in both domains. Specifically, we perform a critical verification matching the acquired phase with the dual gauge period, $A_D$. On the gravity side, our explicit and systematic results agree with standard methods used to analyse the Teukolsky or Regge-Wheeler equations. This procedure provides an effective alternative to instanton and gravity computations. Ultimately, it marks the initial steps toward establishing a dictionary between integrability and gravitational quantities.

\end{abstract}
\vspace{1cm} 
\newpage

\section{Introduction}
\label{intro}
\setcounter{equation}{0}

The Heun equation (HE) and its confluence and reduction limits  -- which will be collectively named here Heun-type equations -- have ample and relevant applications in mathematics and physics. In fact, the HE describes the quantisation/deformation of the Seiberg-Witten (SW) differential \cite {SW1,SW2} of ${\cal N}=2$ supersymmetric Yang-Mills (SYM) with $N_f=4$ flavours in the Nekrasov-Shatashvili (NS) background \cite {NS, MT}; besides its different confluences (and then reductions) yield the same theories with less flavour multiplets: in specific, the Confluent Heun Equation (CHE)  the Doubly Confluent Heun Equation (DCHE) and its reduction into the Modified Mathieu Equation (MME) play the same r\^ole for $N_f=3,2,0$, respectively. In addition, they may appear in many instances of the dimensional reduction (and linearisation) of the Einstein partial differential equations (PDE), when the metric of some massive object, like a black hole (BH), is subjected to the small perturbation of another field  \cite {NOE}. Famous examples are those (radially symmetric) of scalar, electromagnetic and axial (odd-parity) and polar (even-parity) gravitational perturbations of the Schwarzschild metric which are governed, at least at first order, by the Regge-Wheeler \cite {RW} (and Zerilli for polar \cite{Zer1,Zer2}) or the Teukolsky equations \cite {TEU1,TEU2}, which, in their turn, are separated by the standard angular part (of spherical harmonics) and the radial part represented by a CHE  \cite {AGH,FMR}. Without radial symmetry, perturbations of rotating BHs of Kerr metric are described both in their angular and radial parts by two CHE, as well \cite {TEU1,TEU2}. But even the HE finds an application in perturbations of asymptotically Anti-de Sitter (AdS) BHs \cite {STU}. Eventually, Heun-type equations govern some of the cosmological (scalar and tensor) perturbations within the Lambda-CDM (Cold Dark Matter) model ({\it cf.} for instance \cite {BDM} and references therein).

A basis of solutions which is relevant in BH physics is the so-called in/up basis (but the same coincidence happens to other gravitational objects): this is made up of a solution `in' satisfying `incoming' boundary condition at the horizon ({\it i.e.} which `enters' the BH) and another `up' satisfying `outgoing' boundary condition at infinity ({\it i.e.} which very far from the horizon is a wave exiting towards infinity). As crucially noticed in \cite {FGBH}, in integrability parameters both these functions are decaying (the incoming at the horizon, the outgoing at infinity) and constitute the basis usually used in the framework of the Ordinary Differential Equations/Integrable Models 
(ODE/IM) correspondence: for the specific construction of ODE/IM we discuss in this paper, see \cite {FG}, for previous foundational works about different cases, {\it cf.} \cite {ODE-IM1,ODE-IM2,ODE-IM3,LZ,GMN,FR1,FRS} and references therein. On the other hand, from the mathematical and physical point of view fundamental and well studied solutions are the so-called Floquet solutions of index $k$. Of course, the latter can be written (up to a trivial proportionality factor) as the quantum period $a$ of ${\cal N}=2$ SYM and, in analogy with this idea, we identify in the {\it acquired phase} from zero to infinity, $\varphi$, another important exponent of these solutions. In fact, it permits, together with $k$ and the Wronskian of these solutions ({\it cf.} below), the computation of the parallel transport or the so-called Fenchel-Nielsen coordinates (see for instance \cite {CLT}). More in general these new ingredients enlarge considerably the structure of the ODE/IM correspondence and the perspectives of integrability itself\footnote{This was already noted in \cite{FR1} with, for instance, the definition of the prepotential as reported in the conclusive Section.}. For instance, the {\it acquired phase} $\varphi$ plays a central the r\^ole in computing the connection coefficients between the Floquet basis and the in/up basis and, importantly for ${\cal N}=2$ SYM, the dual quantum period $A_D$. In this respect, we find essential contriving a `kink method', vaguely inspired by a homonymous method for computing the conformal limit of an integrable quantum field theory, {\it i.e.} high temperature in statistical mechanics. The noteworthy improvement here is represented by the fact that not only the leading order can be computed, but all of them\footnote{Some progress in this direction was realised in \cite{FRO6} for what concerns the original kink limit on non linear integral equations. The advantage here is due to the presence instead of an ODE, see the {\it Introduction} of \cite{FR-direct} for considerations on the advantages of one or another kind of equation.}, thus realising the systematic computation -- as convergent series  in a small parameter, the instanton coupling of the SYM theory --, of the phase $\varphi$ and then of the connection coefficients between the in/up and the Floquet basis: our results for $A_D$ agree with instanton expansions of gauge theory and this is a non-trivial check of the correctness of our method. Actually, the kink method allows this computation in that it yields explicitly the quantum momentum (the logarithmic derivative) of Floquet solutions (of Heun-like equations) as a single converging series in the small parameter. Through the above change of basis, ODE/IM (or in/up) solutions can be written explicitly as sums of the two Floquet solutions (exponentials of converging series). Furthermore, the same method is effective in producing the quantum momentum of ODE/IM (or in/up) solutions as a {\it unique} series. Needless to mention, the corresponding wave functions encode the linearised metric fluctuations of a black hole background within the framework of gravitational perturbation theory, thereby serving as the foundational dynamical quantities required for the holographic or asymptotic reconstruction of gravitational wave forms.

Differently from the ${\cal N}=2$ SYM based method (for instance \cite {AGH,FMR} and following papers), our approach uses the kink method to directly {\it solve} the ODE and its monodromy and connection problem, thus reproducing, in particular, results both in ${\cal N}=2$ SYM (the quantum periods and then the prepotential) and in gravity (perturbations of BHs). With reference to \cite {AGH} the instanton series for the prepotential of SYM, used to compute quasinormal modes for Schwarzschild and Kerr BHs, is fully computable; and similarly the double series instanton partition functions of the quiver gauge theory of \cite{FMR} as these are Floquet solutions of the CHE. More in detail, the in/up basis assumes a foundational role within the framework of the ODE/IM correspondence and exhibits profound implications for black hole perturbation theory and asymptotic dynamics. Concurrently, the introduction of the Floquet basis constitutes a novel structural paradigm essential for establishing rigorous dualities with gauge-theoretic observables. Consequently, this work rigorously formalizes the analytical properties of the Floquet basis, while retaining a comprehensive derivation of central analytic results concerning the in/up sector (cf. Subsubsection \ref{inc-up}).

Our analysis will begin by considering the paradigmatic example of ODE associated with pure ($N_f=0$) SYM, namely the MME detailed in Section \ref{change}. Despite being one of the simplest realisations within the class of Heun-like equations, this system serves as an ideal prototype, as it fully encapsulates the essential physical and mathematical structures under investigation. In fact, we can introduce in detail the general `kink method' to provide an all order convergent series expansion of the quantum momentum and thus of the acquired phase $\varphi$ and of the connection coefficients. Subsequently, in Section \ref {Nf2}, the same method is exploited on the DCHE, describing ${\cal N}=2$ SYM with $N_f=2$. Section \ref {Nf4} is dedicated with the same aims to the HE, which describes deformation/quantisation of $N_f=4$ and, in the gravity perspective, perturbations of asymptotically $AdS$ BHs. A very important part of this Section is Subsection \ref {CHE-limit} as treating the first confluent limit, the CHE: this ODE emerges in descriptions of $N_f=3$ and Schwarzschild/Kerr BHs. Anew, convergent series expansions of the quantum momentum (and then wave function) and of the acquired phase are derived systematically at all orders. Furthermore, Subsubsection \ref {Rome-MST} analyses the solutions of the CHE and their expressions in terms of converging series and very easily derives from them the explicit coefficients of known solutions in gravity (and mathematical) literature. In this sense it can be regarded as a valid alternative to these methods. Eventually, to test the effectiveness of our new method, it is applied to the computation of the dynamical Tidal Love Number for Schwarzschild BHs in Subsubsection \ref{applgra}.

\section{Case study of the MME: from the ODE/IM to Floquet basis}
\label{change}
\setcounter{equation}{0}

The first move is made from the MME (pure $N_f=0$ SYM):
\be 
\frac {d^2}{dy^2}\psi (y)=\left ( 2e^{2\theta} \cosh y +P^2 \right ) \psi (y) \, ,
\label {mathy}
\ee
which has two irregular singularities at $y=\pm \infty$ resulting from confluence and reduction from the initial HE. As a particular case, this equation was analysed in \cite {FG} as the starting point of the Child-ODE/IM (C-ODE/IM) correspondence to derive the $Q$-function of the self-dual Liouville model as the Wronskian $W[U_0,V_0]$ between its two solutions $V_0(y)$ and $U_0(y)=V_0(-y)$, forming the ODE/IM basis, with $V_0$ uniquely defined by the decaying asymptotic behaviour 
\be
V_0(y)\simeq \frac {1}{\sqrt {2}}\exp \left ( -\frac {\theta}{2}+\frac {y}{4} \right ) \exp \left (-2e^{\theta -\frac {y}{2}} \right )
\, , \label {Vzero}
\ee
when $\textrm {Re}y \rightarrow -\infty$ with $\left |\textrm {Im}\left (\theta - \frac {y}{2} \right ) \right |<\frac {\pi}{2}$. In ODE/IM language one says that $V_0$ is subdominant at $-\infty$ (whilst $U_0$ is subdominant at $+\infty$). The MME is the quantisation \cite {AY} of the SW differential  for ${\cal N}=2$ SYM with $N_f=0$ in the NS background and, by the map $e^{-\frac {y}{2}}=\frac {r}{L}, 2e^\theta =-i\omega L , P=\frac {\ell+2}{2}$, becomes the ODE 
\be
\frac {d^2 \phi }{dr^2}+ \left [ \omega ^2 \left (1+\frac {L^4}{r^4} \right ) -\frac {(\ell+2)^2-\frac {1}{4}}{r^2} \right ] \phi (r)=0
\label {D3brane} \, , 
\ee
which describes the radial wave $\phi (r)=\sqrt {r} \psi (y)$ for scalar perturbations of the D3 brane \cite {MMEGRA} in type IIB theory ($L$ is the radius of both AdS$_5$ and S$_5$, $\omega$ is the energy of the $\ell$-th partial wave). In gravity language $\phi_{up} = \sqrt {r} V_0\stackrel {r \rightarrow \infty} {\sim}  e^{i\omega r}$ is the `outgoing' solution at $r=\infty$, $\phi_{in} = \sqrt {r} U_0 \stackrel {r \rightarrow 0}  {\sim} r e^{i\omega L^2/r}$ is the `incoming' at the origin (the horizon) $r=0$\footnote {The current $J=-i \phi ^\star \partial _r \phi + c.c $, with $\phi$ solution of (\ref {D3brane}), is independent of $r$, under the assumption that the parameters $\omega, L,\ell$ are real. This occurs because, in this case, if $\phi $ is solution of (\ref {D3brane}), then $\phi ^\star$ is also a solution and $J$ is proportional to their Wronskian. 
If $J$ is computed when $\phi =\sqrt {r} V_0$, it equals $2$; if it is computed  when $\phi =\sqrt {r} U_0$, it equals $-2$. Then, $\sqrt {r}V_0$ describes a wave going towards $r=+\infty$, while $\sqrt {r}U_0$ a wave going towards $r=0$.}. Motivated by the relevance of the MME in different areas of physics, we reformulate C-ODE/IM of \cite{FG} by using another pair of solutions, the Floquet basis. To identify two linearly independent Floquet solutions of (\ref {mathy}), it is useful to introduce the logarithmic derivative $\Pi (y)=\frac {d}{dy} \ln \psi (y)$ which satisfies the Riccati equation
\be
\Pi (y)^2 +\frac {d}{dy}\Pi (y)= 2e^{2\theta} \cosh y +  P^2 \, .
\label {Ric}
\ee
In general when $\textrm {Re}\left (\frac {y}{2}+\theta \right )\rightarrow +\infty$ 
there are two possible behaviours  
\be
\Pi (y)\simeq \pm e^\theta e^{\frac {y}{2}}-\frac {1}{4}\pm \frac {16P^2-1}{32e^\theta} e^{-\frac {y}{2}}+\frac {16P^2-1}{64e^{2\theta}}e^{-y}
+O(e^{-3y/2}) \, ,
\label {plus}
\ee
which may each expand, when $\textrm {Re}\left (\frac {y}{2}-\theta \right )\rightarrow -\infty$, in the two alternative ways
\be
\Pi (y)\simeq \pm e^\theta e^{-\frac {y}{2}}+\frac {1}{4}\pm \frac {16P^2-1}{32e^\theta} e^{\frac {y}{2}}- \frac {16P^2-1}{64e^{2\theta}}e^{y}
+O(e^{3y/2})
\label {minus} \, .
\ee
The wave function for generic values of $\theta , P$ cannot decay at both $y=\pm \infty$.
The Floquet solutions satisfy $\psi _\pm(y + 2\pi i) = e^{\pm 2\pi i k}\psi _\pm(y)$ and then they have to be different from both $U_0(y)$ (since, following notations of  \cite {FG},
$U_0(y + 2\pi i) = U_1(y)$) and $V_0(y)$ (since \cite {FG} $V_0(y + 2\pi i) = V_{-1}(y)$). It follows that $\psi _\pm (y)=\alpha _\pm U_0(y)+\beta _\pm V_0(y)$, with $\alpha _\pm, \beta _\pm$ both different from zero. Then, $\psi _\pm (y)$ diverge (i.e. they are 'dominant' in the ODE/IM language) at both $\pm \infty$. The wave function $\psi _+$, corresponds to a solution $\Pi _+(y) =\frac {d}{dy} \ln \psi_+ (y)$ of (\ref {Ric}) which expands when $\textrm {Re}\left (\frac {y}{2}\pm \theta \right )\rightarrow \pm \infty$ as 
\be
\Pi _+(y)\simeq  \pm e^\theta e^{\pm \frac {y}{2}} \mp \frac {1}{4} \pm \frac {16P^2-1}{32e^\theta} e^{\mp \frac {y}{2}}\pm \frac {16P^2-1}{64e^{2\theta}}e^{\mp  y}
+O(e^{\mp 3y/2}) \, , 
\label {plus+}
\ee
and in addition satisfies the periodicity property $\Pi _+(y+2i\pi)=\Pi _+(y)$. 
The corresponding wave function ('ray' of solutions of equation (\ref {mathy}), with proportionality constant $e^c$) is derived from this 'quantum momentum' \cite {FRletter1}
by 'regularising' the integrand
\be
\psi_+(y)=e^c e^{2e^\theta e^{\frac {y}{2}}+2e^\theta e^{-\frac {y}{2}}+\frac {y}{4} } \exp \left [ \int _{-\infty}^y dy' \left ( \Pi _+(y')-e^\theta e^{\frac {y'}{2}}+e^\theta e^{-\frac {y'}{2}}-\frac {1}{4} \right ) \right ]
 \label {floq} \, ,
\ee
so that, when $y \rightarrow -\infty$, 
$\psi_+(y;k) \simeq e^c e^{\frac {y}{4}} e^{2e^\theta e^{-\frac {y}{2}}}$. 
Since  $\Pi _+(y+2i\pi )=\Pi _+(y )$, this is a quasi-periodic solution: $\psi _+(y+2\pi i; k)=e^{2\pi i k}\psi _+(y;k)$, where the Floquet exponent can be given by the integral
\be
2\pi i k= \int _{y}^{y+2i\pi}  dy' \Pi _+(y') \, , \label {floqdef}
\ee
on a straight line parallel to the imaginary axis, of course independent of $y$. 
The $y\rightarrow +\infty$ limit is found by manipulating the integral in the exponent in (\ref {floq}) as follows
\ba 
&& \int _{-\infty}^y dy' \left ( \Pi _+(y')-e^\theta e^{\frac {y'}{2}}+e^\theta e^{-\frac {y'}{2}}-\frac {1}{4} \right )=
\int _{-\infty}^0 dy' \left ( \Pi _+(y')-e^\theta e^{\frac {y'}{2}}+ e^\theta e^{-\frac {y'}{2}}-\frac {1}{4} \right )+ \nonumber \\
&& \int _0^y dy' \left ( \Pi _+(y')-e^\theta e^{\frac {y'}{2}}+e^\theta e^{-\frac {y'}{2}}+\frac {1}{4} \right ) -\frac {y}{2} \nonumber \, ,
\ea
so that, when $y\rightarrow +\infty$
\be
\psi_+(y;k) \simeq e^c e^{2 e^\theta e^\frac {y}{2}-\frac {y}{4}}\exp \left [ \int _{-\infty}^0 dy' \left ( \Pi _+(y')+e^\theta e^{-\frac {y'}{2}}-\frac {1}{4} \right ) +  \int _0^{+\infty} dy' \left (\Pi _+(y')-e^\theta e^{\frac {y'}{2}}+\frac {1}{4} \right ) \right ] 
\label{psi+at+}
\ee
It is then natural to define, independently of the normalisation $e^c$, the acquired 'scattering phase' from $y\rightarrow -\infty$ to $y\rightarrow +\infty$:
\be
\varphi (\theta; P)= \int _{-\infty}^0 dy' \left ( \Pi _+(y')+e^\theta e^{-\frac {y'}{2}}-\frac {1}{4} \right ) +  \int _0^{+\infty} dy' \left (\Pi _+(y')-e^\theta e^{\frac {y'}{2}}+\frac {1}{4} \right )\equiv \varphi _<(\theta ; P)+\varphi _>(\theta ; P)  \, . \label {varphidef}
\ee
This phase satisfies an important functional relation.
Indeed, using the symmetry\footnote {This symmetry is justified by saying that $\Pi _+(y;\theta )$ and $\Pi _+\left (y- i\pi;\theta +\frac {i\pi}{2} \right )$ satisfy the same Riccati equation and have the same asymptotic limits (\ref {plus+}) when $y\rightarrow + \infty$. With a similar reasoning one proves the subsequent identity $\Pi _+(y,\theta)=\Pi _+\left (y+i\pi ;\theta +\frac {i\pi}{2} \right )$.} $\Pi _+(y;\theta )=\Pi _+\left (y-i\pi;\theta +\frac {i\pi}{2} \right )$, for $y>0$, which holds also for the regulator $\left ( e^\theta e^{\frac {y}{2}}-\frac {1}{4} \right ) $ contained in the definition of $\varphi _>$, and then simple arguments of complex analysis, one arrives to the relation
\be 
\varphi _>\left (\theta +\frac {i\pi}{2}; P \right)=\varphi _>(\theta ; P)-i\int _{0}^{\pi}dy' \left ( \Pi _+(iy') - e^\theta e^{\frac {iy'}{2}} +\frac {1}{4} \right ) \label {ctransf} \, .
\ee
Analogously,  
due to the symmetry $\Pi _+(y,\theta)=\Pi _+\left (y+i\pi ;\theta +\frac {i\pi}{2} \right )$, for $y<0$, 
one proves the following property 
\be 
\varphi _<\left (\theta +\frac {i\pi}{2}; P \right )=\varphi _<(\theta ; P)-i\int _{-\pi}^0dy' \left ( \Pi _+(iy') + e^\theta e^{-\frac {iy'}{2}} - \frac {1}{4} \right ) \label {dtransf} \, .
\ee
The sum of the previous two relations gives the transformation property
\be
\varphi \left (\theta +\frac {i\pi}{2}; P \right)=\varphi (\theta; P) -2\pi i k (\theta ; P)\, .
\label  {phifun}
\ee 
Differently from $\varphi$,
the Floquet exponent $k$ does not change under shift $\theta \rightarrow \theta +i\pi/2$:
\be
k\left (\theta +\frac {i\pi}{2};P \right)=\int _0 ^{2\pi} \frac {dy'}{2\pi} \Pi _+\left (iy'; \theta +\frac {i\pi}{2} \right )= 
\int _0 ^{2\pi} \frac {dy'}{2\pi} \Pi _+(iy'\pm \pi ; \theta )=\int _0 ^{2\pi} \frac {dy'}{2\pi} \Pi _+(iy'; \theta )=k(\theta ; P) \, .
\ee
Then, if $\theta $ and $k$ are taken as independent variables, the function $\varphi (\theta ; k)$ satisfies the functional relation
\be
\varphi \left (\theta +\frac {i\pi}{2}; k \right)=\varphi (\theta ; k) -2\pi i k \, .
\label  {phifunk}
\ee
It will be convenient for computations and for physical application to have $k$ and not $P$ as independent variable and then this will be our choice from now on.
Choosing the constant $c= -\frac {\varphi}{2}$, the function $\psi _+$ acquires symmetric asymptotic behaviour at $\pm \infty$. The second Floquet solution, $\psi_{-}(y+2i\pi; k)=e^{-2i\pi k}\psi_{-}(y; k)$, can be defined by the relations $\psi_{\pm}(y; k)= \psi_{\mp}(-y; k)$ \footnote {The solution $\psi _-(y)$ is related to the solution of the Riccati equation (\ref  {Ric}) $\Pi _-(y)=-\Pi _+ (-y)$ by $\Pi _-(y)=d/dy \ln \psi _-(y)$.}. Then, as $\textrm {Re}y \rightarrow -\infty$, with $\left |\textrm {Im}\left (\theta -\frac {y}{2}\right )\right |<\frac {\pi}{2} $, the two Floquet solutions $\psi_{\pm}(y; k)$ behave as
\be
\psi_{\pm}(y;k) \simeq  e^{\mp \frac {\varphi}{2}} e^{\frac {y}{4}}\exp  \left (2e^\theta e^{-\frac {y}{2}} \right )\left [1+O\left ( e^{y/2} \right )  \right ] \, , \label {floq-infty}
\ee
whilst, when $\textrm {Re}y \rightarrow +\infty$, with $\left |\textrm {Im}\left (\theta +\frac {y}{2}\right )\right |<\frac {\pi}{2} $, they behave as
\be
\psi_{\pm}(y;k) \simeq  e^{\pm \frac {\varphi}{2}} e^{-\frac {y}{4}}\exp  \left (2e^\theta e^{\frac {y}{2}} \right )\left [1+O\left ( e^{-y/2} \right )  \right ] \, . \label {floq+infty}
\ee
As proven in \cite {FRletter1}, each element of the ODE/IM basis $\{V_0,U_0\}$ can be expressed as fine-tuned linear combination of elements of the Floquet basis $\{\psi_-,\psi_+\}$ (with coefficients involving the acquired phase since it appears in the asymptotics (\ref{floq-infty}) and (\ref{floq+infty})), in such a way to cancel the diverging contributions at leading order at $y=-\infty$ and $y=+\infty$ for $V_0$ and $U_0$, respectively: 
\be
U_0(y; k) =-\frac {1}{s(\theta,k)} \left [ e^{\frac {\varphi}{2}} \psi _-(y; k)-e^{-\frac {\varphi}{2}}\psi _+ (y; k)\right ] \, , \quad V_0(y; k) =\frac {1}{s(\theta,k)}  \left [ e^{-\frac {\varphi}{2}} \psi _- (y; k)-e^{\frac {\varphi}{2}}\psi _+ (y; k)\right ] \, .
\label{Floq-ode-im-bases-M}
\ee
Automatically, the whole dominant solutions will cancel out in each combination. As explained in \cite {FRletter1}, the constant $s(\theta ,k)$ is exactly determined in terms of the Wronskian between the Floquet solutions: $s(\theta ,k)=W[\psi _+, \psi _-]/(\sqrt {2} e^ {\frac  {\theta} {2}})$. In this way the decaying behaviours (at $y=-\infty$ and $y=+\infty$ for $V_0$ and $U_0$, respectively) emerges exactly:
\be
U_0(y; k) =\frac {\sqrt {2} e^{\frac {\theta}{2}}}{W[\psi _+, \psi _- ]} \left [ e^{-\frac {\varphi}{2}} \psi_+(y; k)-e^{\frac {\varphi}{2}}\psi_-(y; k)\right ] \, , \  V_0(y; k) =\frac  {\sqrt {2} e^{\frac {\theta}{2}}}{W[\psi _+ , \psi _-]}  \left [ e^{-\frac {\varphi}{2}}\psi_-(y; k)-e^{\frac {\varphi}{2}} \psi_+(y; k)\right ] \, . \label {chgbasis}
\ee
In addition, for the MME the Wronskian between the Floquet solutions is simply \cite {FRletter1} $W[\psi _+ ,\psi _-]=-4e^{\theta}\sin 2\pi k $. Immediately, this implies an elegant expression for the connection coefficient 
\be
Q (\theta ; k)\equiv W[U_0,V_0]=\frac {\sinh \varphi (\theta ; k)}{\sin 2\pi k} \, , \label {Qeleg}
\ee
which is also the $Q$-function of Liouville model at self-dual point \cite {FG}. Moreover, (\ref {phifunk}) is equivalent to the 'QQ-system' $1+Q^2\left(\theta +\frac {i\pi} {2}; k\right )=Q(\theta ; k)Q(\theta +i\pi; k)$ and $Q(\theta ; k)$ can be computed through an integrability or gauge Thermodynamic Bethe Ansatz (TBA) \cite {FG}.

\subsection{Large $\theta \rightarrow -\infty$ converging series}

When $\theta $ is negative and very large, the Floquet solutions and then $k$ and $\varphi$ can be computed as converging series. 
In fact, $\Pi _+(y)$ for finite (not too large) $|y|$ allows only for the computation of $k$ (\ref {floqdef}) \cite {GGM} (not of $\varphi$) as integration on the finite segment on the imaginary axis $[0,2\pi i]$: for these values of $y$ and for $\theta \rightarrow -\infty$, the inequality $|2e^{2\theta}\cosh y| < |P^2|$ is valid and then in the Riccati equation (\ref {Ric}) the potential term is a small perturbation. Therefore, the solution of (\ref {Ric}) can be written as the converging series 
\be
\Pi _+(y)=\sum _{n=0}^{+\infty}e^{2n\theta}\Pi _+^{(n)}(y) \, ,
\ee
with $\Pi _+^{(n)}(y)$ fixed recursively by (\ref {Ric}). In fact, $\Pi _+^{(0)}(y)=P$ and then the $\Pi _+^{(n)}(y)$, $n\geq 1$,  solve the ODEs
\be
2P \, \Pi _+^{(1)}(y)+\frac {d}{dy}\Pi _+^{(1)}(y)=2\cosh y \, , \quad  \sum _{m=0}^{n}
\Pi _+^{(m)}(y)\Pi _+^{(n-m)}(y)+\frac  {d} {dy}\Pi _+^ {(n)}(y)=0  \quad \textrm {for} \quad n\geq 2 \, .
\ee
These equations can be solved iteratively by Laurent {\it polynomials} in $e^y$,
\be
 \Pi _+^ {(n)}(y)=\sum \limits _{m=-n}^{n} r_m^{(n)} (P) e^{my} \, , \label {Laupoly}
\ee
where the sum runs over $m$ odd (even) if $n$ is odd (even).  
In fact, the $r_m^{(n)} (P)$ can be written in terms of $r_{m^\prime}^{(n^\prime)} (P)$ with $n^\prime <n$, according to the recursion relations
\ba
&& r_0^{(0)}=P \, ,\quad pr^{(1)}_p +\sum _{m=0}^1 \ \sum _{h={\textrm {max}}\{-m,p+m-1\}}^{{\textrm {min}}\{m,p-m+1\}}r_h^{(m)}r_{p-h}^{(1-m)}=\delta _{p,-1}+\delta _{p,1}  \, , \quad -1 \leq p \leq 1 \, , \\
&& pr^{(n)}_p +\sum _{m=0}^n \ \sum _{h={\textrm {max}}\{-m,p+m-n\}}^{{\textrm {min}}\{m,p-m+n\}}r_h^{(m)}r_{p-h}^{(n-m)}=0
\, , \ \quad -n \leq p \leq n \, ,  \quad n \geq 2 \, , 
\ea
so that at the first orders 
\be
\Pi _+^{(1)}(y)=\frac {e^y}{1+2P}+\frac {e^{-y}}{2P-1} \, , \quad \Pi _+^{(2)}(y)=
-\frac {e^{2y}}{(1+2P)^2(2+2P)}-\frac {1}{P(4P^2-1)}+\frac {e^{-2y}}{(2P-1)^2(2-2P)} \, .
\ee
Furthermore, the convergence is so good that the series can be integrated term by term in the expression  (\ref {floqdef}) (exchange of two limits, integral with series)
\be
k (\theta ; P)=
\sum _{n=0}^{+\infty}e^{4n\theta}r_0^{(2n)}(P)
= P-\frac {e^{4\theta}}{P(4P^2-1)}+\frac {e^{8\theta}(-60P^4+35P^2-2)}{4(4P^2-1)^3(P^2-1)P^3}+O(e^{12\, \theta}) \, . \label {ksmalltheta}
\ee
The acquired phase $\varphi$ brings bring an extra element of challenge, once thought of as expanded at $\theta \rightarrow -\infty$, since its definition contains an integration in $y$ on a non-compact domain, actually the entire real axis. Therefore, even if  $\theta $ is very large and negative, for sufficiently large $y$ the inequality $|2e^{2\theta}\cosh y|< |P^2|$ is not guaranteed and the previous expansion of $\Pi _+(y)$ cannot be used.
To approach the computation of $\varphi$ in the limit $\theta \rightarrow -\infty$ it is useful to gain inspiration from the `kink method' for solving TBA equations\footnote{In that case the computation stops at leading order, here it will be systematically extended to all orders.}. One starts from the decomposition (\ref {varphidef}) $\varphi =\varphi _> +\varphi _<$ and rewrites $\varphi _>$ in terms of 'right kink' quantum momentum
\be
\Pi _>(y)=\Pi _+\left (y-2\theta \right ) \label {rks}
\ee
as
\be
\varphi_>(\theta ; P)=\int _{2\theta }^{+\infty} dy' \left ( \Pi _>(y'; \theta, P) -e^{\frac {y'}{2}} +\frac {1}{4} \right ) \,  .
\label {phi>}
\ee
The shift in $y$ marks the difference with the previous approach.
Now, $\theta $ is present in the integration extreme and, implicitly, in $\Pi _>(y)$, which is subjected to a slightly different Riccati equation
\be
\Pi _>(y;\theta , P)^2 +\frac {d}{dy}\Pi _>(y;\theta , P)=e^{y} +P^2 + e^{4\theta }e^ {-y}\, .
\label {Rictilde}
\ee
Now, in the region $y>2\theta$ (\ref {Rictilde}) can be solved by the convergent series 
\be
\Pi_>(y;\theta ,P)=\sum \limits _{n=0}^{+\infty}\Pi _>^{(n)}(y;P)e^{4n\theta}\, ,
\ee 
since $|e^{4\theta } e^ {-y}|\rightarrow 0$ (and even smaller than $|e^y|$ and $P^2$) and this series can be integrated term by term in (\ref {phi>}), eventually giving $\varphi _>$ as a converging series. In fact, having in mind applications to gauge theories and gravity, one would like to use $\theta$ and $k$ as independent variables. With the aim of rewriting and solving the Riccati equation (\ref {Rictilde}) with $k$ as parameter
one inverts relation (\ref {ksmalltheta}) to find 
$P^2$ in terms of $k^2$, 
\be
P^2=\sum _{n=0}^{+\infty} p_2^{(n)}(k^2) e^{4n\theta} =k^2+\frac {2}{4k^2-1}e^{4\theta}+\frac {20k^4+7k^2}{2k^2(k^2-1)(4k^2-1)^3}e^{8\theta}+O\left (e^{12 \, \theta} \right ) \, 
\label {P2ink}
\ee
and insert this into (\ref {Rictilde}). Expanding the solution depending on $k$ as a converging series
\be
\Pi _>(y;\theta , k )=\sum _{n=0}^{+\infty}\tilde \Pi _>^{(n)}(y;k)e^{4n\theta} \, , \label {thetakexp}
\ee
the function $\tilde \Pi _>^{(n)}$ -- which acquired a tilde to distinguish its dependence on $k$ -- satisfies a first order ODE involving $\tilde \Pi _>^{(m)}$, with $m<n$:
\ba
&& \left (\tilde \Pi _>^{(0)}(y;k) \right )^2 +\frac {d}{dy}\tilde  \Pi _>^{(0)}(y;k)=e^{y} +k^2 \, , \quad 
2\tilde \Pi_>^ {(0)}(y;k)\tilde \Pi _>^ {(1)}(y;k)+\frac  {d} {dy}\tilde \Pi _>^ {(1)}(y;k)={e^ {-y}}+\frac {2}{4k^2-1}\, ,  \nonumber \\
&& \sum _{m=0}^{n}
\tilde \Pi _>^{(m)}(y;k)\tilde \Pi _>^{(n-m)}(y;k)+\frac  {d} {dy}\tilde \Pi _>^ {(n)}(y;k)=p_2^{(n)}(k^2)\, . \label {tildePi>equ}
\ea
Of course, this system can be solved at all orders recursively. In fact, since we are discussing a Floquet solution, we have to choose {\it the zero-th order momentum} $\tilde \Pi _>^{(0)}(y;k)=\frac {d}{dy}\ln J_{2k}(2ie^\frac {y}{2})$, with $J_{2k}$ Bessel function: indeed, $\tilde \Pi _>^{(0)}$ is $2\pi i $ periodic and thus reproduces the required periodicity of the eikonal. Moreover, it behaves as $e^{\frac {y}{2}}-\frac {1}{4}+O(e^{-\frac {y}{2}})$ when $y\rightarrow +\infty$, then entailing, already at leading order, the behaviour (\ref {plus+}). In addition, $\tilde \Pi _>^{(0)}$ goes as $k+\frac {e^y}{2k+1}-\frac {e^{2y}}{(2k+2)(2k+1)^2}+O(e^{3y})$ when $y\rightarrow -\infty$. The others $2\pi i $ periodic $\tilde \Pi _>^{(n)}(y;k)$, $n\geq 1$ are found by separation of variables up to an integration whose constant is fixed by integrability at $+\infty$: for instance, for $n=1,2$,
\ba
\tilde \Pi _>^{(1)}(y;k)&=&\frac {1}{1-4k^2}\frac {d}{dy}\left (e^{-y}+2e^{-y}\frac {d}{dy}\ln J_{2k}(2ie^\frac {y}{2}) \right )=
\frac {d}{dy}\left (\frac {e^{-y}}{1-4k^2}+\frac  {2e^{-y}}{1-4k^2} \tilde \Pi _>^{(0)}(y;k) \right )\, , \label {first-expl} \\
\tilde \Pi _>^{(2)}(y;k)&=&\frac {d}{dy}\Bigl [ \frac {(-32k^4+4k^2+1)e^{-2y}+(20k^2+7)e^{-y}}{4(k^2-1)(1-4k^2)^3}+  \label {sec-expl}  \\
&+& 
\tilde \Pi _>^{(0)}(y;k) \frac {(-16k^4-16k^2+5) e^{-2y}+(40k^2+14) e^{-y}}{4(k^2-1)(1-4k^2)^3} + \frac {2e^{-2y}}{(1-4k^2)^2} \frac {d}{dy} 
\tilde \Pi _>^{(0)}(y;k) \Bigr ] \, . \nonumber 
\ea
This peculiar form allows us to argue the general solution (for $n\geq 1$)
\be
\tilde \Pi _>^{(n)}(y;k)=\frac {d}{dy}  \left  [P_0^{(n)}(e^{-y})+\sum _{m=1}^{n}P_m^{(n)} (e^{-y})\frac {d^{m-1}}{dy^{m-1}} \tilde \Pi _>^{(0)}(y;k) \right ] \, , \label {solpin}
\ee
where $P_m^{(n)}(e^{-y})$ are degree $n$ polynomials: $P_m^{(n)}(e^{-y})=\sum \limits _{h=\textrm {max}\{m,1 \}}^n P_{m,h}^{(n)}(k^2) e^{-hy}$, with $ P_{m,h}^{(n)}(k^2)$ coefficients. These can be found in a recursive manner. One inserts (\ref {solpin}) into (\ref {tildePi>equ}) and, by using only the first of (\ref {tildePi>equ}), expresses the $n$-th equation (\ref {tildePi>equ}), with $n\geq 1$, in the form
\be
\sum _{p=0}^{n+2} C_p^{(n)}\left ( \{ P_m^{(n')}(e^{-y}), n'\leq n  \} \right ) \left ( \tilde \Pi _>^{(0)}(y;k) \right )^p =V^{(n)}(y) \, ,\quad V^{(n)}(y)=\delta _{n,1} e^{-y} + p_2^{(n)}(k^2) \,  , 
\label {finding-pol}
\ee
where the notation means that the 'coefficients' $C_p^{(n)}$ depend on $P_m^{(n')}(e^{-y})$, with $n'\leq n$. The equations  $C_0^{(n)}=V^{(n)}$, $C_p^{(n)}=0$, $p=1,...,n+2$,
determine the various $P_m^{(n)}(e^{-y})$. This procedure implies that $P_m^{(n)}(e^{-y})$ do not depend on the particular choice of the solution $\tilde \Pi _>^{(0)}(y;k)$ of the first of (\ref {tildePi>equ}), but only on the form of $V^{(n)}(y)$. This invariance property for the polynomials holds not only for the MME around the horizon, but also for the MME around infinity and, more in general in all the other cases studied in this paper: the DCHE, the HE, the CHE, both around the horizon and infinity. In other words, if for a specific equation we look for its solutions in the form (\ref {thetakexp}, \ref {solpin}) with different behaviours around a fixed singular point, we find always the same polynomials $P_m^{(n)}(y)$, but different choices of $\tilde \Pi _>^{(0)}(y;k)$ realising the different behaviours.

Eventually, the integral (\ref{phi>}) can be moved inside the series of the solution (\ref{thetakexp}), (\ref{solpin}) and yields a convergent series: explicitly,
\be
\varphi_>(\theta ; k)=\int _{2\theta }^{+\infty} dy' \left ( \Pi _>(y') -e^{\frac {y'}{2}} +\frac {1}{4} \right ) =
-\ln J_{2k}\left (2ie^\theta \right )+i\pi k-\frac {1}{2}\ln \pi -\ln 2 +2e^\theta -\frac {1}{2}\theta +\tilde S_>(\theta ; k) \, , 
\label {phi>2}
\ee
where $\tilde S_>(\theta ; k)=\sum \limits _{n=1}^{+\infty} e^{4n\theta} \int _{2\theta }^{+\infty} dy' \tilde \Pi_>^{(n)}(y';k)$.
Now, from (\ref {solpin}) one deduces the expansion 
\be
\tilde  \Pi _>^{(n)}(y) =\sum \limits
_{m=-n}^{+\infty} a_m^{(n)} e^{my} \, , \quad n\geq 1 \, , \quad a_0^{(n)}=0 \, , 
\ee
with $a_m^{(n)}$ constants\footnote  {The constants $a_ {m}^ {(n)}$ are determined by 
recursion relations:
\be
pa_p^{(n)}+\sum _{m=0}^n \sum _{k=-m}^{p+n-m}a_k^{(m)}a_{p-k}^{(n-m)}=L^{(n)}_p \, , \ n\geq 1 \, , 
\ee
with $L_p^{(1)}=\delta _{p,-1}+\frac {2}{4k^2-1}\delta _{p,0}$ and $L_p^{(n)}=p_2^{(n)}\delta _{p,0}$ for $n\geq 2$.
}, which converges around $y=-\infty$. This series can be integrated term by term for $y$ around $-\infty$, but not for $y\rightarrow +\infty$: its importance, however, is that it allows us to compute exactly all the powers of $e^{2\theta }$ coming from the evaluation of the primitive function of $\tilde \Pi_>^{(n)}(y)$, $n\geq 1$, 
\be
\sum _{\stackrel {m=-n} {m \not=0}}^{+\infty} \frac {a_m^{(n)}}{m}e^{my} \, , 
\label {primitiva}
\ee
at $y=2\theta \rightarrow -\infty$:
\be
e^{4\theta n}\int _{2\theta}^{+\infty} dy \tilde  \Pi _>^{(n)}(y;k)  =e^{4\theta n}\left [ C_n - \sum _{\stackrel {m=-n} {m \not=0}}^{+\infty} \frac {a_m^{(n)}}{m}e^{2\theta m} \right ]= e^{4\theta n} C_n - \sum _{\stackrel {q=n} {q \not=2n}}^{+\infty} \frac {a_{q-2n}^{(n)}}{q-2n}e^{2\theta q}  \, , \label {powerexp}
\ee
where $C_n$ is the value\footnote {An alternative and perhaps more efficient way to compute $C_n$ is the following: since the other primitive function of $\tilde  \Pi _>^{(n)}(y)$, $\left  [P_0^{(n)}(e^{-y})+\sum _{m=1}^{n}P_m^{(n)} (e^{-y})\frac {d^{m-1}}{dy^{m-1}} \tilde \Pi _>^{(0)}(y;k) \right ] $, equals zero if evaluated at $y=+\infty$, then $C_n$ must be minus the constant term in the expansion of such primitive in series of powers of $e^y$.}of the primitive function (\ref {primitiva}) of $\tilde  \Pi _>^{(n)}(y)$ at $+\infty$. Result (\ref {powerexp}) sheds light on the form of $\tilde S_>(\theta ; k)$, which, as a consequence of it, is expressed through positive (non-zero) powers of $e^{2\theta}$.

The reasoning goes in a parallel way for what the concerns the `left kink' quantum momentum
\be
\Pi _<(y)=\Pi _+\left (y+2\theta \right ) \, , \label {lks}
\ee
which works to find  
\be
\varphi _< (\theta ; P) =\int _{-\infty}^{-2\theta } dy' \left ( \Pi _<(y'; \theta, P)+e^{-\frac {y'} {2} }-\frac {1} {4} \right ) \, ,
\label {phi<}
\ee
since $\Pi _<(y;\theta ,P)$ satisfies this novel Riccati equation 
\be
\Pi _<(y;\theta ,P)^2 +\frac {d}{dy}\Pi _<(y;\theta ,P)=e^{-y} +P^2+ e^{4\theta} e^ {y} \, .
\label {Ric<}
\ee
After inserting (\ref {P2ink}) into (\ref {Ric<}), $\theta$ and $k$ are independent parameters and the $2\pi i$-periodic functions entering the series, analogous to (\ref {thetakexp}), 
\be
\Pi _<(y;\theta , k )=\sum \limits _{n=0}^{+\infty}\tilde \Pi _<^{(n)}(y;k)e^{4n\theta} \, , \label {Paiminus-series}
\ee
converging in the region $y<-2\theta$, satisfy the system of ODEs
\ba
&& \left ( \tilde \Pi _<^{(0)}(y;k) \right )^2 +\frac {d}{dy}\tilde  \Pi _<^{(0)}(y;k)=e^{-y} +k^2 \, , \quad 
2\tilde \Pi_<^ {(0)}(y;k)\tilde \Pi _<^ {(1)}(y;k)+\frac  {d} {dy}\tilde \Pi _<^ {(1)}(y;k)={e^ {y}}  +\frac {2}{4k^2-1}\, ,  \nonumber \\
&& \sum _{m=0}^{n}
\tilde \Pi _<^{(m)}(y;k)\tilde \Pi _<^{(n-m)}(y;k)+\frac  {d} {dy}\tilde \Pi _<^ {(n)}(y;k)=p_2^{(n)}(k^2)\, . \label {tildePi<equ}
\ea
As above, the solution of the first equation has to be $\tilde \Pi _<^{(0)}(y;k)=\frac {d}{dy}\ln J_{-2k}(2ie^{-\frac {y}{2}})$, because it is $2\pi i$ periodic and thus reproduces the periodicity of the eikonal. Moreover, it entails the behaviour (\ref {plus+}) in the limit $y\rightarrow -\infty$. The others $\tilde \Pi _<^ {(n)}(y;k)$ are fixed by integrability at $-\infty$. Moreover, the evenness symmetry of the potential entails the connection $\tilde \Pi _<^{(n)}(y;k)=-\tilde \Pi _>^{(n)}(-y;-k)$, from which $\Pi _<(y;\theta , k )=-\Pi _>(-y;\theta , -k )$. This implies the following form
\be
\tilde \Pi _<^{(n)}(y;k)=\frac {d}{dy}  \left  [P_0^{(n)}(e^{y})+\sum _{m=1}^{n}P_m^{(n)} (e^{y})(-1)^m\frac {d^{m-1}}{dy^{m-1}} \tilde \Pi _<^{(0)}(y;k) \right ] \, , \label {solpin<}
\ee 
for the solution of (\ref {tildePi<equ}), with the same polynomials $P_m^{(n)}(x)$ as the ones appearing in (\ref {solpin}), and besides $\varphi _< (\theta ; k) =\int _{-\infty}^{-2\theta } dy' \left ( \Pi _<(y'; \theta, k)+e^{-\frac {y'} {2} }-\frac {1} {4} \right )=-\varphi _>(\theta ; -k)$. Therefore, the use of (\ref {phi>2}) gives
\be 
\varphi (\theta ; k)=\varphi _>(\theta ; k)+\varphi _<(\theta ; k)=\ln J_{-2k}\left (2ie^\theta \right )-\ln J_{2k}\left (2ie^\theta \right )+2i\pi k + \tilde S(\theta ; k) \, , \label {varphismall2bis}
\ee
with $\tilde S(\theta ; k)=\tilde S_>(\theta ; k)-\tilde S_>(\theta ; -k)$ series of positive powers of $e^{2\theta}$. Expanding the Bessel functions at small argument yields
\be
\varphi (\theta ; k)=-4k\theta + \ln \frac {\Gamma (1+2k)}{\Gamma (1-2k)} + S(\theta ; k) \,
\label {varphismall3} \, ,
\ee
with the function $S(\theta ; k) $ a series of positive powers of $e^{4\theta}$. Indeed, the functional relation (\ref {phifun}) imposes that $S(\theta ; k) $ is invariant under shifts $\theta \rightarrow \theta +i\pi/2$ and this prevents the appearance of terms proportional to $e^{(2n+1)\theta}$, which are present in $\tilde S(\theta ; k)$. The expansion of $\varphi$  at $\theta \rightarrow -\infty$ in the variables $\theta$ and $k$ is more convenient for physical applications since these are exactly the gauge variables: the instanton coupling $\Lambda=\hbar \sqrt {2}e^\theta $ and the gauge period \cite {HE} $a=\hbar k$ of ${\cal N}=2$ SYM with $N_f=0$ in the NS background. 
Moreover, it is not difficult to go on in the computation of $\varphi (\theta ,k)$ with respect to the first two terms in the r.h.s. of (\ref {varphismall3}).
From a simple analysis of (\ref {powerexp}) it follows that the $N$-instanton term, i.e. the term in $S(\theta ,k)$ proportional to $e^{4\theta N}$, receives contributions from $2N+1$ integrals: from $\int _{2\theta}^{+\infty}dy [\tilde  \Pi _>^{(0)}(y)- e^{\frac {y}{2}} +\frac {1}{4} ]$ and from integrals (\ref {powerexp}), with $1\leq n\leq 2N$, minus the same terms with $k \rightarrow -k$. In particular, the first order of the function $S(\theta ,k)$, i.e. the `one-instanton' contribution to $\varphi (\theta ,k)$, is the sum of three terms. The first one comes from $\int _{2\theta }^{+\infty} dy' \left [ \tilde \Pi _>^ {(0)}(y)-e^{\frac {y}{2}} +\frac {1}{4} \right ] - (k \rightarrow -k)$:
\be
\ln \frac {J_{-2k}\left (2ie^\theta \right )}{ J_{2k}\left (2ie^\theta \right )}=-4k\theta + \ln \frac {\Gamma (1+2k)}{\Gamma (1-2k)}+\frac {e^{2\theta}}{1-2k} -
\frac {e^{2\theta}}{1+2k}-\frac {e^{4\theta}}{2(2-2k)(1-2k)^2}+\frac {e^{4\theta}}{2(2+2k)(1+2k)^2}+O(e^{6\theta}) \label {1st-contr} \, .
\ee
The second one is
\be
e^{4\theta}\int _{2\theta}^{+\infty }
dy [\tilde \Pi _>^{(1)}(y;\theta ,k) -  [\tilde \Pi _>^{(1)}(y;\theta ,-k) ] =-\frac {4k}{1-4k^2}e^{2\theta}+\frac {8k}{(1-4k^2)^2} e^{4\theta}+O(e^{6\theta}) \label {2nd-contr} \, . 
\ee
The last one is
\be
e^{8\theta}\int _{2\theta}^{+\infty}dy [\tilde \Pi _>^{(2)}(y;\theta ,k) -  [\tilde \Pi _>^{(2)}(y;\theta ,-k) ]=
\frac {e^{4\theta}}{2(2-2k)(1-2k)^2}-\frac {e^{4\theta}}{2(2+2k)(1+2k)^2}+O(e^{6\theta})
\label {3rd-contr} \, .
\ee
Summing all (\ref {1st-contr} -\ref {3rd-contr}) gives 
\be
\varphi (\theta ; k)=-4k\theta + \ln \frac {\Gamma (1+2k)}{\Gamma (1-2k)} + \frac {8k}{(1-4k^2)^2}e^{4\theta}+O(e^{8\theta})
\label {varphismall4} \, .
\ee
The identification $\sqrt {2}e^\theta = \Lambda _0/\hbar$, $k=a/\hbar$ brings to conjecture that
\be
\varphi (\theta ; k)=\frac {A_D(\Lambda _0,a)}{\hbar} \label {PhiAd0}
\ee
with $A_D/\hbar$ the dual gauge period of ${\cal N}=2$ SYM with $N_f=0$ in the NS background. Indeed,
(\ref {varphismall4}) agrees with the instanton expansion \cite {NEK,NO} of $A_D/\hbar$. The proof of conjecture (\ref {PhiAd0}) was addressed in \cite {FRletter1}.
An alternative definition of $A_D$ as Borel resummation of an asymptotic series in the limit $\theta \rightarrow +\infty$ was given in \cite {GGM}.

\medskip

We end this section by remarking that by the techniques discussed before it is possible to construct a converging series in the region $y>2\theta$ also for the ${U_0}_>(y)=U_0(y-2\theta)$, where $U_0(y)$ is the solution of the MME which goes to zero in the limit $y \rightarrow + \infty$ and then, according to the map to gravity variables given before equation (\ref {D3brane}), represents a wave  incoming at the origin of the physical space (the horizon). Defining $\Pi _>^{U_0}(y)=\frac {d}{dy}\ln {U_0}_>(y)$ and expanding
$\Pi _>^{U_0}(y)=\sum \limits_{n=0}^{+\infty}\tilde \Pi _>^{(U_0,n)}(y)e^{4n\theta}$,
the Riccati equation of the MME translates into the system, formally identical to (\ref {tildePi>equ}):
\ba
&& \left ( \tilde \Pi _>^{(U_0,0)}(y) \right )^2 +\frac {d}{dy}\tilde  \Pi _>^{(U_0,0)}(y)=e^{y} +k^2 \, , \quad 
2\tilde \Pi_>^ {(U_0,0)}(y)\tilde \Pi _>^ {(U_0,1)}(y)+\frac  {d} {dy}\tilde \Pi _>^ {(U_0,1)}(y;k)={e^ {-y}}+\frac {2}{4k^2-1}\, ,  \nonumber \\
&& \sum _{m=0}^{n}
\tilde \Pi _>^{(U_0,m)}(y)\tilde \Pi _>^{(U_0,n-m)}(y)+\frac  {d} {dy}\tilde \Pi _>^ {(U_0,n)}(y)=p_2^{(n)}(k^2)\, . \label {tildePiU>equ}
\ea
The choice $\tilde \Pi _>^{(U_0,0)}(y)=\frac {d}{dy}\ln K_{2k}(2e^{\frac {y}{2}})$ implements unambiguously the decaying boundary conditions of the solution $U_0$ at $y=+\infty$. Then, the general solution can be written as a (convergent) series 
\be
\Pi _>^{U_0}(y)=\sum _{n=0}^{+\infty}\tilde \Pi _>^{(U_0,n)}(y)e^{4n\theta} \, , \quad \tilde \Pi _>^{(U_0,n)}(y)=\frac {d}{dy}  \left  [P_0^{(n)}(e^{-y})+\sum _{m=1}^{n}P_m^{(n)} (e^{-y})\frac {d^{m-1}}{dy^{m-1}} \tilde \Pi _>^{(U_0,0)}(y) \right ] \, , \label {solpindec}
\ee
with $P_m^{(n)}(e^{-y})$ polynomials. Since the $P_m^{(n)}(e^{-y})$ are determined only by the potential appearing in the Riccati equation and do not depend on the choice of solution of the first of (\ref {tildePiU>equ}), they coincide with the polynomials used in (\ref {solpin}) to construct the Floquet solution of the Riccati equation $\Pi _>(y)$.

Analogously, the series (\ref {Paiminus-series}, \ref {solpin<}), with $\tilde \Pi  _<^{(0)} (y)=\frac {d}{dy}\ln K_{2k}(2e^{-\frac {y}{2}})$ implementing unambiguously the decaying boundary conditions of the solution $V_0$ at $y=-\infty$, and with the same polynomials $P_m^{(n)} (e^{y})$ used for constructing the Floquet momentum $\Pi _<(y)$, converges in the region $y<-2\theta$ to the logarithmic derivative of the solution $V_0(y+2\theta)$. Indeed, as specified at the beginning of this Section, $V_0$ goes to zero when $y \rightarrow - \infty$, and, in gravity language, represents a wave outgoing at spatial infinity. Eventually, in each of the previous cases, the polynomials stay the same while only the zero-th order momentum has to be changed. Indeed, it inherits and implements each time the specific characteristic of the exact solution.

\section{The DCHE: from ODE/IM to Floquet}
\label{Nf2}
\setcounter{equation}{0}

Without reduction, the DCHE
\be 
\frac {d^2}{dy^2}\psi (y)=\left [ e^{2\theta}(e^{2y}+e^{-2y})+2e^{\theta}q_1 e^y +2e^{\theta} q_2 e^{-y}+P^2\right ] \psi (y) \, 
\label {Nf2eq}
\ee
possesses still two irregular singularities at $y=\pm \infty$, quantises the SW differential for ${\cal N}=2$ SYM with $N_f=2$ flavours in NS background and describes scalar perturbations of the gravitational background generated by the intersection of four stacks of D3-branes in type IIB supergravity, thanks to the map (6.3) of \cite {FGS} (see also \cite {FGBH}). Inspired by the case $N_f=0$, the starting point is the Riccati equation for $\Pi (y)=\frac {d}{dy}\ln \psi (y)$:
\be
\Pi (y)^2 +\frac {d}{dy}\Pi (y)= e^{2\theta}(e^{2y}+e^{-2y})+2e^{\theta}q_1 e^y +2e^{\theta} q_2 e^{-y}+P^2
\label {Ricnf2} \, . 
\ee
Floquet solutions are built from a solution $\Pi _+$ of (\ref {Ricnf2}), which is periodic,
$\Pi_+(y+2\pi i)=\Pi _+(y)$, and expands as   
\ba
\Pi _+(y;\theta , q_1, q_2) &=& e^{\theta +y}+q_1-\frac {1}{2}+\frac {P^2-\left (\frac {1}{2}-q_1 \right )^2}{2 e^{\theta}}e^{-y}+O(e^{-2y}) \, , \quad  \textrm {when} \quad y \rightarrow +\infty  \, ,  \\
\Pi _+(y;\theta , q_1, q_2) &=& -e^{\theta -y}-q_2+\frac {1}{2}-\frac {P^2-\left (\frac {1}{2}-q_2 \right )^2}{2 e^{\theta}}e^{y}+O(e^{2y}) \, , \quad \textrm {when} \quad y \rightarrow -\infty \, .
\ea
These results imply for $y>0$ the invariance 
\be
\Pi _+\left ( y-\frac {i\pi}{2}; \theta +\frac {i\pi}{2}, q_1, -q_2 \right )=\Pi _+(y; \theta, q_1,q_2) \, , \label {inv+}
\ee
whilst, when $y<0$ other symmetry 
\be
\Pi _+\left (y+\frac {i\pi}{2};\theta +\frac {i\pi}{2}, -q_1, q_2 \right )=\Pi _+(y;\theta, q_1,q_2) \, . \label {inv-}
\ee
In terms of $\Pi _+$ the solution of (\ref {Nf2eq}), diverging both at $\pm \infty$, is constructed as
\be 
\psi _+(y; q_1, q_2)=e^{c(\theta ; q_1,q_2)}e^{e^{\theta -y}-\left (q_2-\frac {1}{2}\right ) y}\exp \left [ \int _{-\infty}^{y} dy' \left ( \Pi _+(y';\theta , q_1, q_2)+e^{\theta -y'}+q_2-\frac {1}{2} \right )\right ] \, , \label {floq+}
\ee
where $c(\theta ; q_1,q_2)$ is a constant. Due to the $2\pi i $-periodicity of $\Pi _+(y')$,  $\psi _+$ is quasi-periodic, $\psi _+(y+2\pi i)=e^{2\pi i k}\psi _+(y)$. Then, it is a Floquet solution, with index $k$ (independent of $y$) defined from the relation
\be
2\pi i k=  \int _{y}^{y+2i\pi}  dy' \Pi _+(y';\theta, q_1, q_2) \, , \label {floq2}
\ee
where integration is on a straight line parallel to the imaginary axis. When $y \rightarrow -\infty$, $\psi _+(y; q_1, q_2) \simeq e^{c(\theta ; q_1,q_2)}e^{e^{\theta -y}-\left (q_2-\frac {1}{2}\right ) y}$.
The behaviour of $\psi _+$ when $y\rightarrow +\infty$ is obtained after writing $\psi _+$ for $y>0$ as 
\ba
\psi _+(y; q_1, q_2)&=&e^{c(\theta ,q_1,q_2)}e^{e^{\theta +y}+\left (q_1-\frac {1}{2}\right ) y}\exp \Bigl [ \int _{-\infty}^{0} dy' \left ( \Pi _+(y';\theta , q_1, q_2)+e^{\theta -y'}+q_2-\frac {1}{2} \right )+\nonumber \\
&+&\int _0^y dy' \left ( \Pi _+(y';\theta , q_1, q_2)-e^{\theta +y'}-q_1+\frac {1}{2} \right)  \Bigr ] \, .
\ea
Then, when $y \rightarrow +\infty$, $\psi _+(y; q_1, q_2) \simeq e^{c(\theta ; q_1,q_2)+\varphi (\theta ;  q_1, q_2)}e^{e^{\theta +y}+\left (q_1-\frac {1}{2}\right ) y}$, where 
\ba
\varphi (\theta;  q_1, q_2)  &=&  \int _{-\infty}^{0} dy' \left ( \Pi _+(y';\theta , q_1, q_2)+e^{\theta -y'}+q_2-\frac {1}{2} \right )+\label  {varphidefNf2}\\
&+&\int _0^{+\infty} dy' \left ( \Pi _+(y';\theta , q_1, q_2)-e^{\theta +y'}-q_1+\frac {1}{2} \right) \equiv \varphi _< (\theta ;  q_1, q_2)+ \varphi _>(\theta ; q_1, q_2) \, , \nonumber 
\ea
can be interpreted as the acquired phase in going from $-\infty$ to $+\infty$.
We have defined the important quantities $\varphi, k$. Following the path devised in the case $N_f=0$, symmetries (\ref  
{inv+}, \ref {inv-}) give relations involving them. Relation (\ref {inv+}) brings  
\be
\varphi _>\left (\theta + \frac {i\pi}{2}; q_1, -q_2\right )=\varphi _>(\theta ;  q_1, q_2)-i\int _{0}^{\frac {\pi}{2}}dy\Pi _+(iy;\theta, q_1, q_2) - e^\theta (1-i) -\frac {i\pi}{2}\left (\frac {1}{2}-q_1 \right ) \, . \label {rel>}
\ee
On the other hand, the use of (\ref {inv-}) implies 
\be
\varphi _<\left (\theta + \frac {i\pi}{2}; -q_1, q_2\right )=\varphi _<(\theta ; q_1, q_2)-i\int _{-\frac {\pi}{2}}^{0}dy\Pi _+(iy;\theta, q_1, q_2) + e^\theta (1-i) +\frac {i\pi}{2}\left (\frac {1}{2}-q_2 \right ) \, . \label {rel<}
\ee
A functional relation involving both $\varphi$ and $k$ comes then from 
applying twice (\ref {rel>}) and (\ref {rel<}) and after summing the resulting relations. The result is
\be
\varphi (\theta +i\pi;  q_1, q_2)=\varphi (\theta ; q_1, q_2) -i \int _{-\pi}^{\pi} dy \Pi _+(iy;\theta, q_1, q_2) -i\pi (q_2-q_1)=
\varphi (\theta ; q_1, q_2)-2i\pi k -i\pi (q_2-q_1) \, , \label {relfund}
\ee
which is an exact functional relation we will use in the following.

Now, using the symmetry of the ODE (\ref {Nf2eq}), 
$y \rightarrow -y \, , \quad q_1 \rightarrow q_2 \, , \quad q_2 \rightarrow q_1 \, , \quad P \rightarrow P $, the second Floquet solution,  with Floquet exponent $-k(q_2,q_1)$, can be defined\footnote {The solution $\psi _-(y;q_1,q_2)$ is related to the solution of the Riccati equation (\ref {Ricnf2}) $\Pi _-(y;q_1,q_2)=-\Pi _+ (-y;q_2,q_1)$ by $\Pi _-(y;q_1,q_2)=d/dy \ln \psi _-(y;q_1,q_2)$.} as  
$\psi _-(y; q_1, q_2)= \psi _+ (-y;q_2, q_1)$. 
Within the choice $c (\theta; q_1,q_2)=-\frac {1}{2}\varphi (\theta ;  q_1,q_2)$, when $y \rightarrow -\infty$,
\be
 \psi _+(y; q_1, q_2) \simeq e^{-\frac {1}{2}\varphi (\theta ; q_1,q_2)} \exp \left [ e^{\theta -y}-\left (q_2-\frac {1}{2}\right ) y\right ] \, , \quad \psi _-(y; q_1, q_2) \simeq e^{\frac {1}{2}\varphi (\theta ;  q_2, q_1) } \exp \left [ e^{\theta -y}-\left (q_2-\frac {1}{2}\right ) y \right ]
\ee
and, as $y \rightarrow +\infty$, 
\be
\psi _+(y; q_1, q_2) \simeq e^{\frac {1}{2}\varphi (\theta ; q_1, q_2)}\exp \left [ e^{\theta +y}+\left (q_1-\frac {1}{2}\right ) y\right ] \, , \quad \psi _-(y; q_1, q_2) \simeq e^{-\frac {1}{2}\varphi (\theta ; q_2,q_1)}\exp \left [ e^{\theta +y}+\left (q_1-\frac {1}{2}\right ) y\right ]  \, .
\ee
The change of basis, from $\{ \psi _+,\psi _- \}$ to $\{ \psi _{+,0}, \psi _{-,0} \}$ (the ODE/IM basis) with $\psi _{+,0}\simeq \exp \left [(-\frac {1}{2}- q_1)(\theta + y +\ln 2)-e^{\theta + y} \right ]$ as $y\rightarrow +\infty$ (then subdominant at $+\infty$), $\psi _{-,0}\simeq \exp \left [(-\frac {1}{2}-q_2)(\theta - y +\ln 2)-e^{\theta - y} \right ]$ as $y\rightarrow - \infty$ (then subdominant at $-\infty$)
is \cite {FRletter1} 
\ba
&& \psi _{+,0}(y)=\frac {1}{s(\theta; k,  q_1,q_2)} \left [ e^{-\frac {1}{2}\varphi (\theta; k,  q_1,q_2)}\psi _+(y)- e^{\frac {1}{2}\varphi (\theta; k,  q_2,q_1)}\psi _-(y) \right ] \,  ,\\
&& \psi _{-,0}(y)=-\frac {1}{s(\theta; k,  q_2,q_1)}\left [e^{\frac {1}{2}\varphi (\theta; k,  q_1,q_2)}\psi_+(y)- e^{-\frac {1}{2}\varphi (\theta; k,  q_2,q_1) }\psi _-(y)\right ] \, ,
\ea
where $s$ is a normalisation factor. As for the MME case, the above fine-tuned linear combinations of the Floquet solutions are engineered to cancel their leading diverging behaviour, at $+\infty $ for the linear combination giving $\psi _{+,0}$ and at $-\infty$ for the one giving $\psi _{-,0}$. This cancellation extends at all diverging orders, thus leaving subdominant solutions at $+\infty$ and $-\infty$, respectively.
Again, using results reported in \cite {FRletter1} the constant $s$ can be determined in terms of the Wronskian between the Floquet solutions
\be
\frac {1}{s(\theta; k,  q_1,q_2)}=\left (2e^\theta \right )^{\frac {1}{2}-q_1} \frac {e^{-\frac {1}{2}\varphi (\theta; k,  q_2,q_1)+\frac {1}{2} \varphi (\theta; k, q_1,q_2)}}{W[\psi_+, \psi_-](\theta; k,  q_1,q_2)} \, .
\ee
In this way, we obtain explicit expressions, in terms of quantities related to Floquet solutions, for the change of basis 
\ba
\psi _{+,0}(y;q_1,q_2)&=&\frac {\left (2e^\theta \right )^{\frac {1}{2}-q_1} }{W[\psi_+, \psi_-](\theta; k,  q_1,q_2)} \left [ e^{-\frac {1}{2}\varphi (\theta; q_2,q_1)}\psi _+(y;q_1,q_2)- e^{\frac {1}{2} \varphi (\theta;  q_1,q_2)}\psi _-(y;q_1,q_2) \right ] \, , \nonumber \\
\psi _{-,0}(y;q_1,q_2)&=&\frac {\left (2e^\theta \right )^{\frac {1}{2}-q_2}}{W[\psi_+, \psi_-](\theta; k,  q_2,q_1)}\left [e^{-\frac {1}{2}\varphi (\theta ; q_1,q_2)}\psi _-(y;q_1,q_2)- e^{\frac {1}{2} \varphi (\theta ;  q_2,q_1) }\psi _+(y;q_1,q_2)\right ] \,   \label {nf2psi}
\ea
and, for the connection coefficient,
\be
Q(\theta ; k, q_1,q_2)=W[\psi _{+,0}, \psi _{-,0}](\theta; k, q_1,q_2)=-2\frac{(2e^\theta)^{1-q_1-q_2}}{W[\psi_+, \psi_-](\theta ; k, q_2,q_1)} \sinh \Phi (\theta; k, q_1,q_2) \, , 
\label {Qfornf2}
\ee
with
\be
\Phi (\theta ; k, q_1, q_2)=\frac {\varphi (\theta ; k, q_1,q_2)+\varphi (\theta ; k, q_2,q_1)}{2} \, ,
\ee
which generalise results for the MME derived in \cite {FGBH,FG}. However,
differently from the MME case, the Wronskian $W[\psi _+,\psi _-]$ does not have a simple expression in terms of $\theta, k$. However, it is conveniently computed in $y=0$ after using $\psi _-(y; q_1, q_2)= \psi _+ (-y;q_2, q_1)$. The result is $W[\psi _+,\psi _-](\theta, q_1,q_2,k)=-\psi _+(0;q_1,q_2)\psi _+^\prime (0;q_2,q_1) - 
\psi _+(0;q_2,q_1)\psi _+^\prime (0;q_1,q_2)$, which is symmetric under the exchange of $q_1$ with $q_2$ and by virtue of (\ref {floq+}) becomes
\be
W[\psi_+, \psi_-]=-e^{\frac {1}{2}\varphi _<(\theta; q_1,q_2)+\frac {1}{2}\varphi _<(\theta; q_2,q_1)-\frac {1}{2}\varphi _>(\theta; q_1,q_2)-\frac {1}{2}\varphi _>(\theta; q_2,q_1)+2e^\theta}
\left[\Pi _+(0;q_1,q_2)+\Pi _+(0;q_2,q_1) \right ] \, . \label {Wronsknf2}
\ee
In next subsection we will show that the various terms in (\ref {nf2psi}), -- in particular the Wronskian (\ref{Wronsknf2}) --, are computable in series of powers of $e^{2\theta}$ by using the `kink method' (which is not needed for $\Pi_+(0)$, as it enjoys a direct expansion). Between them it is particularly important the `relative' phase $\Phi $ between the coefficients in front of $\psi _\pm$, which
was shown \cite {FRletter1} to coincide with $A_D/\hbar$, with $A_D$ the dual gauge period of ${\cal N}=2$ SYM with $N_f=2$. Assuming this identification, one can compute $A_D$ and then $\Phi$ also by using the Alday-Gaiotto-Tachikawa (AGT) correspondence \cite {AGT}. However, $\Phi$ is not enough to capture the relation between in/up and Floquet functions. As (\ref {Wronsknf2}) shows, also $\varphi _<, \varphi _>$ and $\Pi _+(0)$ are needed.

\subsection {The $\theta \rightarrow -\infty$ converging series.}

The easiest quantity to study is $\Pi _+(y)$ for finite $y$, since, as for the MME, the quantum momentum \cite {FRletter1} of the Floquet solution develops as 
\be
\Pi _+(y)=\sum _{n=0}^{+\infty} e^{n\theta}\Pi _+^{(n)}(y)
\ee 
and this expansion solves (\ref {Ricnf2}) iteratively starting from  
$\Pi _+^{(0)}=P$: indeed, $\Pi _+^{(n)}(y)$, $n\geq 1$, solve the ODEs
\ba
&& 2P\Pi _+^{(1)}(y)+ \frac {d}{dy} \Pi _+^{(1)}(y)=2q_1e^y+2q_2 e^{-y} \, , \quad \left ( \Pi _+^{(1)}(y) \right )^2+2P \Pi _+^{(2)}(y)+\frac {d}{dy}\Pi _+^{(2)}(y)=2\cosh 2y  \, , \nonumber  \\
&& \sum _{m=0}^{n}
\Pi _+^{(m)}(y)\Pi _+^{(n-m)}(y)+\frac  {d} {dy}\Pi _+^ {(n)}(y)=0  \quad \textrm {for} \quad n\geq 3 \, , \nonumber
\ea
which again are solved iteratively by Laurent {\it polynomials} in $e^y$, $\Pi _+^ {(n)}(y)=\sum \limits _{m=-n}^{n} r_m^{(n)} (P,q_1,q_2) e^{my}$, where the sum runs over $m$ odd (even) if $n$ is odd (even).  The constants $r_m^{(n)}$ are determined by the recursion relations 
\ba
&& r_0^{(0)}=P \, ,\quad pr^{(1)}_p +\sum _{m=0}^1 \ \sum _{h={\textrm {max}}\{-m,p+m-1\}}^{{\textrm {min}}\{m,p-m+1\}}r_h^{(m)}r_{p-h}^{(1-m)}=2q_1\delta _{p,-1}+2q_2 \delta _{p,1}  \, , \quad -1 \leq p \leq 1 \, , \nonumber \\
&&  pr^{(2)}_p +\sum _{m=0}^2 \ \sum _{h={\textrm {max}}\{-m,p+m-2\}}^{{\textrm {min}}\{m,p-m+2\}}r_h^{(m)}r_{p-h}^{(2-m)}=\delta _{p,-2}+\delta _{p,2}  \, , \quad -2 \leq p \leq 2 \, , \label {rec-r-nf2}\\
&& pr^{(n)}_p +\sum _{m=0}^n \ \sum _{h={\textrm {max}}\{-m,p+m-n\}}^{{\textrm {min}}\{m,p-m+n\}}r_h^{(m)}r_{p-h}^{(n-m)}=0
\, , \ \quad -n \leq p \leq n \, ,  \quad n \geq 3 \, . \nonumber 
\ea
The first orders are
\be
\Pi _+^{(1)}(y)=\frac {2q_1e^y}{2P+1}+\frac {2q_2e^{-y}}{2P-1} \, , \\ \Pi _+^{(2)}(y)=\frac {(2P+1)^2-4q_1^2}{(2P+2)(2P+1)^2}e^{2y} -\frac {4q_1q_2}{P(4P^2-1)}    + \frac {(2P-1)^2-4q_2^2}{(2P-2)(2P-1)^2}e^{-2y}
\ee   
This solution gives the last part of (\ref{Wronsknf2}) in square brackets and the Floquet exponent through (\ref {floq2})
\be
k(\theta; P, q_1,q_2)=k(\theta ; P, q_2,q_1)=\sum _{n=0}^{+\infty} e^{2n\theta} r_0^{(2n)} (P,q_1,q_2)=P+\frac {4q_1q_2}{P(1-4P^2)}e^{2\theta}+O(e^{4\theta})
\label {kPexp}
\, . 
\ee
This relation in particular implies that the Floquet exponent of $\psi _-(y; q_1, q_2)=\psi _+(-y;q_2,q_1)$ is $-k(q_1,q_2)$, i.e. the opposite of the Floquet exponent of $\psi _+(y;q_1,q_2)$.

The study of $\varphi $ is not immediate as it involves an exact evaluation of $\Pi _+(y)$ for large $y$ inside the infinite domain integral (\ref {varphidefNf2}). As in the case of the MME, the expansion in a series around $\theta \rightarrow -\infty$ again conflicts with $y\rightarrow \pm \infty$ in the potential of the ODE (\ref {Nf2eq}). The solution is to first consider $\varphi _>$ and then to write it in terms of the right kink quantum momentum
\be
\Pi_> (y)=\Pi _+ (y-\theta) \label {rks2}
\ee
as $\varphi _>=\int _\theta ^{+\infty} dy \left [ \Pi _>(y)-e^y-q_1+\frac {1}{2} \right ] $. The function $\Pi_> (y)$ satisfies the Riccati equation
\be
\Pi _>(y) ^2 +\frac {d}{dy}\Pi _>(y)=e^{2y}+2q_1 e^y+P^2+2e^{2\theta} q_2 e^ {-y}+e^{4\theta}e^{-2y} \, ,
\label {Ricc>}
\ee
which can be solved perturbatively in the domain $y>\theta$. Before doing that, in order to have $k,\theta$ as independent variables, one inverts (\ref {kPexp}) to find $P^2$ in terms of $k^2,q_1,q_2$,
$P^2=\sum _{n=0}^{+\infty}e^{2n\theta} p^{(n)}_2 (k^2,q_1,q_2)$ and plugs this expansion into (\ref {Ricc>}). The  $2\pi i$-periodic functions $\tilde \Pi _>^{(n)} (y)$ entering the converging series  
\be
\Pi _>(y)=\sum _{n=0}^{+\infty} \tilde \Pi _>^{(n)} (y) e^{2n\theta} \label {thetaexpnf2}
\ee
satisfy the first order ODEs
\ba
&& \left (\tilde \Pi _>^{(0)}(y) \right )^2 +\frac {d}{dy}\tilde \Pi _>^{(0)}(y)=e^{2y}+2q_1 e^y+k^2 \, , \quad 
2 \tilde \Pi _>^{(0)}(y) \tilde \Pi _>^{(1)}(y)+\frac {d}{dy} \tilde \Pi _>^{(1)}(y)=2q_2 e^{-y} + \frac {8q_1q_2}{4k^2-1} \, , 
\nonumber  \\
&& 2  \tilde  \Pi _>^{(0)}(y)  \tilde \Pi _>^{(2)}(y)+\frac {d}{dy}  \tilde  \Pi _>^{(2)}(y)+\left (\tilde \Pi _>^ {(1)}(y) \right )^2=e^{-2y} +p^ {(2)}_2(k^2,q_1,q_2) \, , \label {Pi>equ2-n} \\ 
&& 2  \tilde \Pi _>^{(0)}(y)  \tilde \Pi _>^{(n)}(y)+\frac {d}{dy}  \tilde \Pi _>^{(n)}(y)+\sum _{m=1}^{n-1} \tilde \Pi _>^{(m)}(y)  \tilde \Pi _>^{(n-m)}(y)=p^ {(n)}_2 (k^2,q_1,q_2) \, ,  \ n\geq 3 \, , \nonumber 
\ea
which are solved iteratively. Since we are dealing with a Floquet solution, the zero-th order momentum in the first of (\ref {Pi>equ2-n}) has to be chosen as  
\be
\tilde \Pi _>^{(0)} (y)=\frac  {d}{dy} \ln \left [ \exp (-e^y +ky){}_1F_1 \left ( \frac {1}{2}+k+q_1,1+2k; 2e^y \right ) \right ] \, ,
\ee
since the confluent hypergeometric function $_1F_1$ is $2\pi i$-periodic and so $\tilde \Pi _>^{(0)} (y)$. The higher $\tilde \Pi _>^{(n)}(y)$, $n\geq 1$ are found by separation of variables, fixing the integration constant by integrability at $+\infty$. The first of these functions is
\be
\tilde \Pi _>^{(1)} (y)=\frac {2q_2}{1-4k^2} \frac  {d}{dy}\left (e^{-y}+2e^{-y}\tilde \Pi_>^{(0)}(y) \right )
\, 
\ee
and, in general, 
\be
\tilde \Pi _>^{(n)}(y;k)=\frac {d}{dy}  \left  [P_0^{(n)}(e^{-y})+\sum _{m=1}^{n}P_m^{(n)} (e^{-y})\frac {d^{m-1}}{dy^{m-1}} \tilde \Pi _>^{(0)}(y;k) \right ] \, , \label {solpin2}
\ee
with $P_m^{(n)}(e^{-y})$ degree $n$ polynomials\footnote {To avoid heavy notations, we use the same symbol as for the polynomials appearing in (\ref {solpin}) when solving the MME. The same abuse of notation will be used in (\ref {solpin4}).}: $P_m^{(n)}(e^{-y})=\sum \limits _{h=\textrm {max}\{m,1 \}}^n P_{m,h}^{(n)}(k^2) e^{-hy}$, with $ P_{m,h}^{(n)}(k^2)$ coefficients. Similarly to the case of the MME, they can be found in a recursive manner, by inserting (\ref {solpin2}) into (\ref {Pi>equ2-n}).
Eventually, $\varphi _>=\int _\theta ^{+\infty} dy \left [ \Pi _>(y)-e^y-q_1+\frac {1}{2} \right ]$ equals
\be
\varphi _>(\theta; k, q_1, q_2)=2e^\theta +\left (q_1-k-\frac {1}{2} \right )\ln (2e^\theta)+\ln \frac {\Gamma (1+2k)}{\Gamma \left (\frac {1}{2}+k+q_1\right )}-
\ln {}_1F_1 \left ( \frac {1}{2}+k+q_1,1+2k;2e^\theta\right )+ \tilde S_>^{(2)}(\theta ; k, q_1, q_2)  \label {varphi>} \, , 
\ee
with $\tilde S_>^{(2)}(\theta ; k, q_1, q_2) = \sum \limits _{n=1}^{+\infty}  e^{2n\theta} \int  _\theta ^{+\infty} dy \tilde \Pi _>^{(n)} (y) $.  Now, (\ref {solpin2}) implies the series expansion
\be
\tilde  \Pi _>^{(n)}(y) =\sum \limits
_{m=-n}^{+\infty} a_m^{(n)} e^{my} \, , \quad n\geq 1 \, , \quad a_0^{(n)}=0 \, , 
\ee
with $a_m^{(n)}$ constants\footnote {The constants $a_m^{(n)}$, with $n\geq 1$, satisfy recursion relations
\be
pa_p^{(n)}+\sum _{m=0}^n \sum _{h=-m}^{p+n-m}a_h^{(m)}a_{p-h}^{(n-m)}=L^{(n)}_p \, , 
\ee
where 
\be
L_p^{(1)}=\frac {8q_1q_2}{4k^2-1}\delta _{p,0}+2q_2 \delta _{p,-1} \, ,\quad L_p^{(2)}=p_2^{(2)}\delta _{p,0}+\delta _{p,-2} \, ,\quad
L_p^{(n)}=p_2^{(n)}\delta _{p,0} \ \  \textrm {if} \ \ n \geq 3 \, .  
\ee
}. This series can be integrated term by term around $-\infty$, but not for $y\rightarrow +\infty$. As in the MME case, it gives an estimate of the powers of 
$e^{\theta }$ coming from the evaluation of the primitive function 
\be
\sum \limits
_{m=-n}^{+\infty} \frac { a_m^{(n)} } {m}e^{my}  \label {primitiva2}
\ee
of $\tilde \Pi_>^{(n)}(y)$
at $y=\theta \rightarrow -\infty$ for $n\geq 1$: 
\be
e^{2\theta n}\int _{\theta}^{+\infty} dy \ \tilde  \Pi _>^{(n)}(y)  =e^{2\theta n}\left [ C_n - \sum _{\stackrel {m=-n} {m \not=0}}^{+\infty} \frac {a_m^{(n)}}{m}e^{\theta m} \right ]= e^{2\theta n} C_n - \sum _{\stackrel {q=n} {q \not=2n}}^{+\infty} \frac {a_{q-2n}^{(n)}}{q-2n}e^{\theta q}  \, , \label {powerexp2}
\ee
where $C_n$ is the value of the primitive function (\ref {primitiva2}) at $+\infty$. Result (\ref {powerexp2}) sheds light on the form of $\tilde S_>^{(2)}(\theta ,k)$, which, as a consequence of it, is expressed through positive (non-zero) powers of $e^{\theta}$.

The study of $\varphi _<$ proceeds on an analogous path.  It is written as
$\varphi _<(\theta ,k, q_1,q_2)=\int _{-\infty}^{-\theta} dy \left [ \Pi _<(y)+e^{-y}+q_2-\frac {1}{2} \right ]$, where the left kink quantum momentum
\be
\Pi_< (y)=\Pi _+ (y+\theta) \label {lks2}
\ee
satisfies the Riccati equation 
\be
\Pi _<(y)^2 +\frac {d}{dy}\Pi _<(y)=e^{-2y}+2q_2 e^{-y}+P^2+2e^{2\theta} q_1 e^ {y}+e^{4\theta}e^{2y} \, , \label {Ricc<}
\ee
with $P^2$ the function of $k$ coming from the inversion of (\ref {kPexp}). The $2\pi i$-periodic solution of (\ref {Ricc<}) is written for $y<-\theta$ as converging series $\Pi _<(y)=\sum _{n=0}^{+\infty} \tilde \Pi _<^{(n)} (y) e^{2n\theta}$, where $\tilde \Pi _<^{(n)} (y) $ satisfy the ODEs
\ba
&& \left ( \tilde \Pi _<^{(0)}(y) \right )^2 +\frac {d}{dy}\tilde \Pi _<^{(0)}(y)=e^{-2y}+2q_2 e^{-y}+k^2 \, , \quad 
2 \tilde \Pi _<^{(0)}(y) \tilde \Pi _<^{(1)}(y)+\frac {d}{dy} \tilde \Pi _<^{(1)}(y)=2q_1 e^{y}+\frac {8q_1q_2}{4k^2-1} \nonumber \, , \\
&& 2 \tilde \Pi _<^{(0)}(y) \tilde \Pi _<^{(2)}(y)+\frac {d}{dy} \tilde \Pi _<^{(2)}(y)+\left (\tilde \Pi _<^ {(1)}(y) \right )^2=e^{2y} + p^ {(2)}_2 (k^2,q_1,q_2) \, ,  \label {Pi<equ2-n} \\ 
&& 2 \tilde \Pi _<^{(0)}(y)\tilde \Pi _<^{(n)}(y)+\frac {d}{dy}\tilde \Pi _<^{(n)}(y)+\sum _{m=1}^{n-1}\tilde \Pi _<^{(m)}(y) \tilde \Pi _<^{(n-m)}(y)= p^ {(n)}_2 (k^2,q_1,q_2) \, . \nonumber 
\ea
Anew, the periodicity of the zero-th order momentum imposes
\be
\tilde \Pi _<^{(0)}(y)=\frac  {d}{dy} \ln \left [ \exp (-e^{-y} +ky) {}_1F_1 \left (\frac {1}{2}-k+q_2,1-2k;2e^{-y} \right )
  \right]\,  \label {Ric3bis} 
\ee
and the others $\tilde \Pi _<^{(n)}(y)$, $n\geq 1$, integrable at $-\infty$, come from the iterative solutions of the first orders ODEs (\ref {Pi<equ2-n}). From the form of (\ref {Ric3bis}) and of (\ref {Pi<equ2-n}), it follows\footnote {This relation implies that
\be
\tilde \Pi _<^{(n)}(y;k,q_1,q_2)=\frac {d}{dy}  \left  [P_0^{(n)}(e^{y};q_2,q_1)+\sum _{m=1}^{n}(-1)^m P_m^{(n)} (e^{y};q_2,q_1)\frac {d^{m-1}}{dy^{m-1}} \tilde \Pi _<^{(0)}(y) \right ] \, , \label {solpin2<}
\ee
with $P_m^{(n)} (e^{y};q_2,q_1)$ the same polynomials appearing in (\ref {solpin2}), with $q_1$ exchanged with $q_2$.} that $\tilde \Pi _<^{(n)} (y;k,q_1,q_2)=-\tilde \Pi _>^{(n)} (-y;-k,q_2,q_1)$ and then that $\varphi _<(\theta; k, q_1, q_2)=-\varphi _>(\theta;-k, q_2, q_1)$. 
Then, relation (\ref {varphi>}) implies
\ba
\varphi (\theta; k, q_1, q_2)&=&\varphi _>(\theta ; k, q_1,q_2)+\varphi _<(\theta ; k, q_1,q_2)=-(2k+q_2-q_1)\theta - (2k+q_2-q_1)\ln 2+\nonumber \\
&+&\ln \frac {\Gamma (1+2k)\Gamma \left (\frac {1}{2}-k+q_2 \right )}{\Gamma (1-2k)\Gamma \left (\frac {1}{2}+k+q_1 \right )}+S^{(2)}(\theta ; k, q_1, q_2)  \, , 
\label {phifinalbis}
\ea
where 
$S^{(2)}(\theta ; k, q_1, q_2) =\tilde S_>^{(2)}(\theta ; k, q_1, q_2) - \tilde S_>^{(2)}(\theta ; -k, q_2, q_1)
+ \ln \frac {{}_1F_1 \left (\frac {1}{2}-k+q_2,1-2k;2e^{\theta} \right )}{{}_1F_1 \left ( \frac {1}{2}+k+q_1,1+2k;2e^\theta\right )}$.
The function $S^{(2)}$, which contains positive powers of $e^\theta$, is however
bounded from functional relation (\ref {relfund}) to satisfy the relation $S^{(2)}(\theta +i\pi;k, q_1,q_2)=S^{(2)}(\theta ; k, q_1,q_2)$. This property implies that $S^{(2)}$ must expand only with positive powers of $e^{2\theta}$. 

From the analysis of (\ref {powerexp2}), it follows again that the $N$-instanton term, i.e. the term in  $S^{(2)}$ proportional to $e^{2\theta N}$ comes from $2N+1$ integrals: to be precise, from $\int _\theta ^{+\infty} dy \left [ \tilde \Pi _>^{(0)}(y)-e^y-q_1+\frac {1}{2} \right ]$ and from integrals (\ref {powerexp2}), with $1\leq n\leq 2N$, minus the same terms with $k\rightarrow -k$ and $q_1 \leftrightarrow q_2$.
In particular, to derive $S^{(2)}$ up to order $e^{2\theta}$, only $n=0,1,2$ are needed. The terms $n=0$ contributes through the expansion of $ \ln \frac {{}_1F_1 \left (\frac {1}{2}-k+q_2,1-2k;2e^{\theta} \right )}{{}_1F_1 \left ( \frac {1}{2}+k+q_1,1+2k;2e^\theta\right )}$ up to order $e^{2\theta}$, while $n=1$ through the integration of the exact solution of $\tilde \Pi _>^{(1)}$. Finally, the asymptotics $\tilde \Pi _>^{(2)}\simeq \frac {(2k-1)^2-4q_2^2}{(2k-2)(2k-1)^2}e^{-2y}$ when $y\rightarrow -\infty$ gives what comes from $n=2$. Summing everything gives
\be
S^{(2)}(\theta ; k, q_1, q_2) =\left [ \frac {4(q_1-q_2)}{4k^2-1}+\frac {32kq_1q_2}{(4k^2-1)^2}\right ]e^{2\theta}+O(e^{4\theta}) \, .
\ee
Therefore, the symmetrised phase $\Phi$ reads 
\be
\Phi (\theta; k, q_1,q_2)=-2\theta k  -2k\ln 2 +\ln \frac {\Gamma (1+2k)}{\Gamma (1-2k)}+\frac {1}{2}\ln \frac {\Gamma (1/2+q_1-k)\Gamma (1/2+q_2-k)}{\Gamma (1/2+q_1+k)\Gamma (1/2+q_2+k)}+  \frac {32kq_1q_2}{(4k^2-1)^2}e^{2\theta}+O(e^{4\theta})
\label {Phiformulakbis} \, .
\ee
The map $4e^\theta \hbar =\Lambda _2$ (instanton coupling), $\hbar q_j=m_j$, $\hbar k=a$ brings to conjecture the equality
\be
\Phi (\theta;k,q_1,q_2)= \frac {A_D (\Lambda _2,a,m_1m_2)}{\hbar} \label {relPhiAD2} \, , 
\ee
with $A_D$ the dual gauge period of ${\cal N}=2$ SYM with $N_f=2$, masses $m_j$ and gauge period $a$ in the NS background. 
Indeed, (\ref {Phiformulakbis}) reproduces the small instanton expansion of $A_D$.  The conjecture (\ref {relPhiAD2}) will be proved in an incoming publication \cite {FR-next1} .

Interestingly, as computing $\Pi$ at any finite value does not need the kink method, the computations of the $\varphi$s in this Section give also the only non trivial part of the Wronskian between Floquet solutions at $y=0$ (\ref {Wronsknf2}). Hence, it can be computed (and converges) at all orders; in particular at one-loop it reads
\be
W[\psi _+ ,\psi _-]=- (2e^\theta )^{1-q_1-q_2}\frac {\sin 2\pi k}{\pi }\left ( \prod _{i=1}^2 \Gamma \left (\frac {1}{2}+q_i+k \right )\Gamma \left (\frac {1}{2}+q_i-k \right )  \right )^{\frac {1}{2}}+\ldots \, ,
\label{Wronsknf2-expan}
\ee
from which the $Q$ function (\ref {Qfornf2}) enjoys the elegant (convergent) expansion
\be
Q(\theta; k, q_1, q_2)=2\pi \left (  \left ( \prod _{i=1}^2 \frac {1}{\Gamma \left (\frac {1}{2}+q_i+k \right )\Gamma \left (\frac {1}{2}+q_i-k \right ) } \right )^{\frac {1}{2}} +\ldots \right )\frac { \sinh \Phi (\theta; k,  q_1,q_2) } {\sin 2\pi k} \, . 
\ee
In both these formul{\ae} dots stand for computable instanton corrections.

\section{The HE: from decaying to Floquet solutions}
\label{Nf4}
\setcounter{equation}{0}

The HE reads
\begin{equation}\label{ODEy}
    \frac {d^2}{dy^2 } \psi (y)=V(y) \psi(y)  \, , 
\end{equation}
where the potential 
\ba
&&    V(y) = -\frac{1}{4 \left(e^{ \theta }-4 e^{\frac {\theta}{2}}\cosh y+4\right)^2}\Biggl[
    -16 \left(e^{\theta }+4\right)P^2 -16 e^{\theta } + 24 e^{\theta } \left(q_1 q_2+q_3 q_4\right)-e^{2\theta } \left(q_1^2+q_2^2+q_3^2+q_4^2\right) +\nonumber \\
    &&+4 e^{\frac {\theta}{2} +y}\left(\frac{e^{ \theta }}{2}+e^{\theta } q_1^2-\left(e^{\theta
    }+8\right) q_2 q_1+e^{\theta } q_2^2-e^{ \theta } q_3 q_4+8 P^2 +2\right)+4 e^{\frac {\theta}{2} -y} \left(\frac{e^{ \theta }}{2}+e^{\theta } q_3^2-\left(e^{
    \theta }+8\right) q_4 q_3 + \right. \nonumber \\
    &&\left. +e^{ \theta } q_4^2-e^{ \theta } q_1 q_2+8
    P^2+2\right) -4 \left(q_1-q_2\right)^2 e^{\theta +2 y}-4
    \left(q_3-q_4\right)^2 e^{\theta -2 y}\Biggr]\, \label {Vy} 
\ea
has four regular singular points at $y = \pm \infty , \pm (\frac {\theta}{2} -\ln 2)$.
This equation is the quantisation of SW differential for ${\cal N}=2$ SYM with $N_f=4$ in the NS background. On the gravity side, it is worth mentioning that the (angular and radial) Teukolsky equations for Kerr
Newman-de Sitter geometry can be transformed \cite {STU} into the HE. 
Defining $q_+ = (q_1, q_2)$ and $q_- = (q_3, q_4)$,
the equation (\ref{ODEy}) has the Carter symmetry
$y\mapsto -y$ and $q_+\leftrightarrow q_-$. Other important symmetries are $\hat \Omega: q_3 \rightarrow -q_3, q_4 \rightarrow -q_4$ and 
$\hat \Lambda: q_1 \rightarrow -q_1, q_2 \rightarrow -q_2$. 
Choosing $q_1>q_2, q_1+q_2>1,q_3>q_4, q_3+q_4>1$, the ODE/IM solutions are uniquely defined by their asymptotics around the singular points 
\ba
\label{asymptotics}
   &&\psi _{-\infty,0}  \simeq \exp\left (\frac{q_3 - q_4}{2}y\right ) \ \textrm{for } y \to -\infty\,, \quad
    \psi _{+\infty,0} \simeq \exp\left (-\frac{q_1 - q_2}{2}y\right )\  \textrm{for } y \to +\infty\, , \nonumber \\
    &&\psi _{-y_0,0} \simeq (e^{-y} - e^{y_0})^{\frac{1}{2}(1+q_3+q_4)} \ \text{for } y \to -y_0\,, \quad
    \psi _{y_0,0} \simeq (e^{y} - e^{y_0})^{\frac{1}{2}(1+q_1+q_2)} \ \text{for } y \to y_0\, , 
\ea
where $y_0 = |\frac {\theta}{2} -\ln 2|$. Floquet solutions are different from the ODE/IM ones. Their property is that they are modified only by a phase factor when the variable $e^y$ rotates around two regular singularity. For instance the Floquet solution $\psi _+$ satisfies
\begin{equation}
\psi_+(y+2\pi i ; q_+, q_-) = e^{2\pi i k(q_+,q_-)}\psi_+(y; q_+, q_-) 
\quad \text{for } |y| < y_0\, ,
\end{equation}
where $k$ is the Floquet index in the region $[-y_0,y_0]$. It 
is constructed from the solution $\Pi _+(y)=\frac {d}{dy}\ln \psi _+(y) $ of the Riccati equation
\be
\Pi (y)^2 +\frac {d}{dy}\Pi(y) = V(y) \, , \label  {Ricnf4}
\ee
which satisfies $\Pi _+(y+2\pi i)=\Pi _+(y)$ and has the asymptotics
\be
 \Pi_+ \stackrel {y\rightarrow -y_0}{\simeq}\frac{\frac{ e^{-y_0}}{2}(1-q_3-q_4)}{e^y - e^{-y_0}} \, , \quad 
  \Pi_+ \stackrel {y\rightarrow y_0}{\simeq}\frac{\frac{e^y}{2} (1-q_1-q_2)}{e^y - e^{y_0}} \, .
\ee
As usual, 
\be
2\pi i k=  \int _{y}^{y+2i\pi}  dy' \Pi _+(y') \, 
\label {floqnf4}
\ee
and ($e^c$ is a normalisation constant)
\begin{equation}
    \psi_{+}(y; q_+, q_-) = e^c \left (e^{-y}-e^{y_0}\right ) ^{\frac {1-q_3-q_4}{2}} \exp \left \{ \int_{-y_0}^y d{y^\prime}\left [\Pi_+(y^\prime; q_+, q_-)- \frac {\frac {e^{-y_0}}{2}(1-q_3-q_4)}{e^{y'}-e^{-y_0}}\right ] \right \} \, .
\end{equation}
As $\psi _+$ is different from the functions of the ODE/IM basis, it diverges at 
$\pm y_0$. In particular 
\be
 \psi_+  \simeq e^c \left (e^{-y}-e^{y_0}\right ) ^{\frac {1-q_3-q_4}{2}}\text{for } y \to -y_0\,,\quad \psi_+  \simeq e^{\varphi} e^c \left (e^y-e^{y_0}\right ) ^{\frac {1-q_1-q_2}{2}}\ \text{for } y \to y_0\, , \label {floqasy}
\ee
where $\varphi$, which has the meaning of acquired phase when going from $-y_0$ to $y_0$, reads  
\begin{multline}
    \varphi(\theta; k,q_+,q_-) = 
    \int_{-y_0}^{0}dy \left [\Pi_+(y; k, q_+, q_-)-\frac{\frac {e^{-y_0}}{2}(1-q_3-q_4)}{e^{y} - e^{-y_0}}\right]+\frac {1}{2}(1-q_3-q_4)\ln (1-e^{y_0}) +
    \\+\int_{0}^{y_0}dy \left [\Pi_+(y; k , q_+, q_-)-\frac{\frac {e^{y}}{2}(1-q_1-q_2)}{e^{y} - e^{y_0}}\right ]- \frac {1}{2} (1-q_1-q_2) \log(1-e^{y_0}) \, . \label {varphiHeun}
\end{multline}
The first line identifies $\varphi _<$, the second $\varphi _>$.
From the symmetry $\Pi _+(y-i\pi; \theta +2\pi i)=\Pi _+(y,\theta )$, valid for $y>0$, we find the property 
\be
\varphi _>(\theta +2\pi i;k,q_+,q_-)=\varphi _>(\theta;k,q_+,q_-)-\int _0^{i\pi} dy \Pi _+(y) +\frac {1-q_1-q_2}{2} \, \ln \left ( \frac {1-e^{y_0}}{e^{y_0}-1} \right ) \, .
\ee
From the symmetry $\Pi _+(y+i\pi, \theta +2\pi i)=\Pi _+(y,\theta )$, valid for $y<0$, we find the property 
\be
\varphi _<(\theta +2\pi i;k,q_+,q_-)=\varphi _<(\theta;k,q_+,q_-)-\int _{-i\pi}^0 dy \Pi _+(y) +\frac {1-q_3-q_4}{2}\, \ln \left ( \frac {e^{y_0}-1}{1-e^{y_0}} \right ) \, .
\ee
Summing the last two expressions furnishes
\be
\varphi (\theta +2\pi i;k,q_+,q_-)=\varphi (\theta; k,q_+,q_-)-2i\pi k +\frac {q_3+q_4-q_1-q_2}{2} \, \ln \left ( \frac {1-e^{y_0}}{e^{y_0}-1} \right )\, .
\label {shiftnf4}
\ee
The other Floquet solution is defined\footnote {The solution $\psi _-(y;q_+,q_-)$ is related to the solution of the Riccati equation (\ref {Ricnf4}) $\Pi _-(y;q_+,q_-)=-\Pi _+ (-y;q_-,q_+)$ by $\Pi _-(y;q_+,q_-)=\frac {d}{dy} \ln \psi _-(y;q_+,q_-)$.} by 
\be
\psi_-(y; q_+, q_-) = \psi_+(-y; q_-, q_+) \label {Floq-def}
\ee
and then satisfies $\psi_-(y +2\pi i; q_+, q_-) = e^{-2\pi i k(q_-,q_+)}\psi_-(y; q_+, q_-)$ 
for $|y |< y_0$.
To have more symmetric formul{\ae}, it is convenient to choose from now on $c=-\frac {1}{2} \varphi$. Then, the connection between the functions of the Floquet basis and two functions of the ODE/IM basis (\ref {asymptotics}) is done in the same spirit of cancelling diverging orders as (\ref {chgbasis}) for the MME and (\ref {nf2psi}) for the DCHE. For instance, the relation between $\{ \psi _{y_0,0} ,  \psi_{-y_0,0} \}$ and $\{ \psi _+, \psi _-\} $ is   
\ba
\psi _{y_0,0} (y)&=&\frac {1}{s(q_-,q_+)}\left [ e^{\frac {\varphi (\theta ; k, q_-,q_+)}{2}}\psi _-(y;q_+,q_-)- e^{-\frac {\varphi (\theta;k,  q_+,q_-)}{2}}\psi _+(y;q_+,q_-)  \right  ] \, , \nonumber \\
\psi _{-y_0,0}(y)&=&\frac {1}{s(q_+,q_-)}\left [e^{\frac {\varphi (\theta ;k,  q_+,q_-)}{2}}\psi _+(y;q_+,q_-)- e^{-\frac {\varphi (\theta ;k,  q_-,q_+)}{2}}\psi _-(y;q_+,q_-)  \right  ] \, ,  \label {nf4psi-one}
\ea
where $s$ is a constant. To determine this constant, it is useful to write the connection coefficient between $\psi _{y_0,0}$ and $\psi _{-y_0,0}$,  $W[\psi _{y_0,0},\psi _{-y_0,0}]$, by means of (\ref {nf4psi-one}):
\be
W[\psi _{y_0,0},\psi _{-y_0,0}]=-\frac {2 W[\psi _+,\psi _-] (q_+,q_-) \sinh \Phi (q_+,q_-)}{s(q_+,q_-)s(q_-,q_+)} \, ,  \quad \Phi (q_+,q_-)=\frac {1}{2} [\varphi (q_+,q_-)+\varphi (q_-,q_+) ] \label {wrone} \, .
\ee
Then, by using the basis $\{ \psi _{-y_0,0}, \psi _{-y_0,1}=\hat \Omega \psi _{-y_0,0}\}$, 
we can write
\be
\psi _{y_0,0}(y)=\Upsilon _0 \, \psi _{-y_0,0}(y)+\Upsilon _1 \, \psi _{-y_0,1}(y) \label {nf4-chng} \, , 
\ee
with $\Upsilon _0, \Upsilon _1$ constants. $\Upsilon _1 $ can be computed in terms of $s$ by comparing the first of (\ref {nf4psi-one}) with (\ref {nf4-chng}) in the limit $y \rightarrow -y_0$ and 
using the facts that, when $y \rightarrow -y_0$, $\psi _{-y_0,0}\rightarrow 0$, $\psi _{-y_0,1}\simeq (e^{-y} - e^{y_0})^{\frac{1}{2}(1-q_3-q_4)}$, whilst (see (\ref {floqasy})) $\psi _\pm \simeq e^{\mp \frac {\varphi (q_\pm, q_\mp)}{2}}(e^{-y} - e^{y_0})^{\frac{1}{2}(1-q_3-q_4)}$:
\be
\Upsilon _1=\frac {e^{\varphi (q_-,q_+)}-e^{-\varphi (q_+,q_-)}}{s(q_-,q_+)} \, .
\ee
Then, we have the chain of equalities 
\be
W[\psi _{y_0,0},\psi _{-y_0,0}]=\Upsilon _1 \,  W[\psi _{-y_0,1},\psi _{-y_0,0}]=-\frac {e^{\varphi (q_-,q_+)}-e^{-\varphi (q_+,q_-)}}{s(q_-,q_+)} (q_3+q_4) e^{y_0} \label {wrtwo} \, ,
\ee
where we used the relation $ W[\psi _{-y_0,1},\psi _{-y_0,0}]=-(q_3+q_4)e^{y_0}$. 
Comparing (\ref {wrone}) with (\ref {wrtwo}), we find 
\be
\frac {1}{s(q_+,q_-)}=\frac {(q_3+q_4)e^{y_0}e^{\frac {\varphi (q_-,q_+)-\varphi (q_+,q_-)}{2}}}{W[\psi _+,\psi _-](q_+,q_-)} \, , 
\ee
which allows us to write the connection formul{\ae}
\ba
\psi _{y_0,0} (y)&=&\frac {(q_1+q_2)e^{y_0}}{W[\psi _+,\psi _-](q_-,q_+)}\left [ e^{\frac {\varphi (\theta ; k, q_+,q_-)}{2}}\psi _-(y;q_+,q_-)- e^{-\frac {\varphi (\theta; k, q_-,q_+)}{2}}\psi _+(y;q_+,q_-)  \right  ] \, , \nonumber \\
\psi _{-y_0,0}(y)&=&\frac {(q_3+q_4)e^{y_0}}{W[\psi _+,\psi _-](q_+,q_-)}\left [e^{\frac {\varphi (\theta ; k, q_-,q_+)}{2}}\psi _+(y;q_+,q_-)- e^{-\frac {\varphi (\theta ; k, q_+,q_-)}{2}}\psi _-(y;q_+,q_-)  \right  ] \, ,  \label {nf4psi}
\ea
and the connection coefficient
\be
Q(\theta; k,  q_+,q_-)=W[\psi _{y_0,0},\psi _{-y_0,0}]=-(q_3+q_4)(q_1+q_2)e^{2y_0} \frac {2 \sinh \Phi(\theta; k,  q_+,q_-)}{W[\psi _+,\psi _-](q_-,q_+)} \,  \label {Qnfequal4}
\ee
in terms of the Wronskian $W[\psi _+,\psi _-]$ between Floquet solutions.
Similarly to the DCHE case, the Wronskian between Floquet solutions can be given the form
\be
W[\psi _+,\psi _-](q_+,q_-)=-e^{\frac {1}{2}\varphi _<(\theta ; q_+,q_-)+\frac {1}{2}\varphi _<(\theta ; q_-,q_+)-\frac {1}{2}\varphi _>(\theta ; q_+,q_-)-\frac {1}{2}\varphi _>(\theta ; q_-,q_+)}[\Pi _+(0;q_+,q_-)+\Pi _+(0;q_-,q_+)] \, ,
\label{Wronsknf4}
\ee
which in particular shows its symmetry under the exchange of $q_+$ with $q_-$.
As for the previous cases, in (\ref {nf4psi}) the relative phase factor between the coefficients in front of $\psi _\pm$ depends on the symmetrised phase $\Phi (\theta; k,q_+,q_-)$, whilst the complete change of basis needs also the knowledge of $\varphi _>, \varphi _<$ and $\Pi _+(0)$.
Everything is computable in converging series when $\theta \rightarrow -\infty$, as results of next subsection will show with the important difference that the computation of $\Pi _+(0)$ is easier as it does not need the kink method. 

We discussed the change from the basis given by $\{ \psi _{y_0,0} ,  \psi_{-y_0,0} \}$ to the Floquet basis and found the connection coefficient $Q=W[\psi _{y_0,0} ,  \psi_{-y_0,0}]$. The same computations can be repeated {\it mutatis mutandis} starting from another arbitrary pair from the ODE/IM functions (\ref {asymptotics}).  In this way we can compute all the connection coefficients of the Heun equations. It is important to remark that in the paper \cite {BILT} some of the connection coefficients of Heun equations are computed by using AGT correspondence. Our findings agree with their formul{\ae} at first orders in $e^\theta$ as $\theta \rightarrow -\infty$.

\subsection {Converging expansion around $\theta \rightarrow -\infty$}
\label{Nf4-convexp}

For $\theta \rightarrow -\infty$ and finite $y$ inside the interval $|y|<y_0$, $y_0=\ln 2 -\frac {\theta}{2}$, the Floquet momentum expands as
\ba
\Pi _+(y)&\simeq & P+\left [\left (\frac {2P-1}{8}+\frac {q_1q_2}{4P+2}\right )e^y+\left (\frac {2P+1}{8}+\frac {q_3q_4}{4P-2}\right )e^{-y}\right ]e^{\theta/2} + \nonumber \\
&+& \left [C e^{2y}+De^{-2y}+\frac {1}{2P}\left (\frac {4P^2+1}{32}-\frac {q_1q_2q_3q_4}{2(4P^2-1)} \right ) \right ]e^\theta +O(e^{3\theta/2}). 
\ea
The form of $C,D$ is not relevant for our aim, which is the use of (\ref {floqnf4}) to find the Floquet index in the region $[-y_0,y_0]$ as a function of $P$:
\be
k(q_+,q_-)=k(q_-,q_+)=  \frac {1}{2\pi i}\int _{0}^{2i\pi}  dy' \Pi _+(y')=P+\frac {1}{2P}\left (\frac {4P^2+1}{32}-\frac {q_1q_2q_3q_4}{2(4P^2-1)} \right )e^\theta +O(e^{2\theta}) \, . \label {knf4exp}
\ee
Of course, a more efficient recursive relation of the form (\ref{rec-r-nf2}) can be easily derived in this case as well. Immediately, the previous expression implies that the Floquet exponent of $\psi_-(y, q_+, q_-) = \psi_+(-y, q_-, q_+)$ is $-k(q_+,q_-)$, i.e. the opposite of the Floquet exponent of $\psi _+$.
The inversion of (\ref {knf4exp}) gives $P^2=k^2+\left (\frac {q_1q_2q_3q_4}{2(4k^2-1)}-\frac {4k^2+1}{32} \right ) e^\theta +O(e^{2\theta}) $ and, using this expression, $V(y)$ is made dependent parametrically on $k^2$. The quantity $\varphi _>(q_+,q_-)$ is written in terms of the `right kink' quantum momentum
\be
\Pi_>(y)=\Pi _+ \left (y-\frac {\theta}{2} \right )
\ee
as 
\be
\varphi _>(\theta;k, q_+,q_-)= \int_{\frac {\theta}{2}}^{\ln 2}dy \left [\Pi_>(y; k, q_+, q_-)-\frac{\frac {e^{y}}{2}(1-q_1-q_2)}{e^{y} - 2}\right ]- \frac {1}{2} (1-q_1-q_2)\log(1-e^{y_0}) \label {phi4>}
\ee
and $\Pi _>$ satisfies the modified Riccati equation
\be
\Pi_>(y)^2 + \frac {d}{dy}\Pi _>(y)=V_>(y) \equiv V\left (y-\frac {\theta}{2} \right ) \, . \label {kinknf4>}
\ee
We are interested (see (\ref {phi4>})) to the region $\frac {\theta }{2} <y<\ln 2 $, where we can expand in converging series 
$V_>(y)=\sum \limits _{n=0}^\infty V _>^{(n)}(y) e^{n\theta}$ and 
\be
\Pi_>(y)=\sum _{n=0}^\infty \tilde \Pi _>^{(n)}(y) e^{n\theta} \label  {thetaexpnf4} \, .
\ee
The zero-th order momentum $ \tilde \Pi _>^{(0)}(y)$ is the $2\pi i$-periodic solution to
\be
\left ( \tilde \Pi_>^{(0)}(y) \right )^2+\frac {d}{dy} \tilde \Pi _>^{(0)}(y)=V_>^{(0)}(y) \, , 
\quad
V_>^{(0)}(y)=\frac {16k^2-2e^{y}(1+4k^2-4q_1q_2)+e^{2y}(q_1-q_2)^2}{4(e^{y}-2)^2} \label {nf4Ricc>}
\, 
\ee
and then is expressed in terms of an hypergeometric function:
\begin{equation}
 \tilde \Pi _> ^{(0)}(y )= \frac {d}{dy} \ln \left [ (e^{y}-2)^{\frac {1-q_1-q_2}{2}}e^{ky} {}_2F_1\left (\frac {1}{2}+k-q_1,\frac {1}{2}+k-q_2;1+2k;\frac {e^{y}}{2} \right ) \right ] \, . \label  {nf4lead}
\end{equation}
The first correction $ \tilde \Pi _> ^{(1)}(y )$ satisfies 
\be
2  \tilde \Pi _> ^{(0)}(y )  \tilde \Pi _> ^{(1)}(y ) +  \frac {d}{dy} \tilde \Pi _> ^{(1)}(y ) =V_>^{(1)}(y) \, , \quad
V_>^{(1)}(y)=\frac {(4k^2-1+4q_3q_4)\left (1-4e^{-y}-\frac {4q_1q_2}{4k^2-1}\right )}{16(e^y-2)} \label {Pi>equ4-1}
\ee
and has the form
\be
 \tilde \Pi _> ^{(1)}(y )= \left (-\frac {1}{4}+\frac {q_3q_4}{1-4k^2}\right )\frac {d}{dy}  \left [\frac {1}{2}e^{-y}+\left (e^{-y}-\frac {1}{2} \right ) \tilde \Pi _> ^{(0)}(y ) \right ] \, .  \label  {nf4subl}
\ee
For $n\geq 2$ the equation is 
\be
2  \tilde \Pi _>^{(0)}(y)  \tilde \Pi _>^{(n)}(y)+\frac {d}{dy}  \tilde \Pi _>^{(n)}(y)+\sum \limits_{m=1}^{n-1} \tilde \Pi _>^{(m)}(y)  \tilde \Pi _>^{(n-m)}(y)=V^ {(n)}_> (y) \, \label 
{Pi>equ4-n}
\ee
 and the claim is that for $n\geq 1$
\be
\tilde \Pi _>^{(n)}(y;k)=\frac {d}{dy}  \left  [P_0^{(n)}(e^{-y})+\sum _{m=1}^{n}P_m^{(n)} (e^{-y})\frac {d^{m-1}}{dy^{m-1}} \tilde \Pi _>^{(0)}(y;k) \right ] \, , \label {solpin4}
\ee
with $P_m^{(n)}(e^{-y})$ polynomials of degree $n$ in the variable\footnote {The correct behaviour of $\Pi _>(y)$ when $y\rightarrow \ln 2$ is $\Pi _>(y) \simeq \frac {1-q_1-q_2}{e^y-2}$. This behaviour is already reproduced by the zero instanton term (\ref {nf4lead}). Then, instanton corrections $\sum _{n=1}^\infty \tilde \Pi _>^{(n)}(y) e^{n\theta}$ should be regular when $y=\ln 2$. Since the $(m-1)$-th derivative of $ \tilde \Pi _>^{(0)}$ has a $m$-th order pole at $y=\ln 2$, it follows that $P_m^{(n)}(e^{-y})$ has an $m$-th order zero when $y=\ln 2$. Then, we can claim the form $P_m^{(n)}(e^{-y})= \sum \limits_{h=\textrm {max}\{m,1 \}}^{n} P_{m,h}^{(n)}(k^2,q_+,q_-) \left (e^{-y}-\frac {1}{2} \right )^h$, with $P_{m,h}^{(n)}$ coefficients.} $\left (e^{-y}-\frac {1}{2} \right )$. Again, they
can be found in a recursive manner, by inserting (\ref {solpin4}) into (\ref {Pi>equ4-n}).
Then, the quantity $\varphi_>$ (\ref  {phi4>}) can be written as 
\be
\varphi _>=k\left (\ln 2 -\frac {\theta}{2} \right )-  \frac {1}{2} (1-q_1-q_2)\ln (1-2e^{-\frac {\theta}{2}})+\ln \frac {\Gamma (1+2k)\Gamma (q_1+q_2)}{\Gamma \left (\frac {1}{2}+k+q_1 \right ) \Gamma \left (\frac {1}{2}+k+q_2 \right )}+ \tilde S^{(4)}_>(\theta ; k) \, , 
\ee
with $ \tilde S^{(4)}_>(\theta ; k) =\sum \limits _{n=1}^{+\infty }e^{n\theta} \int _{\frac {\theta}{2}} ^{\ln 2} dy \tilde \Pi _>^{(n)}(y) $.
In general, for $n\geq 1$ the expression (\ref {solpin4}) implies that the expansion $\tilde  \Pi _>^{(n)}(y) =\sum \limits
_{m=-n}^{+\infty} a_m^{(n)} e^{my}$, with $a_m^{(n)}$ constants and $a_0^{(n)}=0$, holds. 
This series can be integrated term by term around $y=-\infty$, but not for $y\rightarrow \ln 2$
\be
e^{\theta n}\int _{\frac {\theta}{2}}^{\ln 2} dy \, \tilde  \Pi _>^{(n)}(y)  =e^{\theta n}\left [ C_n - \sum _{\stackrel {m=-n} {m \not=0}}^{+\infty} \frac {a_m^{(n)}}{m}e^{\frac {\theta} {2}  m} \right ]= e^{\theta n} C_n - \sum _{\stackrel {q=n} {q \not=2n}}^{+\infty} \frac {a_{q-2n}^{(n)}}{q-2n}e^{\frac {\theta}{2}  q}  \, , \label {powerexp4}
\ee
where $C_n$ is the value of the primitive function $\sum \limits
_{m=-n}^{+\infty} \frac  {a_m^{(n)}}{m} e^{my}$ at $y=\ln 2$. This result implies that $\tilde S^{(4)}_>(\theta ,k) $ expands in positive powers of $e^{\frac {\theta}{2}}$. 
On the other hand, $\varphi _<(q_+,q_-)$ is written in terms of the left kink quantum momentum
\be
\Pi _<(y)=\Pi _+ \left (y+\frac {\theta}{2} \right) 
\ee
 as
\be
\varphi _<(\theta; k, q_+,q_-)=
\int_{-\ln 2}^{-\frac {\theta}{2}}dy \left [ \Pi_<(y; k , q_+, q_-)-\frac{\frac {1}{2}(1-q_3-q_4)}{2e^{y}- 1}\right ]+ \frac {1}{2} (1-q_3-q_4)\log\left (1-e^{y_0} \right ) \, .
\ee
The function $\Pi _<(y)$ satisfies the other modified Riccati equation 
\be
\Pi_<(y)^2 + \frac {d}{dy}\Pi _<(y)=V\left (y+\frac {\theta}{2} \right )=V_<(y) \label {nf4Ricc<} \, .
\ee
Similarly to the other cases, it is solved by the simple symmetry 
\be
\Pi _< (y;k, q_+,q_- )=-\Pi _> (-y;-k,q_-,q_+ ) \, , \label {SYMM}
\ee
which implies that $\varphi _<(\theta; k,q_+,q_-)=-\varphi _>(\theta; -k, q_-,q_+)$. The sum $\varphi _> + \varphi _<$ gives
the acquired phase
\ba
\varphi (\theta ; k,q_+, q_-) &=&-\theta k+ 2k\ln 2 +\ln \frac {\Gamma (1+2k) \Gamma (q_1+q_2) \Gamma (1/2-k+q_3) \Gamma (1/2-k+q_4)}{\Gamma (1-2k) \Gamma (q_3+q_4) \Gamma (1/2+k+q_1) \Gamma (1/2+k+q_2)}+\nonumber \\
&+& \frac {q_1+q_2-q_3-q_4}{2}\ln \left (1-2e^{-\frac {\theta}{2}} \right )+ S^{(4)}(\theta ; k,q_+,q_-) \, , 
\ea 
where, using that $\tilde S^{(4)}_< (\theta ; k,q_+,q_-)=-\tilde S^{(4)}_> (\theta ;- k, q_-,q_+)$,
\be
S^{(4)}(\theta ; k,q_+,q_-)=\tilde S^{(4)}_>(\theta ; k,q_+,q_-)+\tilde S^{(4)}_< (\theta ; k,q_+,q_-)+\ln \frac {  {}_2F_1\left (\frac {1}{2}-k-q_3,\frac {1}{2}-k-q_4;1-2k; \frac {e^{\frac {\theta}{2}}}{2}    \right ) }{{}_2F_1\left (\frac {1}{2}+k-q_1,\frac {1}{2}+k-q_2;1+2k;\frac {e^{\frac {\theta}{2}}}{2} \right ) } \label {Sfour} \, .
\ee
Now, the functional relation (\ref  {shiftnf4}) implies that $S^{(4)}(\theta ; k)+\frac {q_3+q_4-q_1-q_2}{2}\ln \left (1-\frac {1}{2}e^{\frac {\theta}{2}}\right )$ is $2\pi i$-periodic and then must contain only integer powers of $e^\theta$. Then, the symmetrised phase 
\be
\Phi (\theta;k,q_+,q_-)=\frac {1}{2} [\varphi (\theta;k,q_+,q_-)+\varphi (\theta;k,q_-,q_+) ] \label {symm-phase-Nf4}
\ee
reads
\be
\Phi (\theta; k, q_+,q_-)=-\theta k +2k \ln 2+\ln \frac {\Gamma (1+2k)}{\Gamma (1-2k)}+\frac {1}{2}\sum _{i=1}^4 \ln \frac {\Gamma \left (\frac {1}{2}-k+ q_i \right )}{\Gamma \left (\frac {1}{2}+k+ q_i \right )}+\frac {1}{2}[S^{(4)}(\theta ;k,q_+,q_-)+S^{(4)}(\theta ; k,q_-,q_+) ] \, ,  \label  {nf4Phigen}
\ee
with $\frac {1}{2}[S^{(4)}(\theta ;k,q_+,q_-)+S^{(4)}(\theta ;k,q_-,q_+) ]$ containing only integer positive powers of $e^\theta$.
For the quantity $\Phi$, the term $e^{N\theta}$ ($N\geq 1$), the so-called $N$-instanton contribution, comes from the last term in (\ref {Sfour}) (symmetrised upon the exchange $q_+ \leftrightarrow q_-$) and from a finite number ($2N$) of terms, coming from $\sum \limits_{n=1}^{2N}e^{n\theta} \int _{\frac {\theta}{2}} ^{\ln 2} dy \tilde \Pi _>^{(n)}(y) $, contained in $\tilde S^{(4)}_>(\theta ,k)$.
In particular, the computation of the symmetrised phase $\Phi (q_+,q_-)=\frac {1}{2} [\varphi (q_+,q_-)+\varphi (q_-,q_+) ]$ up to one instanton needs the knowledge of $\tilde \Pi _>^{(1)}$ and of the limit
for $y\rightarrow -\infty$ of $ \tilde \Pi _> ^{(2)}(y): a_>e^{-2y}$, with
\be
a_>=\frac {q_3q_4}{4(2k-1)}+\frac {12k^2-4k-9}{64(2k-2)}+\frac {k(q_3^2+q_4^2)}{8(2k-1)^2}
+\frac {q_3^2+q_4^2-4q_3^2q_4^2}{16(2k-2)(2k-1)^2} \, .
\ee
The result is
\be
\Phi (\theta; k, q_+,q_-)=-\theta k +2k \ln 2+\ln \frac {\Gamma (1+2k)}{\Gamma (1-2k)}+\frac {1}{2}\sum _{i=1}^4 \ln \frac {\Gamma \left (\frac {1}{2}-k+ q_i \right )}{\Gamma \left (\frac {1}{2}+k+ q_i \right )}+\left (-\frac {k}{8}+\frac {2kq_1q_2q_3q_4}{(1-4k^2)^2} \right )e^\theta + O(e^{3\theta/2})   \, . \label  {nf4Phi1inst}
\ee
The replacements $\frac {e^\theta \hbar}{4}=\Lambda _4 $ (instanton coupling), $\hbar q_j=m_j$, $\hbar k=a$ led to conjecture that
\be
\Phi (\theta;  k, q_+,q_-)=\frac {A_D(\Lambda _0,a,m_j)}{\hbar} \, , \label {PhiAD4} 
\ee
with ${A_D} $ the dual gauge period of ${\cal N}=2$ SYM with $N_f=4$, with masses $m_j$ and gauge period $a$, in the NS background. Indeed, expression (\ref {nf4Phi1inst}) reproduces the small instanton expansion of $A_D/\hbar$.  The proof of relation (\ref {PhiAD4}) will be addressed in an incoming publication \cite {FR-next1}.

As above, the computations of the $\varphi$s in this Section give the only non trivial part of the  Wronskian between Floquet solutions at $y=0$ (\ref {Wronsknf4}) (as computing $\Pi$ at any finite value does not need the kink method). It becomes a converging series at all orders which starts at one-loop as
\be
W[\psi _+ ,\psi _-]=- (-2e^{- \frac {\theta}{2}})^{1-\frac {\sum _{i=1}^4 q_i}{2}}\frac {\sin 2\pi k}{\pi }\frac {1}{\Gamma (q_1+q_2)\Gamma (q_3+q_4)}\left ( \prod _{i=1}^4 \Gamma \left (\frac {1}{2}+q_i+k \right )\Gamma \left (\frac {1}{2}+q_i-k \right )  \right )^{\frac {1}{2}}+\ldots 
\label{Wronsknf4-expan}
\ee
and entails for the $Q$ function (\ref{Qnfequal4}) the elegant (convergent) expansion
\ba
Q(\theta; k,  q_+,q_-)&=&2\pi  \left (    \frac {\Gamma (1+q_1+q_2) \Gamma (1+q_3+q_4) \left (-2e^{-\frac {\theta}{2}}\right )^{1+\sum _{i=1}^4 \frac {q_i}{2}}}{\left [ \prod \limits_{i=1}^4  \Gamma \left (\frac {1}{2}+q_i+k \right )\Gamma \left (\frac {1}{2}+q_i-k \right ) \right ] ^{\frac {1}{2}} }  +\ldots \right ) \cdot \nonumber \\
&\cdot & \frac { \sinh \Phi (\theta; k,q_+,q_-) } {\sin 2\pi k} \, . 
\ea
In both these formul{\ae} dots stand for (computable) instanton corrections.

\subsection {The confluent limit to the CHE}
\label {CHE-limit}

The 'confluence' limit from the HE (\ref {ODEy}, \ref {Vy}) to the CHE is made through the following steps. First, one defines the new independent variable $y'=y+\frac {\theta}{2}$, in which the two finite (for $\theta $ finite) singularity points occur at $y'=\theta -\ln 2$ and $y'=\ln 2$, respectively. The second step is to perform on the (Heun) equation in the variable $y'$ the scaling limit  $\theta \rightarrow -\infty, |q_4| \rightarrow +\infty$ with fixed $q_4 e^\theta \equiv 4e^{\theta _3}$, which will be called for simplicity $\lim \limits _{H->CH}$. This limit sends the regular singularity at $y'=\theta -\ln 2$ onto the other (regular) singularity at $y'=-\infty$, thus creating an irregular singularity at $y'=-\infty$, whilst the other two regular singularities, at $y'=\ln 2, +\infty$, stay untouched. Following this prescription, $\psi ^{(3)}(y^\prime)= \lim \limits _{H->CH} \psi \left (y=y'-\frac {\theta}{2}\right )$ satisfies the equation
\be
\frac {d^2}{d{y'}^2 } \psi ^{(3)}(y')=V^{(3)}(y') \psi ^{(3)} (y') \,\, ,
\label{CHE1}
\ee 
where 
\ba
V^{(3)}(y')&=&\frac {1}{(e^{y'}-2)^2}\Bigl [ \frac {(q_1-q_2)^2}{4}e^{2y'}+ (2q_1q_2+q_3 e^{\theta _3}-2P^2-\frac {1}{2})e^{y'} + \nonumber  \\
&+& (4P^2-6q_3 e^{\theta _3}+e^{2\theta _3} ) +(8q_3e^{\theta _3}-4e^{2\theta _3})e^{-y'}+4e^{2\theta _3}e^{-2y'} \Bigr ] \, ,
\label{CHE2}
\ea
which is a CHE. 
Interestingly, the CHE has relevance in gravity: using the maps
\begin{align}
R &=\frac{r\sqrt{r -2M}}{\sqrt{2M}} \psi\,, \qquad     r = 4 M e^{-y'}\, , \qquad 
e^{\theta _3} = - 4 i M \omega\,, \qquad P^2= \ell (\ell+1)-8M^2 \omega^2+\frac{1}{4}  \,, \label  {maptonf3}\\
 q_1 &=  2 -2i M \omega \,, \qquad q_2 =  -2i M \omega \,, \qquad q_3 = -2 -2i M \omega\, ,\nonumber
 \end{align}
it maps into the Teukolsky equation for Schwarzschild background (in its homogeneous version, with spin $s=-2$ and integer $\ell \geq 2$) \cite{TEU1,TEU2}:
\be
r^4 f^2(r) \frac{d}{dr} \left[ \frac{\frac{d}{dr}R(r)}{r^2 f(r)} \right] + \left( \frac{\omega^2 r^2 + 4 i \omega (r-M)}{f(r)} - 8 i \omega r - (\ell+2)(\ell-1) \right) R(r) = 0\,, \quad  \quad f(r) = 1 - \frac {2M}{r} \, .
\ee

\medskip

The confluence limit on the Floquet index in the region $[-y_0,y_0]$ of the HE (\ref {knf4exp}) gives the corresponding interval $[-\infty,\ln 2]$ Floquet index of the CHE:
\be
k_{3}=\lim \limits _{H->CH} k=P-\frac {q_1q_2q_3}{P(4P^2-1)}e^{\theta _3}+O(e^{2\theta _3}) \label {Floq-CHE} \, .
\ee
Anew, a recursive relation of the type (\ref{rec-r-nf2}) can be easily derived and gives immediately all the coefficients above. The same confluence limit on (\ref {nf4Phi1inst}) gives the acquired phase on the interval $]-\infty,\ln 2]$
\be
\lim \limits _{H->CH}  \Phi =\Phi _3=-\theta _3 k_3 +\ln \frac {\Gamma (1+2k_3)}{\Gamma (1-2k_3)}+\frac {1}{2}\sum _{i=1}^3 \ln \frac {\Gamma \left (\frac {1}{2}-k_3+ q_i \right )}{\Gamma \left (\frac {1}{2}+k_3+ q_i \right )}+\frac {8k_3q_1q_2q_3}{(1-4k_3^2)^2} e^{\theta _3}+ O(e^{2\theta _3}) \, . \label {nf3Phi1inst}
\ee
With the identification $4 e^{\theta_3} \hbar=\Lambda _3 $ (instanton coupling), $\hbar q_j=m_j$, $\hbar k_3=a$, we conjecture that
\be
\Phi _3(\theta _3; k_3, q_1,q_2,q_3)=\frac {A_D(\Lambda _3,a,m_1,m_2,m_3)}{\hbar} \, , \label  {PhiAd3}
\ee
with $A_D $ the dual gauge period for ${\cal N}=2$ SYM with $N_f=3$, masses $m_j$ and gauge period $a$, in the NS background. In fact, (\ref {nf3Phi1inst}) reproduces the small instanton expansion of $\frac {A_D}{\hbar}$, up to one instanton. Actually, the validity of (\ref {PhiAd3}) hinges upon the validity of its HE equivalent (\ref {PhiAD4}): if this relation holds, then (\ref {PhiAd3}) follows immediately by applying on both sides the confluence limit from the HE to the CHE. 

\medskip


Coming to wave functions, Floquet solutions of the CHE can be derived from those of the HE in subsection (\ref {Nf4-convexp}); however, the main idea and the safest procedure is again to expand the quantum momentum or eikonal rather than the wave function. To this aim, it is convenient to interpret in a more general way the two regions of $y$, as argument of the Floquet solution $\psi _+(y)$, where the application of the kink method gives converging expansions for the wave functions. In fact, in the last Section these two regions are $(-y_0,0)$ and $(0,y_0)$, separated by zero, which is, however, arbitrary.

Indeed, there is much more freedom to choose the left and the right region: in general, the right region can be ${\cal {A}}_r\leq y<\left |\ln 2 -\frac {\theta}{2}\right |$, the left region $- \left |\frac {\theta}{2}-\ln 2 \right |<y\leq {\cal {A}}_l$, with ${\cal {A}}_r, {\cal {A}}_l$ two arbitrary points between the two singularities at $y=\pm y_0$, i.e. $|{\cal {A}}_l|<|\ln 2 -\frac {\theta}{2}|$, $|{\cal {A}}_r|<|\ln 2 -\frac {\theta}{2}|$. In particular, then, the left and right region can overlap. With the limit to the CHE in mind, one can suppose $\theta $ large and negative. Then, in the right region the argument of the `right kink' $\Pi _>(y)=\frac {d}{dy} \ln \psi _>(y)$ in equation  (\ref {kinknf4>}), with $ \psi _>(y)=\psi _+\left (y-\frac {\theta}{2}\right )$, ranges in the interval ${\cal {A}}_r+\frac {\theta}{2} \leq y<\ln 2$. In the left region the argument of the `left kink' $\Pi _<(y)=\frac {d}{dy}\ln \psi _<(y)$ in equation  (\ref {nf4Ricc<}), with $ \psi _<(y)=\psi _+\left (y+\frac {\theta}{2}\right )$, sits in the interval $-\ln 2 <y\leq {\cal {A}}_l -\frac {\theta}{2}$.  
Now, a possibility for studying the CHE limit is to choose 
${\cal {A}}_r=a_r-\frac {\theta}{2} $, with $a_r<\ln 2$, which gives the convergence $\theta$-expansion for any interval $a_r\leq y<\ln 2 $ of the $\psi _>(y)$, and the choice ${\cal {A}}_l=a_l-\frac {\theta}{2} $ with $a_l< \ln 2$ which entails the convergence of the $\psi _<(y)$ $\theta$-expansion for any $-\ln2 <y\leq a_l-\theta$. In this way, the right region for the HE Floquet function $\psi _+(y)$ is $a_r-\frac {\theta}{2} <y<\ln 2 -\frac {\theta}{2}$ which entails for the CHE variable $a_r \leq y'<\ln 2$, for any finite $a_r<\ln 2$; in this region the quantum momentum of the CHE Floquet function $\psi _+^{(3)}(y')$ is given by $\Pi _+^{(3)}(y') =\frac {d}{dy'} \ln \psi _+^{(3)}(y')= \lim \limits _{H->CH} \Pi _+(y'-\frac {\theta}{2})=
\lim \limits _{H->CH} \Pi _>(y')$ and then directly by the ${H->CH}$ limit on (each term of) the $\theta \rightarrow -\infty$ expansion (\ref{thetaexpnf4}, \ref  {nf4lead}, \ref  {nf4subl}, \ref{solpin4}) regarding the `right kink' quantum momentum of the HE Floquet solutions. In other words, the following series is readily established by taking the term-by-term limit
\be
\Pi _+^{(3)}(y') =\sum _{n=0}^\infty \tilde \Pi _{+,right}^{(3,n)}(y') e^{n\theta_3} \, ,
\label{paiexp-hor}
\ee
where, at first orders, 
\begin{equation}
\tilde \Pi _{+,right}^{(3,0)}(y')= \frac {d}{dy^\prime} \ln \left [ (e^{y'}-2)^{\frac {1-q_1-q_2}{2}}e^{k_3y'} {}_2F_1\left (\frac {1}{2}+k_3-q_1,\frac {1}{2}+k_3-q_2;1+2k_3;\frac {e^{y'}}{2} \right ) \right ] \, , \label  {nf4lead2}
\end{equation}
(easily the H$\rightarrow$CH limit  of the solution $ \tilde \Pi _>^{(0)}(y')$ (\ref  {nf4lead})), and 
\be
\tilde \Pi _{+,right}^{(3,1)}(y')= \frac {4q_3}{1-4k_3^2} \frac {d}{dy'}\left [\frac {e^{-y'}}{2}+\left (e^{-y'}-\frac {1}{2} \right )
 \tilde \Pi _{+,right}^{(3,0)}(y') \right ] \label {nf3exp} \, ,
\ee
(again the H$\rightarrow$CH limit of $ \tilde \Pi _>^{(1)}(y')$ (\ref {nf4subl})). In general, (\ref {solpin4}) easily entails the same form for $n\geq 1$ 
\be
\tilde \Pi _{+,right}^{(3,n)}(y')=\frac {d}{dy'}  \left  [P_{right,0}^{(n)}(e^{-y'})+\sum _{m=1}^{n}P_{right,m}^{(n)} (e^{-y'})\frac {d^{m-1}}{d{y'}^{m-1}} \tilde \Pi _{+,right}^{(3,0)}(y') \right ] \, , \label {solpin2bis}
\ee
with the CHE limit degree $n$ polynomials
\be
P_{right,m}^{(n)}(e^{-y'})=\sum \limits_{h=\textrm {max}\{m,1 \}}^n P_{right,m,h}^{(n)}(k_3^2,q_1,q_2,q_3)\left (e^{-y'}-\frac {1}{2} \right )^h \label {Pmrightform} \, .
\ee

Analogously, the left region for the HE function $\psi _+(y)$ is $\frac {\theta}{2}-\ln 2 <y\leq a_l-\frac {\theta}{2}$, which means in the CHE variable $\theta -\ln 2<y'\leq a_l$, and eventually, upon  the H$\rightarrow$CH limit the left interval $y'\leq a_l$, with $a_l <\ln 2$. As a consequence, the left convergence interval can be considered to be $y'<\ln 2$. But now, the quantum momentum $\Pi _+^{(3)}(y') =\frac {d}{dy'} \ln \psi _+^{(3)}(y')= \lim \limits _{H->CH} \Pi _+(y'-\frac {\theta}{2})$ of the CHE Floquet function $\psi _+^{(3)}(y')$ (same expression as before) undergoes a less obvious expansion as term by term limit since it does not involve a kink momentum only, but a shift of it: $\lim \limits _{H->CH} \Pi _<(y'-\theta)$. By direct inspection, this function does not admit an expression as the series of $e^{\theta_3}$ powers (by performing the term-by-term limit), but yet its shifted value 
\be 
\Pi _+^{(3)}(y'+\theta _3)=\Pi _<^{(3)}(y')=\sum \limits _{n=0}^{+\infty} \tilde \Pi _{<,left}^{(3,n)}(y')  e^{n\theta _3} \, , \quad y'<\ln 2 -\theta _3 \, 
\label{paiexp}
\ee 
does around $\theta_3\rightarrow -\infty$. The first two orders are easily found
\be
\tilde \Pi _{<,left}^{(3,0)}(y')=\frac {d}{dy'} \ln \left [\exp \left (-e^{-y'}+k_3y' \right ) {}_1F_1\left (\frac {1}{2}-k_3+q_3,1-2k_3;2 e^{-y'} \right )\right ] \label {Pi3+1} \, , 
\ee
\be
\tilde \Pi _{<,left}^{(3,1)}(y')=\left (-\frac {1}{4}+\frac {q_1q_2}{1-4k_3^2} \right ) \frac {d}{dy'} \Bigl [ \frac {1}{2}e^{y'}-e^{y'}\tilde \Pi _{<,left}^{(3,0)}(y') \Bigr ] \label {Pi3+2} \, .
\ee
As we will rigorously see just below, the kink method is instead necessary for the left interval and this is in fact the `left kink' CHE solution. In details, the CHE limit makes the ${}_2F_1$ hypergeometric function, in $\tilde \Pi _<^{(n)}(y'-\theta)$, loose one parameter, as $q_4 e^\theta \rightarrow 4e^{\theta _3}$ and become the confluent hypergeometric function  ${}_1F_1$ (this is ostensive in the series representation or in the differential equation). Moreover, the limit on the full series is under control and yields the general solution  
\be
\tilde \Pi _{<,left}^{(3,n)}(y')=\frac {d}{dy'}  \left  [P_{left, 0}^{(n)}(e^{y'})+\sum _{m=1}^{n}P_{left,m}^{(n)} (e^{y'})\frac {d^{m-1}}{d{y'}^{m-1}}  \tilde \Pi _{<,left}^{(3,0)}(y') \right ] \, , \label {solpin2ter}
\ee
with the (limiting) degree $n$ polynomials $P_{left,m}^{(n)}(e^{y'})=\sum \limits _{h=\textrm {max}\{m,1 \}}^n P_{left,m,h}^{(n)}e^{hy'}$.

\subsubsection {On the CHE {\it ab initio}}
\label {CHE}
Actually, the same limiting results, in particular expressions (\ref {paiexp-hor}-\ref {Pmrightform}), can be obtained directly by applying our general procedure to the CHE {\it ab initio}. Indeed, for what concerns the right region {\bf $a_r \leq y' <\ln 2$} the potential  (\ref{CHE2}) can be expanded at $\theta_3\rightarrow -\infty$ (without any kink procedure)
\be
V^{(3)}(y')=\sum _{n=0}^{+\infty} V^{(3,n)}_{right}(y')e^{n\theta _3} \,\, ,
\ee
with $P^2$ function of $k_3$ (arising from the inversion of (\ref {Floq-CHE})). In this way, the Riccati equation for $\Pi _<^{(3)}(y')$ (\ref {paiexp-hor}) can be expanded in positive integer powers $e^{n\theta_3}$ and brings the non linear ODE system
\be
\left ( \tilde \Pi _{+,right} ^{(3,0)}(y') \right )^2+\frac {d}{dy'} \tilde \Pi _{+,right} ^{(3,0)}(y')=V^{(3,0)}_{right}(y')
=\frac {1}{(e^{y'}-2)^2}\Bigl [ \frac {(q_1-q_2)^2}{4}e^{2y'}+ (2q_1q_2-2k_3^2-\frac {1}{2})e^{y'} +4k_3^2) \Bigr ] \, , 
\label {Pi-in-0}
\ee
 \be
2  \tilde \Pi  _ {+,right} ^{(3,0)}(y')  \tilde \Pi  _ {+,right} ^{(3,1)}(y') +  \frac {d}{dy'} \tilde \Pi  _ {+,right} ^{(3,1)}(y') =V^{(3,1)}_{right}(y')=\frac {q_3\left (1-4e^{-y'}-\frac {4q_1q_2}{4k_3^2-1}\right )}{e^{y'}-2} \label {Pi-in-1}
\ee
and for $n\geq 2$ 
\be
2  \tilde \Pi  _ {+,right}^{(3,0)}(y')  \tilde \Pi  _ {+,right}^{(3,n)}(y')+\frac {d}{dy'}  \tilde \Pi _ {+,right}^{(3,n)}(y')+\sum \limits_{m=1}^{n-1} \tilde \Pi _ {+,right}^{(3,m)}(y')  \tilde \Pi _ {+,right}^{(3,n-m)}(y')=V^ {(3,n)}_{right} (y') \, . \label  {Pi-in-n} 
\ee

Anew, expressions (\ref {paiexp}-\ref {solpin2ter}) can be obtained by applying our procedure directly to the CHE as well, but via our new found  kink method. In fact, the {\it shifted} CHE potential
\be
V^{(3)}(y'+\theta _3)=\sum _{n=0}^{+\infty} V^{(3,n)}_{left} (y') e^{n\theta _3}
\ee 
has to be expanded at $\theta_3\rightarrow -\infty$ (in the region $y'+\theta _3 \leq a_l$ which means $y'+\theta _3 < \ln 2$) with $P^2$ function of $k_3$ (arising from the inversion of (\ref {Floq-CHE})) inside the (shifted) Riccati equation for $\Pi _<^{(3)}(y')$ (\ref {paiexp}). In this way, the latter expands in positive integer powers $e^{n\theta_3}$ into a system of ODEs for $y'<\ln 2-\theta _3$:
\be
\left ( \tilde \Pi _{<,left} ^{(3,0)}(y') \right )^2+\frac {d}{dy'} \tilde \Pi _{<,left} ^{(3,0)}(y')=V^{(3,0)}_{left}(y')
=k_3^2 +2q_3 e^{-y'}+e^{-2y'} \label {Pi-up-0} \, , 
\ee
 \be
2  \tilde \Pi  _ {<,left} ^{(3,0)}(y')  \tilde \Pi  _ {<,left} ^{(3,1)}(y') +  \frac {d}{dy'} \tilde \Pi  _ {<,left} ^{(3,1)}(y') =V^{(3,1)}_{left}(y') 
= \frac {4q_1q_2+4k_3^2-1}{8}e^{y'}+\left (\frac {q_3}{2}-\frac {q_1q_2q_3}{k_3(4k_3^2-1)} \right ) 
\label {Pi-up-1}
\ee
and for $n\geq 2$ 
\be
2  \tilde \Pi  _ {<,left}^{(3,0)}(y')  \tilde \Pi  _ {<,left}^{(3,n)}(y')+\frac {d}{dy'}  \tilde \Pi _ {<,left}^{(3,n)}(y')+\sum \limits_{m=1}^{n-1} \tilde \Pi _ {<,left}^{(3,m)}(y')  \tilde \Pi _ {<,left}^{(3,n-m)}(y')=V^ {(3,n)}_{left} (y') \, \label 
{Pi-up-n} \, .
\ee

\medskip

Besides, the logarithmic derivative of the other Floquet solution, $\Pi _-^{(3)}(y')=\frac {d}{dy'}\ln \psi _-^{(3)}(y')$ is obtained from $\Pi _+^{(3)}(y')$ by replacing $k_3\rightarrow -k_3$ and hence the this holds at all orders in (\ref{paiexp-hor}, \ref{solpin2bis}) and (\ref{paiexp}, \ref{solpin2ter}) 
\ba
\Pi _-^{(3)}(y'; k_3)&=& \Pi _+^{(3)}(y'; -k_3)=\sum \limits _{n=0}^{+\infty} \tilde \Pi _{+,right}^{(3,n)}(y';-k_3)  e^{n\theta _3} \quad \textrm {if} \quad a_r\leq y'<\ln 2  \, , \nonumber \\
&& \label{paiexp-} \\
\Pi _-^{(3)}(y'; k_3)&=& \Pi _+^{(3)}(y'; -k_3)=\sum \limits _{n=0}^{+\infty} \tilde \Pi _{+,left}^{(3,n)}(y'; -k_3)  e^{n\theta _3} \quad \textrm {if} \quad -\infty<y'<\ln 2 \, . \nonumber 
\ea
A proof of all these statements is provided in Appendix \ref {floqCHE}.

Now, the advantage of starting immediately from the CHE enters the stage. While quantum momenta of CHE solutions are easily obtained by the CHE limit of HE momenta, the same limiting procedure on wave-functions and related quantities has to be applied with circumspection. 
For instance the acquired phase of the HE Floquet solutions in going from $-y_0$ to $y_0$, $\varphi (\theta , q_+,q_-)$, is affected by a diverging CHE limit (which surprisingly cancel out in the sum (\ref{symm-phase-Nf4}) as proven by (\ref{nf3Phi1inst})). Therefore, it is more convenient to find the wave functions with a specific normalisation directly as particular solutions of the CHE (\ref {CHE1}, \ref {CHE2}). 

Let us start from the Floquet solutions $\psi _\pm ^{(3)}(y')$, which behave up to normalisations $\rho_\pm, \eta_\pm$, when $y ' \rightarrow -\infty$ as 
\be
\psi _\pm ^{(3)}(y')\simeq \rho _\pm  e^{-\left (-q_3+\frac {1}{2}\right )(\theta _3+i\pi)+\left (-q_3+\frac {1}{2}\right )y'}\exp \left (  e^{\theta _3  -y'} \right ) \label {floq3infin} \, 
\ee
and, when $y'\rightarrow \ln 2$ as 
\be
\psi _\pm ^{(3)}(y') \simeq \eta _\pm \frac {1}{\sqrt {2}} (e^{y'}-2)^{\frac {1-q_1-q_2}{2}} \, \quad 
\label {floq3ln2} \, .
\ee
As detailed in \cite {DAN}, the decaying solution as $y'\rightarrow \ln 2$ may be normalised by the requirement
\be
\psi _{0,0}(y')\simeq \frac {1}{\sqrt {2}}(e^{y'}-2)^{\frac {1+q_1+q_2}{2}} \,\, ,\label {psi-00dan}
\ee
and is proportional to the {\it incoming} wave function of the gravity set-up $\psi _{in}(y')=\hat C \psi _{0,0}(y')$. Similarly, the subdominant solution as $y'\rightarrow -\infty$ may be chosen by the normalisation 
\be
\psi_{-,0}(y') \simeq e^{-\left (q_3+\frac {1}{2}\right )\theta _3+\left (q_3+\frac {1}{2}\right )y'}\exp \left ( - e^{\theta _3  -y'} \right) \, .  \quad \label {psi-dan}
\ee 
It is proportional to the {\it outgoing} gravity wave function $\psi _{up}(y')$: $ \psi _{up}(y')=\hat D \psi _{-,0}(y')$.

Of course, the acquired phase of $\psi _\pm ^{(3)}$, appearing in the asymptotics when going from $-\infty$ to $\ln 2$, can be defined, as above, in terms of $\Pi ^{(3)}_\pm (y')=\frac {d}{dy'}\ln \psi _\pm ^{(3)}(y')$ with suitable subtractions in the integral ($a<\ln 2$):
\ba
\varphi _\pm ^{(3)}&=& \int _{-\infty}^a dy' \left [\Pi ^{(3)}_\pm (y')+e^{\theta _3-y'}-\frac {1}{2}+q_3 \right ] +
\int _a ^{\ln 2 } dy' \left [ \Pi _\pm ^{(3)}(y')-\frac {(1-q_1-q_2) e^{y'}}{2(e^{y'}-2)} \right  ] + \nonumber \\
&+&e^{\theta _3 -a}+
\left (q_3-\frac {1}{2} \right ) (\theta _3+i\pi -a) -\frac {1-q_1-q_2}{2}\ln (e^a-2) +\frac {\ln 2}{2} \, . \label {acq-phase-che}
\ea
In the previous expression $\Pi ^{(3)}_\pm (y')$ may be borrowed from (\ref{paiexp}) in both integrals or alternatively from (\ref{paiexp}) (or (\ref{paiexp-})) in the first integral and from (\ref{paiexp-hor}) (or (\ref{paiexp-})) in the second, the latter way being more efficient that the former since it resums infinite terms of it. Besides, it is also given by the ratio
\be
\frac {\eta _\pm}{\rho _\pm} \equiv e^{\varphi _\pm ^{(3)}}.
\ee
Thus, a symmetric choice of the normalisation constants in (\ref {floq3infin}, \ref {floq3ln2}) in terms of acquired phases may well be:
\be
\rho _-=e^{-\frac {\varphi _-^{(3)}}{2}} \, , \quad  \rho _+=e^{-\frac {\varphi _+^{(3)}}{2}}  \, , \quad \eta _-=e^{\frac {\varphi _-^{(3)}}{2}} \, , \quad \eta _+=e^{\frac {\varphi _+^{(3)}}{2}} \, . \label {symm-choice}
\ee

Now,  the substitutes of (\ref  {nf4psi}) (and previous cases (\ref {nf2psi}, \ref {chgbasis})) express the decaying solutions (at $\ln 2$ and $-\infty$, respectively)
\ba
\psi _{0,0}(y')&=&\check C  \left [e^{-\frac {\varphi _-^{(3)}}{2}}  \psi _-^{(3)}(y')-e^{-\frac {\varphi _+^{(3)}}{2}} \psi _+^{(3)}(y')\right] \, , 
\label {00-+}  
\\
\psi _{-,0}(y')&=& \check D  \left [ e^{\frac {\varphi _+^{(3)}}{2}} \psi _+^{(3)}(y')-e^{\frac {\varphi _-^{(3)}}{2}} \psi _-^{(3)}(y')\right ] \, ,
\label {-0-+} \, 
\ea
as the linear combination of Floquet solutions, tuned to cancel the divergence between them (as before).\footnote{Although not impossible, it is rather delicate and cumbersome to obtain these connection formul\ae\,  as suitable regularisations of  (\ref {nf4psi}).} Their normalisations $\check C, \check D$ can be found in terms of Floquet related quantities as 
\be
\check C= \frac {(q_1+q_2) e^{\frac {\varphi _+^{(3)}+\varphi _-^{(3)}}{2}}}{W[\psi _+^{(3)}, \psi _-^{(3)}] }\, , \quad
\check D= \frac {2i e^{i\pi q_3} e^{-\frac {\varphi _-^{(3)}+\varphi _+^{(3)}}{2}} }{W[\psi _+^{(3)}, \psi _-^{(3)}] } \label {checkCD} \, ,
\ee
by applying our procedure above. Let us find $\check C$. The ODE/IM approach prescribes \cite{DAN} to write $\psi _{-,0}(y')$ in terms of a basis of solutions defined by their behaviour at $y'=\ln 2$ as the subdominant $\psi _{0,0}(y')$ and the dominant
\be
\psi _{0,1}(y') = {\bf \Omega}_+ \psi _{0,0}(y') \simeq  \frac {1}{\sqrt {2}}(e^{y'}-2)^{\frac {1-q_1-q_2}{2}} \, , \quad y'\rightarrow \ln 2 \, . 
\ee
given uniquely by this symmetry of the equation (\ref {CHE1}): $ {\bf \Omega}_+ :q_1,q_2 \rightarrow -q_1, -q_2$\footnote{Clearly, this is not a symmetry for a {\it vacuum}, {\it i.e.} a solution.}. Now, following \cite {DAN}, let us define the {\it fundamental} connection coefficient
\be
Q_1=W[\psi_{0,0}, \psi_{-,0}]\, . \label {Q1fundam}
\ee
Since $W[\psi _{0,0}, \psi _{0,1}]=-(q_1+q_2)$ once computed at $y'=\ln2$, it allows writing the aforementioned linear combination as
\be
\psi _{-,0}(y')= \delta _0 \psi _{0,0}(y') - \frac{Q_1}{(q_1+q_2)}    \psi _{0,1}(y')\,,
\label{conn---0}
\ee
where also $\delta_0$ could be written in terms of $Q_1$ \cite{DAN}. On one hand, (\ref  {00-+}, \ref  {-0-+}) immediately compute 
\be
Q_1=\check C \check D W[\psi _+^{(3)}, \psi _-^{(3)}] \left (e^{\frac {\varphi _-^{(3)}-\varphi _+^{(3)}}{2}}-e^{\frac {\varphi _+^{(3)}-\varphi _-^{(3)}}{2}} \right ) \label {wronsk-second} \, .
\ee
On the other hand, the limit $y' \rightarrow \ln 2$ on (\ref{conn---0}) can be compared with (\ref {-0-+}), provided by the behaviour (\ref {floq3ln2}, \ref {symm-choice}) of the Floquet solutions at $\ln 2$,  and then this yields
\be
Q_1=-\check D (e^{\varphi _+^{(3)}}-e^{\varphi _-^{(3)}}) (q_1+q_2)  \label {wronk-first} \, .
\ee
Equating these two results gives the first of (\ref {checkCD}).

To find $\check D$, our approach prescribes to write $\psi _{0,0}(y')$ as linear combination in the basis of the subdominant solution as $y'\rightarrow -\infty$ (at spatial infinity), $\psi_{-,0}(y')$ (\ref {psi-dan}), and the particular dominant
\be
\psi_{-,1} (y')={\bf \Omega}_- \psi _ {-,0}(y')\simeq e^{-\left (-q_3+\frac {1}{2}\right )(\theta _3+i\pi)+\left (-q_3+\frac {1}{2}\right )y'}\exp \left (  e^{\theta _3  -y'} \right ) \,\,, \quad  y'\rightarrow -\infty \, , \label {psi-dan-2}
\ee
obtained with the symmetry of (\ref {CHE1}) ${\bf \Omega} _-: \theta _3 \rightarrow \theta _3 +i\pi$, $q_3 \rightarrow -q_3$ \cite{DAN}. Again the fundamental connection coefficient  (\ref {Q1fundam}) gives this linear combination in the form
\be
\psi _{0,0}(y')=\gamma _0 \psi _{-,0}(y')- \frac  {Q_1}{2ie^{i\pi q_3}}\psi _{-,1}(y') \, , 
\label {Dan2}
\ee
since $W[\psi _{-,0}, \psi _{-,1}]=2ie^{i\pi q_3}$ once computed at $y'=-\infty$ and where also $\gamma_0$ can be written in terms of $Q_1$ \cite{DAN}. Now, the comparison of the $y'\rightarrow -\infty$ limit of this relation with (\ref {00-+}), provided by (\ref {floq3infin}, \ref {symm-choice}), furnishes another value for $Q_1$
\be
- \frac  {Q_1}{2ie^{i\pi q_3}}=\check C \left ( e^{-\varphi _-^{(3)}} -e^{-\varphi _+^{(3)}}\right )
\label {wronsk-first2} \, . 
\ee
Eventually, equating this value to (\ref {wronsk-second}) gives the second of (\ref {checkCD}).

Now, it is not difficult to find an expression of the limit (\ref{nf3Phi1inst}) in terms of the acquired phases $\varphi _\pm^{(3)}$, conjectured to be proportional to the dual gauge period $A_D$ in (\ref {PhiAd3}). First, (\ref {varphiHeun}) for Heun equation converts into the relatively compact formula
\be
2\Phi (q_+,q_-)=\varphi (q_+,q_-)+\varphi (q_-,q_+)= \int _{-y_0}^{y_0} dy \left [ \Pi _+(y, q_+,q_-)-\Pi _-(y, q_+,q_-) \right ] \, . 
\ee
Upon introducing in the integral the CHE variable $y'=y+\frac {\theta}{2}$ and dividing the integration interval in the regions $[-y_0+\frac {\theta}{2}=\theta -\ln 2, a ]$ and $[a, y_0+\frac {\theta}{2}=\ln 2]$, the latter becomes
\be
2\Phi (q_+,q_-)=\int _{\theta -\ln 2}^a dy' \left [\Pi _+\left (y'-\frac {\theta}{2} \right )- \Pi _-\left (y'-\frac {\theta}{2} \right ) \right ]+ \int _a^{\ln 2} dy' \left [\Pi _+\left (y'-\frac {\theta}{2} \right )- \Pi _-\left (y'-\frac {\theta}{2} \right ) \right ] \,\, .
\ee
Now, the H$\rightarrow$CH limit on both sides of this equality reaches the goal
\be
2\Phi _3=\int _{-\infty}^a dy' [\Pi _+^{(3)}(y')-\Pi _-^{(3)}(y') ] + \int _a^{\ln 2} dy' [\Pi _+^{(3)}(y')-\Pi _-^{(3)}(y') ]= 
\varphi _+^{(3)}-\varphi _-^{(3)} \,\, ,\label {acq-phase-Phi}
\ee
where the last equality comes from (\ref {acq-phase-che}).
This equality extends, to the CHE, (\ref {symm-phase-Nf4}), derived from Heun equation, and the second of (\ref {Qfornf2}), valid for the DCHE. And also in this case, it can be computed exactly as a convergent series expansion at $\theta _3 \rightarrow -\infty$ from that of $\Pi_3$. As a consequence, the fundamental connection coefficient can be computed by inserting (\ref {checkCD}) into (\ref {wronsk-second}) to produce again this general form ({\it cf.} (\ref{Qfornf2}) and (\ref{Qnfequal4}))
\be
Q_1(\theta _3; k,  q_1,q_2,q_3)=-4i e^{i\pi q_3} (q_1+q_2)  \frac {\sinh \Phi _3 (\theta _3; k,  q_1,q_2,q_3)}{W[\psi _+^{(3)}, \psi _-^{(3)}]}\, ,  \label {Q1finalformula}
\ee
 involving the acquired phase (or, better, $\Phi_3=A_D$) and the Wronskian, which can be easily computed at any value of $y'$. In particular for $y'=0$ the general form (analogous to (\ref {Wronsknf2},  \ref {Wronsknf4})) can be easily derived
\ba
&& W[\psi _+^{(3)}, \psi _-^{(3)}]=e^{-\frac {\varphi _-^{(3)}+\varphi _+^{(3)}}{2}}\exp \Bigl [ \int _{-\infty}^0 dy' \left ( \Pi _+^{(3)}(y')+\Pi _-^{(3)}(y')+2 e^{\theta _3-y'}+2q_3-1 \right ) + \nonumber \\
&+& 2e^{\theta _3}+(2q_3-1)(\theta _3+i\pi ) \Bigr ] \left ( \Pi _+^{(3)}(0)- \Pi _-^{(3)}(0) \right )
\label{Wronsknf3}
\ea
and allows a computation at all orders as in the previous cases (\ref{Wronsknf2-expan}, \ref{Wronsknf4-expan}) (details in \cite{OFR}). For instance, it gives at one loop order the result (\ref{wronsktheta3-infty}), although derived there by an alternative approach.

Now, with the aim of providing exact formul{\ae} useful for applications in gravity, we can rephrase (\ref {00-+}, \ref {-0-+}) in terms of functions $\psi _{in},\psi _{up}$:
\ba
\psi _{in}(y')&=& \hat C \psi _{0,0}(y') =\tilde C \left [e^{-\frac {\varphi _-^{(3)}}{2}}  \psi _-^{(3)}(y')-e^{-\frac {\varphi _+^{(3)}}{2}} \psi _+^{(3)}(y')\right]
\label {00-floq} \, ,  \\
\psi _{up}(y')&=& \hat D \psi _{-,0}(y')= \tilde D  \left [ e^{\frac {\varphi _+^{(3)}}{2}} \psi _+^{(3)}(y')-e^{\frac {\varphi _-^{(3)}}{2}} \psi _-^{(3)}(y')\right ]
\label {0--floq} \, , 
\ea
with $\tilde C= \check C \hat C, \tilde D= \check D \hat D$. 
Thanks to the expansion of $\Phi_3$  (\ref {nf3Phi1inst}) around $\theta_3\rightarrow -\infty$ and those of the two quantum momenta (\ref {paiexp}, \ref{paiexp-}), given at all orders for all $y'<\ln 2$ ($r>2M$), both terms in the right hand side of (\ref {00-floq}, \ref {0--floq}) can be written as exponentials of a single series of integer powers of $e^{\theta _3}$.
Actually, to make explicit these calculations, it is necessary to compute the proportionality factors $\hat C, \hat D$ and then $\tilde C, \tilde D$  in (\ref {00-floq}, \ref {0--floq}). To normalise $\psi _{in}(y')$ one constructs the gravity function which has the physical meaning of a wave incoming in the horizon 
\be
R_{in}(r)=\frac{r\sqrt{r -2M}}{\sqrt{2M}} \psi _{in}(y') \, . \label {Rinpsiin}
\ee
Then, following the conventions of \cite {FMR}, one requires that its behaviour at $r\rightarrow +\infty$ is exactly  
\be
R_{in}(r)\simeq B r^3 e^{i\omega r^\ast}+\frac {e^{-i\omega r^\ast}}{2i\omega r}
\, , \quad r^\ast =r +2M \ln \frac {r}{2M} \, , \label {asyform}
\ee
thus fixing the normalisation of $R_{in}$ and then the constant $\hat C$ (and, as a by-product, the constant $B$ above). To detail this computation, we follow \cite {DAN} where in the style of the ODE/IM correspondence the coefficients of the linear combination
$\psi _{0,0}(y')=\gamma _0 \psi _{-,0}(y')+\gamma _1 \psi _{-,1}(y')$ are expressed in terms of $Q_1=W[\psi_{0,0}, \psi_{-,0}]$ as follows
\be
\psi_{0,0}(y')=-\frac{i e^{-i \pi q_3}}{2}Q_1(\theta _3+i\pi,q_1,q_2,-q_3)\psi_{-,0}(y')+\frac{i e^{-i \pi q_3}}{2}Q_1(\theta _3,q_1,q_2,q_3)\psi_{-,1}(y') \, . \label {psi00dan}
\ee
Then, upon comparing the $y' \rightarrow -\infty$ limits of $\psi_{0,0}(y')$, coming from (\ref {psi00dan}) and  (\ref {psi-dan}, \ref  {psi-dan-2}) with the same limit, descending from 
(\ref {Rinpsiin}, \ref {asyform}), for $\psi _{in}(y')$, one eventually finds the exact ratio 
\be
\hat C=\frac {\psi _{in}(y')}{\psi _{0,0}(y')}=- \frac {2^{q_3-\frac {1}{2}}e^{-\theta _3 \left (q_3+\frac {1}{2}\right )}}{MQ_1(\theta _3,q_1,q_2,q_3)}=\frac {i \omega (-2iM\omega )^{\frac {1}{2}+2iM\omega} }{Q_1 (\theta _3,q_1,q_2,q_3) } \, , \label {hatCexpr}
\ee
from which the exact behaviour at the horizon,
\be
R_{in}(r)\simeq 2^{5/2}(iM\omega )^{\frac {3}{2}+2iM\omega}\frac {1}{Q_1(\theta _3,q_1,q_2,q_3)} \left (1-\frac {2M}{r} \right )^{2-2iM\omega} \, \label {incoming} \quad \textrm {as} \quad r\rightarrow 2M 
\ee
and, as announced, the exact value for the coefficient $B$:
\be
B= -8 \omega ^3 e^{-2(2+q_3)\theta _3}e^{-i\pi q_3}4^{q_3}\frac {Q_1(\theta _3+i\pi,q_1,q_2,-q_3)}{Q_1(\theta _3,q_1,q_2,q_3)} =   -\frac {\omega ^3}{2}(2M\omega)^{4iM\omega}\frac {Q_1(\theta _3+i\pi,q_1,q_2,-q_3)}{Q_1(\theta _3,q_1,q_2,q_3)}
 \label {Bexact} \, .
\ee

Coming now to (\ref  {0--floq}) and to
the proportionality constant $\hat D$, we remember that 
the function $\psi _{up}(y')$ is defined, again following the normalisation of \cite {FMR}, by the fact that, as $r \rightarrow +\infty$,
\be
R_{up}(r)=\frac{r\sqrt{r -2M}}{\sqrt{2M}} \psi _{up}(y') \simeq r^3 e^{i\omega r^\ast} \, , 
\label  {Rupinfinity}
\ee
i.e. its gravity equivalent $R_{up}$ satisfies outgoing boundary conditions at infinity.
Then, writing in (\ref {Rupinfinity}) $ \psi _{up}(y')=\hat D  \psi _{-,0}(y')$ and using (\ref {psi-dan}), it is simple to find that 
\be
\hat D= 2^{\frac {3}{2}-q_3}M^2 e^{\theta _3 \left ( q_3+\frac {1}{2}\right )} =4M^2 (-2iM\omega)^{-\frac {3}{2}-2iM\omega} \, . \label {hatDexpr}
\ee
Eventually, the exact expressions for the constants $\tilde C=\check C \hat C, \tilde D=\check D \hat D$ come from the combined use of (\ref {checkCD}, \ref {hatCexpr}, \ref {hatDexpr}): 
\ba
\tilde C&=&\hat C \check C=  \frac {2^{q_3-\frac {5}{2} }e^{-\theta _3 \left (q_3+\frac {1}{2} \right )-i\pi q_3}e^{\frac {\varphi _+^{(3)}+\varphi _-^{(3)}}{2}}}{iM \sinh \Phi _3}   =
-\frac {\omega (-2iM\omega)^{\frac {1}{2}+2iM\omega}e^{\frac {\varphi _+^{(3)}+\varphi _-^{(3)}}{2}}}{4e^{2\pi M \omega}\sinh \Phi _3} \, , \nonumber \\
\tilde D&=&\hat D \check D= \frac {iM^2 2^{\frac {5}{2}-q_3}e^{i\pi q_3}e^{\theta _3 \left (q_3+\frac {1}{2} \right)}e^{-\frac {\varphi _-^{(3)}+\varphi _+^{(3)}}{2}}}{W[\psi _+^{(3)}, \psi _-^{(3)}]}    =
\frac {8iM^2e^{2\pi M\omega}e^{-\frac {\varphi _-^{(3)}+\varphi _+^{(3)}}{2}}}{(-2iM\omega )^{\frac {3}{2}+2iM\omega}W[\psi _+^{(3)}, \psi _-^{(3)}]}  \, . \label {tildeCD}
\ea
Up to now, we gave exact formul{\ae}. In the following we discuss their $\theta_3\rightarrow -\infty$ limit.

The Floquet solutions follow from the quantum momentum and then enjoy two possible   expansions intervals around $e^{\theta _3}=0$: one from (\ref{paiexp}) gives $\psi _\pm ^{(3)}(y')$ for all $y'<\ln 2$ (more effective for $y'\rightarrow -\infty$, {\it i.e.} large distance $r\rightarrow \infty$ from the horizon), while the other from (\ref{paiexp-hor}) gives  $\psi _\pm ^{(3)}(y')$ on the finite interval $y'\in [a_r,\ln 2)$,  $a_r<\ln 2$ (more effective around the horizon $y'=\ln 2$, {\it i.e.} $r=2M$). In the first, $\psi _\pm^{(3)}(y')$ are dominated  by the leading orders (\ref {Pi3+1}):
\be
\psi _\pm ^{(3)}(y')\simeq \tilde A_\pm \exp \left (-e^{\theta _3-y'}\pm k_3(y'-\theta _3) \right ) {}_1F_1\left (\frac {1}{2}\mp k_3+q_3,1 \mp 2k_3;2 e^{\theta _3-y'} \right )\, , 
\label{lead-Flo-1}
\ee
while in the second the two expansions start with (\ref{nf4lead2}):
\be
\psi _\pm ^{(3)}(y')\simeq \tilde B_\pm  (e^{y'}-2)^{\frac {1-q_1-q_2}{2}}e^{\pm k_3y'} {}_2F_1\left (\frac {1}{2}\pm k_3-q_1,\frac {1}{2}\pm k_3-q_2;1\pm 2k_3;\frac {e^{y'}}{2} \right )
\label{lead-Flo-2} \, .
\ee
In (\ref {lead-Flo-1}, \ref {lead-Flo-2}) $\tilde A_\pm$ and $\tilde B_\pm$ are quantities independent of $y'$, which fix the normalisation of  $\psi _\pm ^{(3)}(y')$. For this reason, they have to be related to $\rho_\pm$ and $\eta _\pm$ and then to the acquired phase $\varphi _\pm ^{(3)}$. For instance, 
comparison of (\ref {floq3infin}) with (\ref {lead-Flo-1}) entails
\be
 \tilde A_\pm e^{-i\pi \left (q_3-\frac {1}{2}\right )} 2^{-\frac {1}{2}\pm k_3 +q_3} \frac {\Gamma (1\mp 2k_3)}{\Gamma \left (\frac {1}{2}\mp k_3 +q_3 \right )}\left (1+O(e^{\theta _3}) \right )=\rho _\pm =e^{-\frac {\varphi _\pm ^{(3)}}{2}} \, .
\label {rhoArel}
\ee
On the other hand, comparison of (\ref {floq3ln2}) with (\ref {lead-Flo-2}) entails 
\ba
&& \tilde B_\pm 2^{\pm k_3} {}_2F_1\left (\frac {1}{2}\pm k_3-q_1,\frac {1}{2}\pm k_3-q_2;1\pm 2k_3;1 \right ) \left (1+O(e^{\theta _3}) \right )= \nonumber \\
&=&\tilde B_\pm 2^{\pm k_3}\frac {\Gamma (1\pm 2k_3) \Gamma (q_1+q_2)}{\Gamma \left (\frac {1}{2}\pm k_3 +q_1 \right )
\Gamma \left (\frac {1}{2}\pm k_3 +q_2 \right )  } \left (1+O(e^{\theta _3}) \right )=\frac {\eta _\pm}{\sqrt {2}}=
\frac {e^{\frac {\varphi _\pm ^{(3)}}{2}}}{\sqrt {2}}
\, \label {etaBrel} \, .
\ea
An immediate consequence is an alternative computation to the Wronskian (\ref{Wronsknf3}). In fact, it can be obtained in two ways: on the one side by using (\ref {lead-Flo-1}) and (\ref {rhoArel}) at one-loop order and neglecting instanton corrections
\be
W[\psi _+^{(3)}, \psi _-^{(3)}]= -2k_3  \tilde A_+ \tilde A_- +\dots=2k_3 e^{-\frac {\varphi _+^{(3)}+\varphi _-^{(3)}}{2}} e ^{2i\pi q_3 }2^{1-2q_3} \frac {\Gamma \left ( \frac {1}{2}+k_3+q_3 \right ) \Gamma \left ( \frac {1}{2}-k_3+q_3 \right )}{\Gamma (1-2k_3) \Gamma (1+2k_3)} +\dots \, .
\ee
On the other side, it comes out from (\ref {lead-Flo-2}) and (\ref {etaBrel}):
\be
W[\psi _+^{(3)}, \psi _-^{(3)}]= -2k_3 (-2)^{1-q_1-q_2}\tilde B_+ \tilde B_- +\dots=-2k_3 (-2)^{1-q_1-q_2}\frac {e^{\frac {\varphi _+^{(3)}+\varphi _-^{(3)}}{2}}}{2}\frac {\prod \limits _{i=1}^{2} \Gamma \left (\frac {1}{2}+k_3+q_i \right ) \Gamma \left (\frac {1}{2}-k_3+q_i \right ) }{\Gamma (1+2k_3) \Gamma (1-2k_3) \Gamma ^2 (q_1+q_2)}+\dots \label {wronskianeta} \, .
\ee
Then, the square root of their product yields the one-loop result
\be
W[\psi _+^{(3)}, \psi _-^{(3)}]= \frac {e^{i\pi q_3} 2^{\frac {1}{2} -q_3-\frac {q_1+q_2} {2} }(-1)^{-\frac {q_1+q_2} {2} }  \sin 2\pi k_3} {\pi \Gamma (q_1+q_2)} \sqrt {\prod \limits _{i=1}^{3} \Gamma \left (\frac {1}{2}+k_3+q_i \right ) \Gamma \left (\frac {1}{2}-k_3+q_i \right )} +\dots \label {wronsktheta3-infty} \, ,
\ee
which could be systematically completed by all (converging) instanton corrections, giving the same results as (\ref{Wronsknf3}). Eventually, this expansion of the Wronskian can be plugged into (\ref {Q1finalformula}) and gives the elegant converging expansion
\be
Q_1(\theta _3; k_3, q_1,q_2,q_3)=2\pi \left( -\frac {i (-1)^{\frac {q_1+q_2} {2}} 2^{\frac {1+q_1+q_2+2q_3}{2}}\Gamma (1+q_1+q_2)}{\left [ \prod \limits_{i=1}^3 \Gamma \left (\frac {1}{2}+q_i+k_3 \right ) \Gamma \left (\frac {1}{2}+q_i-k_3 \right ) \right ]^{\frac {1}{2}}} +\dots \right)\frac {\sinh \Phi _3(\theta _3; k_3, q_1,q_2,q_3)}{\sin 2\pi k_3} \, ,
\label {Q1finaldecoupling}
\ee
where 
\be
\Phi _3(\theta _3; k_3, q_1,q_2,q_3)=-\theta _3 k_3 +\ln \frac {\Gamma (1+2k_3)}{\Gamma (1-2k_3)}+\frac {1}{2}\sum _{i=1}^3 \ln \frac {\Gamma \left (\frac {1}{2}-k_3+ q_i \right )}{\Gamma \left (\frac {1}{2}+k_3+ q_i \right )}+O(e^{\theta _3}) \, 
\ee
is exact (and converging) at all orders, while the dots come from the instanton corrections to the rest.

After discussing the Floquet solutions, we arrive at the study of the expansions in the limit  $\theta _3 \rightarrow -\infty$ of the incoming and outgoing solutions. Staring first with $\psi _{in}$, which, by virtue of (\ref {00-floq}), is written as 
\be
\psi _{in}(y') = \tilde C e^{-\frac {\varphi _-^{(3)}}{2}}  [ \psi _-^{(3)}(y') -e^{-\Phi _3(\theta _3, k_3)} \psi _+^{(3)}(y') ] \, , 
\label {in-floq}
\ee
let us notice that for large negative $\theta _3$ the coefficient $e^{-\Phi _3(\theta _3, k_3)} \sim e^{k_3\theta _3} \cdot \exp (\textrm  {series of } e^{n\theta _3})$,  see (\ref {nf3Phi1inst}), is a subdominant (non perturbative) contribution and can be approximated to zero. Yet, if $y'\rightarrow \ln 2$, {\it i.e.} in physical terms in the vicinity of the horizon, both Floquet functions are needed in (\ref {00-floq}) to reproduce the decaying behaviour of $\psi _{in}$ at $y'=\ln 2$, because the diverging contributions of the Floquet solutions have to cancel each other (at leading order and then automatically at all orders), as in the very same construction of all the other formul{\ae}, (\ref {chgbasis}, \ref {nf2psi}, \ref {nf4psi}), connecting Floquet to decaying solutions. As physical consequence, far from the horizon, at $y' \ll \ln 2$ or at least not too close to it, we may neglect the second term in the r.h.s. of (\ref {in-floq}) and approximate
\be 
\psi _{in}(y') \simeq \tilde C e^{-\frac {\varphi _-^{(3)}}{2}}  \psi _-^{(3)}(y') 
\label{Floq-approx-in}
\ee
up to non-perturbative contributions; this is the so-called Floquet approximation {\it not too close to the horizon}. Now, as $\theta _3 \rightarrow -\infty$ the Floquet solution allows two different expansions (\ref{paiexp-}) we are going to consider.

Without further limitations, the second of (\ref{paiexp-}) with (\ref{paiexp}) is usable (and convergent) from large distances up to the horizon, $y'\in (-\infty,\ln 2)$, which means that, not too close to the horizon, at leading order only the lower subscript $-$ of (\ref{lead-Flo-1}) contributes:
\be
\psi _{in}(y')\simeq \tilde C \tilde A_- e^{-\frac {\varphi _-^{(3)}}{2}} \exp \left (-e^{\theta _3-y'}-k_3(y'-\theta _3) \right ) {}_1F_1\left (\frac {1}{2}+k_3+q_3,1+2k_3;2 e^{\theta _3-y'} \right ) \, . \label  {psi-in-far}
\ee
Otherwise, for generic values of $y'$ closer to the horizon, both Floquet solutions $\psi _\pm ^{(3)}(y')$ would contribute. In gravity variables the solution (\ref  {psi-in-far}) becomes, after using the property ${}_1F_1(a,b;z)=e^z {}_1F_1(b-a,b;-z)$,
\be
R_{in}(r) \simeq \frac {\tilde C \tilde A_- e^{-\frac {\varphi _-^{(3)}}{2}}}{\sqrt {2M}} \, r^{\frac {3}{2}}  (-i\omega r)^{k_3}e^{-i\omega r}{}_1F_1 \left (\frac {5}{2}+k_3+2iM\omega,1+2k_3; 2i\omega r \right ) \, 
\label  {limit-in2} \, .
\ee
In gravity variables (\ref {maptonf3}), the above approximation $\theta _3 \rightarrow -\infty$ and $y'\ll \ln 2 $ translates into $\frac {M}{r}\ll \omega r$ and (\ref {limit-in2}) becomes (using the explicit values for $\tilde C, \tilde A_-, e^{-\frac {\varphi _-^{(3)}}{2}}$)
\be
R_{in}(r)\simeq \frac {1}{2}(-2)^{\frac {5}{2}+k_3}\frac {\Gamma \left (-\frac {3}{2}+k_3-2iM\omega \right )}{\Gamma (1+2k_3)}(-4iM\omega)^{2iM\omega}
(i\omega r)^{\frac {3}{2}+k_3}e^{-i\omega r}{}_1F_1 \left (\frac {5}{2}+k_3+2iM\omega,1+2k_3; 2i\omega r \right ) \, , \label  {limit-in3}
\ee
that correctly satisfies the asymptotic behaviour (\ref {asyform}) at $r\rightarrow +\infty$. In fact, we know there are corrections from $e^{-\Phi _3}\psi _+^{(3)}(y')$ in the exact formula (\ref {00-floq}) only getting closer to the horizon. With the map $k_3=\hat \ell+\frac {1}{2}$, formula (\ref {limit-in3}) can be compared with the r.h.s. of (33) of \cite {FMR}: this  expression is obtained by using the quiver gauge partition function which gives the Floquet solution $Z_-(r)=\tilde C e^{-\frac {\varphi _-^{(3)}}{2}} R_-(r)$, with 
\be
R_-(r)=\frac{r\sqrt{r -2M}}{\sqrt{2M}}\psi _-^{(3)}(y') \, .
\ee
The approximation (\ref  {limit-in3}) reproduces (33) of \cite {FMR}, apart from an extra prefactor $1/2$, which seems to be missing in \cite {FMR}, and a little improvement, as we found a resummation to all instantons of the first two arguments of the ${}_1F_1$ function: indeed, in notations of \cite {FMR}, $k_3=P+O(e^{\theta _3})=\ell +\frac {1}{2}+O(e^{\theta _3})$. However, (\ref  {psi-in-far}) is the leading term in the series expansion of only one of the terms, $\tilde C e^{-\frac {\varphi _-^{(3)}}{2}}\psi _-^{(3)}(y') $, of the exact expression (\ref {00-floq}): then, this series does not give the correct result since it misses the subleading contribution $-\tilde C e^{-\frac {\varphi _+^{(3)}}{2}} \psi _+^{(3)}(y')$. In our approach, this correction can be evaluated explicitly by computing $\varphi _\pm^{(3)}$ and $\psi _+^{(3)}$: its leading order is an additional hypergeometric function (to (\ref {limit-in3}) with subleading coefficient).

A last computation based on the expression (\ref {limit-in3}) is the match with the $e^{i\omega r^\ast}$ term of (\ref {asyform}): this gives the leading expansion of the coefficient $B$ in (\ref {asyform}) for $M\omega \ll 1$:
\be
B\simeq -8\omega ^3 (4 M\omega )^{4iM\omega}e^{-i\pi k_3}\frac {\Gamma \left (-\frac {3}{2}+k_3-2iM\omega \right )}{\Gamma \left (\frac {5}{2}+k_3+2iM\omega \right )} \, . \label {Bapprox}
\ee
Besides and more effectively, the exact expression (\ref {Bexact}) can be computed at leading order using (\ref {Q1finaldecoupling}, \ref {nf3Phi1inst}). Then, it can be verified that it reproduces this leading contribution (which also follows by functional methods in \cite {DAN}).  

\medskip

On the contrary, the implementation of Floquet approximation (\ref{Floq-approx-in}) is more restricted with the first of (\ref{paiexp-}) as given by (\ref {paiexp-hor} - \ref {solpin2bis}). In fact, it possesses the further limitation to be an exact (converging) expansion as $\theta _3 \rightarrow -\infty$ when $y'$ stays finite (not too far from the horizon). For instance, physically the leading order (\ref {lead-Flo-2}, \ref {etaBrel}) is effective when $e^{\theta _3} \ll e^{2y'}$, {\it i.e.} $\omega r \ll \frac {M}{r}$ in gravity variables (\ref {maptonf3}) and furnishes ($k_3= \hat \ell+\frac {1}{2}$) 
\ba
&& \psi _-^{(3)}(y') \simeq \frac {e^{\frac {\varphi _-^{(3)}}{2}}}{\sqrt {2}}
\frac {\left  (\frac {r}{2M} \right )^{\hat \ell+\frac {1}{2}}\left (\frac {4M}{r}-2 \right )^{-\frac {1}{2}+2iM\omega}\Gamma \left (2-\hat \ell -2iM\omega\right )\Gamma \left (-\hat \ell -2iM\omega\right )}{ \Gamma (-2\hat \ell)\Gamma \left (2 -4iM\omega\right )} \cdot \nonumber  \\
&& \cdot  {}_2F_1\left (-\hat \ell -2+2iM\omega,-\hat \ell+2iM\omega;-2\hat \ell; \frac {2M}{r}\right ) \, . \nonumber 
\ea
Therefore, not too far from the horizon, but even not too close to it (so that Floquet approximation (\ref{Floq-approx-in}) works), the incoming solution writes, at leading order, as 
\ba
&& R_{in} (r) \simeq \frac {\tilde C}{\sqrt {2}}  \, 
\frac {r^{\hat \ell+2}(-2)^{-1/2+2iM\omega} \left (1-\frac {2M}{r} \right )^{2iM\omega}\Gamma \left (2-\hat \ell -2iM\omega\right )\Gamma \left (-\hat \ell -2iM\omega\right )}{(2M )^{\hat \ell+1}\Gamma (-2\hat \ell)\Gamma \left (2 -4iM\omega\right )} \cdot \nonumber \\
&& \cdot  {}_2F_1\left (-\hat \ell -2+2iM\omega,-\hat \ell+2iM\omega;-2\hat \ell; \frac {2M}{r}\right ) \, . \label {Rinfar}
\ea
Upon inserting the specific value for $\tilde C$ (\ref {tildeCD}) and using (\ref  {wronskianeta}, \ref {wronsktheta3-infty}), it becomes
\ba
&& R_{in} (r) \simeq -\omega e^{\left (\hat l +\frac {1}{2} \right ) \theta _3}
\frac {(-2)^{1+2iM\omega}(iM\omega )^{\frac {1}{2}+2iM\omega}\Gamma \left (\hat \ell -1-2iM\omega\right )}{(2M )^{\hat \ell+1}\Gamma (2+2\hat \ell) 2^{2\hat l}} \cdot \nonumber \\
&& \cdot r^{\hat \ell+2}  \left (2-\frac {4M}{r} \right )^{2iM\omega} {}_2F_1\left (-\hat \ell -2+2iM\omega,-\hat \ell+2iM\omega;-2\hat \ell; \frac {2M}{r}\right ) \, . \label {Rinfar2}
\ea
Because of its proportionality to the Floquet solution, (\ref {Rinfar2}) can be compared with (33) of \cite {FMR}, which, actually, reproduces the Floquet solution $Z_-(r)$. It coincides with (33) of \cite {FMR} apart the same factor $1/2$ as before, the usual improvement given by the resummation at all instantons in the arguments of the $ {}_2F_1$ and the $\Gamma $ functions and a different exponent for the binomial $\left (1-\frac {2M}{r} \right )$. We stress again that the validity of (\ref {Rinfar2}) as approximation of $R_{in}$ in the limit $\theta _3 \rightarrow -\infty$ cannot be extended close to the horizon, $r=2M$. The exact behaviour of $R_{in}$ around the horizon is given by (\ref {incoming}) and this is reproduced by the specific linear combination (\ref {00-floq}) of the two Floquet functions. Alternatively,
in next Subsubsection \ref {inc-up} we will provide a single convergent series in powers of $e^{\theta _3}$ for $\psi _{in}$, which works around the horizon: at leading order (in the limit $\theta _3 \rightarrow -\infty$) this will give a different hypergeometric function that corrects (\ref {Rinfar2}).

\medskip

Analogous considerations can be developed about the outgoing solution, which, following (\ref {0--floq}), takes the form 
\be 
\psi _{up}=\tilde D e^{\frac {\varphi _+^{(3)}}{2}} \left [\psi _+^{(3)}(y')- e^{-\Phi _3(\theta _3, k_3)}\psi _-^{(3)}(y')  \right  ]. 
\ee
Around infinity ($y'\rightarrow -\infty$), both Floquet solutions are needed to reproduce the decaying behaviour of $\psi _{up}$, in the spirit of the cancellation in (\ref {0--floq}). Yet, at fixed $y'$ the term $e^{-\Phi _3(\theta _3, k_3)} \sim e^{k_3\theta _3}$ is suppressed for $\theta _3 \rightarrow -\infty$; then, not too close to $-\infty$ (spatial $r$ infinity), the approximate proportionality
\be 
\psi _{up}(y')\simeq  \tilde D e^{\frac {\varphi _+^{(3)}}{2}} \psi _+^{(3)}(y')\,\, ,
\label{Floq-approx-up}
\ee 
holds. Anew, this is called Floquet approximation which holds {\it not too close to infinity}. 

Therefore, the approximation (\ref{Floq-approx-up}) can be applied efficiently by using just the expansion of $\psi_+^{(3)}$ around the horizon, {\it i.e.} when {\bf $a_r\leq y'<\ln 2$,} where the leading order of the Floquet for $\theta _3 \rightarrow -\infty$ (in gravity variables these two requests mean $\omega r \ll \frac {M}{r}$) is given by (\ref {lead-Flo-2}) and then is expected to give already a good approximation as 
\be
\psi _{up}(y') \simeq \tilde D \tilde B_+ e^{\frac {\varphi _+^{(3)}}{2}}  (e^{y'}-2)^{\frac {1-q_1-q_2}{2}}e^{k_3y'} {}_2F_1\left (\frac {1}{2}+ k_3-q_1,\frac {1}{2}+ k_3-q_2;1+ 2k_3;\frac {e^{y'}}{2} \right ) \, .
\ee
By using the expressions for $\tilde D$ (\ref {tildeCD}) and the analogous one, coming from (\ref {etaBrel}) for $\tilde B_+$, this finally entails, in gravity variables $R_{up}(r)= \frac{r\sqrt{r -2M}}{\sqrt{2M}}\psi _{up}(y')$
\be
R_{up}(r) \simeq  - \frac {i 2^{k_3-\frac {1}{2} }(-i\omega r)^{\frac {3}{2}-k_3}\left (1-\frac  {2M}{r} \right )^{2iM\omega} \Gamma (2k_3) {}_2F_1\left (-\frac {3}{2}+ k_3+2iM\omega,\frac {1}{2}+ k_3+2iM\omega;1+2k_3;\frac {2M}{r} \right )}{\pi \omega ^{3} (-4iM\omega )^{2iM\omega}\Gamma \left ( -\frac {3}{2}+k_3-2iM\omega \right ) } \, .
\ee

On the contrary, the expansion (\ref{paiexp}) around infinity has to be used with care as it is usable (and convergent) in principle from arbitrary large distances up to the horizon, $y'\in (-\infty,\ln 2)$. But it is allowed inside the approximation (\ref{Floq-approx-up}) only {\it not too close to infinity}, where it is less accurate. Nevertheless, upon using the plus sign of (\ref {lead-Flo-1}), the Floquet approximation gives at leading order
\be
\psi _{up}(y')\simeq  \tilde D \tilde A_+ e^{\frac {\varphi _+^{(3)}}{2}}\exp \left (-e^{\theta _3-y'}+k_3(y'-\theta _3) \right ) {}_1F_1\left (\frac {1}{2}- k_3+q_3,1 - 2k_3;2 e^{\theta _3-y'} \right )\, .
\ee
Inserting the $\theta _3 \rightarrow -\infty$ limits of $\tilde D, \tilde A_+, e^{\frac {\varphi _+^{(3)}}{2}}$, extracted from (\ref {tildeCD}) and (\ref {rhoArel}), respectively, and passing to gravity variables, the up solution $R_{up}(r)=\frac{r\sqrt{r -2M}}{\sqrt{2M}} \psi _{up}(y')$ for large $r$ (large negative $y$), which more correctly means $\frac {M}{r} \ll \omega r$, acquires the form
\be
R_{up}(r)\simeq \frac {1}{2\sqrt {2} \omega ^3}\frac {(-2)^{-k_3}\Gamma \left (2k_3 \right )}{(-4iM\omega)^{2iM\omega}\Gamma \left (- \frac {3}{2}+k_3-2iM\omega \right )}
(i\omega r)^{\frac {3}{2}-k_3}e^{-i\omega r}{}_1F_1 \left (\frac {5}{2}-k_3+2iM\omega,1-2k_3; 2i\omega r \right ) \, .
\label {Rupclose}
\ee
Again, we reiterate that (\ref {Rupclose}) cannot be trusted for $r\rightarrow +\infty$. Moreover, 
in Subsubsection \ref {inc-up} we will provide a single convergent series in powers of $e^{\theta _3}$ for $\psi _{up}$, which works for all $y'\leq a_l$ and then also when $y'\rightarrow -\infty$: at leading order (in the limit $\theta _3 \rightarrow -\infty$) this will give a different hypergeometric function that corrects (\ref{Rupclose}).

\subsubsection {A single series for incoming ({\it in}) and outgoing ({\it up}) solutions (with Floquet's form)}
\label {inc-up}

With relations (\ref {00-floq}, \ref {0--floq}) the incoming and the outgoing functions are expressed as sum of two terms, proportional to Floquet functions, each expressed as single convergent series in powers of $e^{\theta _3}$. Actually, in the same fashion as for the Floquet solutions, also the incoming and the outgoing functions can be given alternative expressions as single convergent series in powers of $e^{\theta _3}$. The single series for $\psi _{in}(y')=\hat C \psi _{0,0}(y')$ holds around the horizon  ($a_r\leq y'<\ln 2$), the one for $\psi _{up}(y')= \hat D \psi _{-,0}(y') $ for $y'<\ln 2$. 
We now give a brief proof of this statement. 

Starting from the incoming solution, which is uniquely defined by the behaviour 
\be
\psi _{in}(y') \simeq - \frac {2^{q_3-1}e^{-\theta _3 \left (q_3+\frac {1}{2} \right )}}{MQ_1(\theta _3; k_3, q_1,q_2,q_3)}(e^{y'}-2)^{\frac {1+q_1+q_2}{2}} \, , \quad y' \rightarrow \ln 2 \, , \label {psi-in-ln2}
\ee
one considers its quantum momentum $\Pi _{in}(y') =\frac {d}{dy'} \ln \psi _{in}(y')$, in the finite interval {\bf $y'\in [a_r,\ln 2)$}, and builds the expansion, for $\theta_3\rightarrow -\infty$,
\be
\Pi _{in}(y') =\sum _{n=0}^\infty \tilde \Pi _{in}^{(n)}(y') e^{n\theta_3} \, .
\label{inc-hor}
\ee
Since $\Pi _{in}(y')$ satisfies the Riccati equation of the CHE, the various `components' $ \tilde \Pi _{in}^{(n)}(y') $ must satisfy the system of ODEs (\ref {Pi-in-0}-\ref {Pi-in-n}), with the obvious replacement $\tilde \Pi _{+,right}^{(3,n)}(y') \rightarrow \tilde \Pi _{in}^{(n)}(y')$.
The function
\be
\tilde \Pi _{in} ^{(0)}(y' )= \frac {d}{dy^\prime} \ln \left [ (e^{y'}-2)^{\frac {1+q_1+q_2}{2}}e^{-y'\left (\frac {1}{2}+q_2 \right )} {}_2F_1\left (\frac {1}{2}+k_3+q_2,\frac {1}{2}-k_3+q_2;1+q_1+q_2;1-\frac {2} {e^{y'}}\right ) \right ] \,  \label  {nf4lead2inc}
\ee
is the solution of (\ref {Pi-in-0}) which implements\footnote {We successfully compared (\ref {nf4lead2inc}) with the form for the solution incoming at the horizon given in (6.9) of \cite {CDDGS}.} the boundary condition for $\tilde \Pi _{in} ^{(0)}(y' )$, coming from (\ref {psi-in-ln2}).
After making this choice for $\tilde \Pi _{in} ^{(0)}(y' )$, the others equations (\ref {Pi-in-1}, \ref {Pi-in-n}) are solved by 
\be
 \tilde \Pi _{in} ^{(1)}(y' )=\frac {4q_3}{1-4k_3^2} \frac {d}{dy'}\left [\frac {e^{-y'}}{2}+\left (e^{-y'}-\frac {1}{2} \right )
 \tilde \Pi _{in} ^{(0)}(y') \right ]  \label {nf3expinc} \, , 
\ee
and, for $n\geq 2$, 
\be
\tilde \Pi _{in}^{(n)}(y')=\frac {d}{dy'}  \left  [P_{right,0}^{(n)}(e^{-y'})+\sum _{m=1}^{n}P_{right,m}^{(n)} (e^{-y'})\frac {d^{m-1}}{d{y'}^{m-1}} \tilde \Pi _{in}^{(0)}(y') \right ] \, , \label {solpin2bisin}
\ee
with $P_{right,m}^{(n)}(e^{-y'})$ degree $n$ polynomials: these coincide with the polynomials appearing in the expression (\ref {solpin2bis}) for the Floquet solution, since they are determined only by the form of the potential and not by the particular choice of the solution of (\ref {Pi-in-0}). This property of generality of the polynomials was proved for the MME after equation (\ref {finding-pol}), but, as remarked there, it can be easily extended to all the equations discussed in this paper. 

\medskip

The solution outgoing to infinity is more conveniently studied in its `left kink' version $\psi _{<,up}(y')=\psi _{up}(y'+\theta _3)$.
This is univocally defined by the decaying behaviour
\be
\psi _{<,up}(y') \simeq 2^{\frac {3}{2}-q_3}M^2 e^{\theta _3 \left (q_3 +\frac {1}{2}\right )}e^{\left (q_3+\frac {1}{2}\right )y'}\exp \left (-e^{-y'} \right ) \label {psi-up-infi}
\ee 
as $y' \rightarrow -\infty$. Its quantum momentum $\Pi _{<,up}(y') =\frac {d}{dy'} \ln \psi _{<,up}(y')$ is expanded, for $\theta_3\rightarrow -\infty$, in the region $y'<a_l-\theta _3$
\be
\Pi _{<,up}(y') =\sum _{n=0}^\infty \tilde \Pi _{<,up}^{(n)}(y') e^{n\theta_3} \, .
\label{up-infin}
\ee
Then, the components $ \tilde \Pi _{<,up}^{(n)}(y') $ must satisfy the system of (kink) ODEs (\ref {Pi-up-0}-\ref {Pi-up-n}), with the obvious replacement $\tilde \Pi _{<,left}^{(3,n)}(y') \rightarrow \tilde \Pi _{<,up}^{(n)}(y')$. The function
\be
\tilde \Pi _{<,up}^{(0)}(y')=\frac {d}{dy^\prime} \ln \left [\exp \left (-e^{-y'}+k_3y'  \right ) U\left (\frac {1}{2}-k_3+q_3,1-2k_3;2e^{-y'} \right ) \right ] \, , \label {nf4lead2up}
\ee
where $U(a,b;z)$ is the Tricomi confluent hypergeometric function\footnote {The Tricomi confluent hypergeometric function is this linear combination of the two ${}_1F_1 $-equation (Kummer's) 
\be
U(a,b;z)=\frac {\Gamma (1-b)}{\Gamma (a-b+1)}{}_1F_1 (a,b;z)+\frac {\Gamma (b-1)}{\Gamma (a)}z^{1-b}{}_1F_1 (a-b+1,2-b;z),
\ee
which enjoys the right asymptotics since $U(a,b;z) \simeq z^{-a}$ when $|z| \rightarrow +\infty$.
}, is the solution of (\ref {Pi-up-0}) which satisfies the boundary condition at $y'\rightarrow -\infty$ coming from (\ref {psi-up-infi}).
After making this choice for $\tilde \Pi _{<,up} ^{(0)}(y' )$, the others equations (\ref {Pi-up-1}-\ref {Pi-up-n}) are solved by 
\be
\tilde \Pi _{<,up}^{(1)}(y')=\left (-\frac {1}{4}+\frac {q_1q_2}{1-4k_3^2} \right ) \frac {d}{dy'} \Bigl [ \frac {1}{2}e^{y'}-e^{y'}\tilde \Pi _{<,up}^{(0)}(y') \Bigr ] \, 
\ee
and, in general, by  
\be
\tilde \Pi _{<, up}^{(n)}(y')=\frac {d}{dy'}  \left  [P_{left, 0}^{(n)}(e^{y'})+\sum _{m=1}^{n}P_{left,m}^{(n)} (e^{y'})\frac {d^{m-1}}{d{y'}^{m-1}}  \tilde \Pi _{<,up}^{(0)}(y') \right ] \, , \label {solp-up2ter}
\ee
with $P_{left,m}^{(n)}(e^{y'})$ degree $n$ polynomials. They coincide with the polynomials appearing in the expression (\ref {solpin2ter}), due to their independence of 
the choice of the particular solution of (\ref {Pi-up-0}).

\subsubsection {Proving Rome II \cite {CDFMPP} and solving MST \cite{MST} Ans\"atze for gravity and beyond}
\label {Rome-MST}

An important remark concerns the structure of the wave functions as entailed by our exponential form above, so that it holds, upon fixing the Heun-like equation, for any kind of solution, Floquet, incoming ({\it in}) and outgoing ({\it up}), and at any fixed expansion point, {\it i.e.} the radial infinity or the horizon. For instance, regarding the latter point expansion, let us recall the `right kink' Floquet solution $\psi _>(y)=\psi _+(y-2\theta)$ of the MME, 
\be
\psi _>(y)=\exp \left [ \int ^y dy' \Pi _>(y') \right ]=\exp \left [ \sum _{n=0}^{+\infty} e^{4n\theta }\int ^y dy' \tilde \Pi _>^{(n)} (y')  \right ] \, ,
\label {psi>final}
\ee
with $\Pi _>(y')$ given by (\ref {thetakexp}-\ref {solpin}). It can be manipulated in a so general manner to be replicable in all the other cases as well. In fact, upon using the explicit form (\ref {solpin}) for $\tilde \Pi _>^{(n)} (y)$, it reads, within the two instanton expansion, 
\ba
\psi _>(y)&=& \left [1+e^{4\theta}P_0^{(1)}(e^{-y})+e^{8\theta} \left ( P_0^{(2)}(e^{-y}) +\frac {1}{2}\left (P_0^{(1)}(e^{-y}) \right )^2 \right )  +O(e^{12\, \theta})\right ]  \psi _>^{(0)} (y) + \nonumber \\
&+&  \left [e^{4\theta}P_1^{(1)}(e^{-y})+e^{8\theta} \left ( P_1^{(2)}(e^{-y}) +P_0^{(1)}(e^{-y}) P_1^{(1)}(e^{-y})  \right )  +O(e^{12\, \theta})\right ]
\frac {d \psi  _>^{(0)} (y)}{dy}+ \label  {final-psi_>} \\
&+&  \left [ e^{8\theta} \left ( P_2^{(2)}(e^{-y})\frac {d}{dy}\tilde \Pi _>^{(0)}(y) +\frac {1}{2}\left ( P_1^{(1)}(e^{-y})\right )^2 \left (\tilde \Pi _>^{(0)}(y)\right )^2\right ) +O(e^{12\, \theta})  \right ]  \psi _>^{(0)} (y) \nonumber \, ,
\ea
where {\it the zero-th order wave-function} is one of the two Floquet solutions to the zero-th order ODE
\be
\psi  _>^{(0)} (y)=e^{\int^y \tilde \Pi ^{(0)}_>(y')} = 2\sqrt {\pi} e^{-i\pi k }e^{\frac {\theta +\varphi}{2}}
  J_{2k}\left (2i e^{\frac {y}{2}}\right ) \, ,  \label {psizero-right}
\ee
with the normalisation of $\psi _>(y)$ dictated by (\ref {floq+infty}). If the Heun-like equation and the expansion point are fixed, -- the MME and the horizon in this case --, then the other wavefuntion {\it i.e.} the `right kink' incoming solution ($U_{0,>}(y)=U_0(y-2\theta)$) shares the same polynomials, but it is given by only changing {\it the zero-th order wave-function}  with the one showing the suitable decaying asymptotic behaviour (as $y\rightarrow+\infty$): 
\be 
\psi  _>^{(0)} (y) \rightarrow  U_{0,>}^{(0)} (y)=2 (2\pi )^{-\frac {1}{2}} e^{-\frac {\theta}{2}} K_{2k}\left (2e^{\frac {y}{2}}\right ).
\label {Uzero-right}
\ee 
In other words, the zero-th order wave-function has to be chosen in such a way to inherit and implement the characterising property of the exact solution. For both solutions, we can  use the first of (\ref {tildePi>equ}) and substitute the derivative $\frac {d}{dy}\tilde \Pi _>^{(0)}(y)=e^y+k^2-\left ( \tilde \Pi _>^{(0)}(y)\right )^2$ inside the last square bracket. This leads to {\it the surprising cancellation} of the terms proportional to $\left (\tilde \Pi _>^{(0)}(y)\right )^2$, thanks to the relevant relation $P_2^{(2)}(e^{-y})=\frac  {1} {2}\left ( P_1^ {(1)}(e^{-y})\right ) ^2$; hence the third square bracket in (\ref{final-psi_>}) contains only polynomials in $e^{-y}$ and then can be grouped together with the first one. Analogous cancellations of powers $\left ( \tilde \Pi _>^{(0)}(y)\right ) ^n$ repeat with the same pattern at all computed orders in $e^ {4\theta}$ thanks to peculiar identities among the polynomials $P_m^{(n)}(e^{-y})$, only preserving $\psi  _>^{(0)} (y)$ and its first derivative. In mathematical terms, this leads to conjecture the exact wave function in the form
\ba
\psi _>(y)&=& \left ( \sum \limits_{n=0}^{+\infty} e^{4n\theta } {\cal P}^{(n)}(e^{-y}) \right ) \psi _>^{(0)} (y)+ \left ( \sum \limits_{n=1}^{+\infty} e^{4n\theta } {\cal Q}^{(n)}(e^{-y}) \right ) \frac {d \psi  _>^{(0)} (y)}{dy} \, , \label {psiform} \\
U_{0,>}(y)&=& \left ( \sum \limits_{n=0}^{+\infty} e^{4n\theta } {\cal P}^{(n)}(e^{-y}) \right )  U_{0,>}^{(0)} (y)+ \left ( \sum \limits_{n=1}^{+\infty} e^{4n\theta } {\cal Q}^{(n)}(e^{-y}) \right ) \frac {d U_{0,>}^{(0)}(y)} {dy}  \label {psiformU0} \, , 
\ea
where ${\cal P}^{(n)},{\cal Q}^{(n)}$ are polynomials in $e^{-y}$ of degree $n$, depending polynomially only on $P^{(n)}_0$ and $P^{(n)}_1$: ${\cal P}^{(0)}=1, {\cal P}^{(1)}=P^{(1)}_0, {\cal P}^{(2)}=\frac {1}{2}\left ( P_0^{(1)}\right )^2+P_0^{(2)}+\frac {1}{2}\left ( P_1^{(1)}\right )^2(e^y+k^2), ...$, ${\cal Q}^{(1)}=P^{(1)}_1, {\cal Q}^{(2)}=P_0^{(1)}P_1^{(1)}+P_1^{(2)}, ... $. In general for $n\geq 1$, 
\be
{\cal P}^{(n)}(e^{-y}) =\sum _{m=1}^n {\cal P}^{(n)}_m  e^{-my} \, , \quad {\cal Q}^{(n)}(e^{-y}) =\sum _{m=1}^n {\cal Q}^{(n)}_m e^{-my} \, . 
\ee
In other words, expanding around the horizon (up to any finite distance) the two wavefuntions can be treated on an equal footing by obtaining the same polynomials and changing only the zero-th order wave-function. 

Analogously, (still keeping the MME and) expanding around infinity, the Floquet and outgoing solutions can be treated on an equal footing by solving  `the left kink equation' (\ref {Ric<}) with the same form as above (\ref{psi>final}) the identical polynomials for both solutions, differing from the ones of the right kink. Anew, the two are given by only choosing two different zero-th order wave functions: the other zero-th order Floquet solution
\be
\psi^{(0)} _+ (y+2\theta)=\psi _<^{(0)} (y)=2\sqrt {\pi} e^{i\pi k }e^{\frac {\theta -\varphi}{2}}  J_{-2k}\left (2i e^{-\frac {y}{2}}\right )
\ee
for the `left kink' Floquet solution, whilst the zero-th order decaying one (as $y\rightarrow-\infty$)
\be
V^{(0)}_0(y+2\theta)=V_{0,<}^{(0)} (y)=2 (2\pi )^{-\frac {1}{2}} e^{-\frac {\theta}{2}}  K_{2k}\left (2e^{-\frac {y}{2}}\right )
\ee 
to give the `left kink' outgoing solution. Actually, (\ref {solpin<}) teaches us that the latter case is obtained from the `right-kink' by the replacement $P_m^{(n)}(e^{-y}) \rightarrow (-1)^m P_m^{(n)}(e^{y})$\footnote{In the last analysis, this is due to the evenness symmetry of the potential of the MME, therefore it is lost in the more physical CHE case.}, which implies
${\cal P}^{(n)}(e^{-y}) \rightarrow {\cal P}^{(n)}(e^{y})$ and ${\cal Q}^{(n)}(e^{-y})\rightarrow - {\cal Q}^{(n)}(e^{y})$. Eventually, the forms for the `left kink' Floquet and outgoing solution are, respectively
\ba
\psi _<(y)&=& \left ( \sum \limits_{n=0}^{+\infty} e^{4n\theta } {\cal P}^{(n)}(e^{y}) \right ) \psi _<^{(0)} (y)- \left ( \sum \limits_{n=1}^{+\infty} e^{4n\theta } {\cal Q}^{(n)}(e^{y}) \right ) \frac {d \psi  _<^{(0)} (y)}{dy} \, , \label {psiform<} \\
V_{0,<}(y)&=& \left ( \sum \limits_{n=0}^{+\infty} e^{4n\theta } {\cal P}^{(n)}(e^{y}) \right )  V_{0,<}^{(0)} (y)- \left ( \sum \limits_{n=1}^{+\infty} e^{4n\theta } {\cal Q}^{(n)}(e^{y}) \right ) \frac {d V_{0,<}^{(0)}(y)} {dy}  \label {psiformV0<} \, , 
\ea    
with ${\cal P}^{(n)}(x)$ and ${\cal Q}^{(n)}(x)$ the same polynomials as in (\ref {psiform}, \ref {psiformU0}).

\medskip

By applying this method, we found the same form as (\ref {psiform}, \ref  {psiformU0}) for solutions around the horizon and (\ref {psiform<}, \ref {psiformV0<}) for solutions around infinity, {\it i.e.}  a linear combination of the zero-th order solution (which is always a hypergeometric function) and its first derivative only, in all the cases of the DCHE, CHE and HE, thanks to the aforementioned cancellations of terms proportional to higher powers $\left ( \tilde \Pi ^{(0)}(y)\right ) ^n$. Moreover, the coefficients of the linear combination are easily expressed via our original polynomials and then can be truncated at a precise instanton order consistently, thus becoming polynomials in both the exponential variables $e^\theta$ and $e^{-y}$ (for horizon), $e^y$ (for infinity). As in the MME prototype, in each Heun-like equation these polynomials are the same upon fixing the expansion point: for Floquet and incoming solution at the horizon on one side, and for Floquet and outgoing solution at the spatial infinity on the other, since the equations determining the polynomials $P^{(n)}_m$ are by construction the same around the same point. In general, at fixed expansion point the difference lies only in the choice of the zero-th order solution which appears in a {\it universal} form with only its first derivative, {\it cf.} expressions above. For instance in the CHE case: around the horizon (\ref {nf4lead2}) for the Floquet, while (\ref{nf4lead2inc}) for incoming solution; similarly around infinity (\ref{Pi3+1}) for the Floquet, while (\ref {nf4lead2up}) for the outgoing. In other words, by solving the Riccati equation order by order, we can prove the validity of expressions (\ref{psiform}, \ref {psiformU0}) around horizon and (\ref {psiform<}, \ref  {psiformV0<}) around infinity for solutions of the Heun equation and its confluences and reductions ({\it cf.} also \cite {OFR} for details, the systematic solutions and proofs). Very interestingly, limitedly to Floquet solutions of the CHE the same form (\ref {psiform}) is present in Section IV of the recent \cite {CDFMPP}, as apparently guessed by trials and errors. Crucially, as already argued above, the series appearing in the exponent of (\ref {psi>final}) should have some non zero convergence radius as a function of $e^{4\theta}$, and then, since the exponential converges everywhere, each series in (\ref {psiform}) should share this radius. This fact is rather important in applications. 

\medskip

Eventually, a general pattern seems to show that the general form (\ref{psiform}-\ref{psiformV0<}) can be recast into a well-known one for exact solutions of CHE at small $M\omega$ \cite{MST}. In fact, we are going to extend the latter form to all Heun-like equations, thus confirming the validity of (\ref{psiform}-\ref{psiformV0<}). Let us analyse this derivation in some detail for the prototype MME around the horizon and subsequently for the original CHE \cite{MST}, both around infinity and around horizon. Actually, extensions to other Heun-like equations can be pursued along the same path, but for brevity their detailed description will not be discussed here.

Summarising, for the MME the zero-th order in (\ref {psiform}) is given by (\ref {psizero-right}), while the one in (\ref {psiformU0}) is insured by (\ref{Uzero-right}). Then, at order $e^{4n\theta}$, $n\geq 1$\footnote {For $n=0$ only the simple contribution $\psi _>^{(0)} (y)$ is present.}, $\psi _>(y)$ is given by 
\be
\left  ( \sum _{m=1}^n {\cal P}^{(n)}_m e^{-my} \right ) J_{2k}\left (2i e^{\frac {y}{2}}\right ) +
\left ( \sum _{m=1}^n {\cal Q}^{(n)}_m  e^{-my} \right ) \frac {d J_{2k}\left (2i e^{\frac {y}{2}}\right ) }{dy}   
\label {polyform-1} \, , 
\ee
as ${\cal P}^{(n)}(e^{-y})=\sum _{m=1}^n {\cal P}^{(n)}_m e^{-my}$ and ${\cal Q}^{(n)}(e^{-y})=\sum _{m=1}^n {\cal Q}^{(n)}_m  e^{-my}$; and similarly $U_{0,>}(y)$ is given by 
\be
\left  ( \sum _{m=1}^n {\cal P}^{(n)}_m e^{-my} \right ) K_{2k}\left (2 e^{\frac {y}{2}}\right ) +
\left (  \sum _{m=1}^n {\cal Q}^{(n)}_m  e^{-my} \right ) \frac {d K_{2k}\left (2 e^{\frac {y}{2}}\right ) }{dy}   
\label {polyform-2} \, . 
\ee
In both cases, the repeated use of the functional relations 
\be
z^{-1}J_{2k}(z) =\frac {J_{2k-1}(z)+J_{2k+1}(z) } {4k}  \, , \quad \frac {d}{dz}J_{2k}(z) =\frac  {J_{2k-1}(z)-J_{2k+1}(z)}  {2}  \, , \quad \label {Bess-prop-1}
\ee
\be
z^{-1}K_{2k}(z)=\frac {K_{2k+1}(z)-K_{2k-1}(z) } {4k}  \, , \quad \frac {d}{dz}K_{2k}(z) = -\frac {K_{2k-1}(z)+K_{2k+1}(z)}  {2}  \, , \quad \label {Bess-prop-2}
\ee
allows us to write all the terms appearing in (\ref {polyform-1}, \ref {polyform-2}), in terms of constants multiplied by Bessel functions with shifted order $2k+2h$, $|h|\leq n$. Explicitly, (\ref {psiform}, \ref {psiformU0}) acquire the form
\ba
\psi _>(y)&=&2\sqrt {\pi} e^{-i\pi k +\frac {\theta +\varphi}{2}} \sum _{n=0}^{+\infty} e^{4n\theta } \left ( \sum _{h=-n}^{n}A_h^{(n)}(k) J_{2k+2h}\left (2i e^{\frac {y}{2}} \right ) \right ) \, , \label {pre-MST-like-0}\\
U_{0,>}(y)&=&\sqrt {\frac {2}{\pi}} e^{-\frac {\theta}{2}}  \sum _{n=0}^{+\infty} e^{4n\theta } \left ( \sum _{h=-n}^{n}B_h^{ (n)}(k) K_{2k+2h}\left (2 e^{\frac {y}{2}} \right ) \right ) \, ,  \label {pre-MST-like} \, 
\ea
with $A_h^{ (n)}(k)$, $B_h^{(n)}(k)$ ($A_0^{ (0)}(k)=B_0^{(0)}(k)=1$) constant coefficients, or, equivalently,
\be
\psi _>(y)=2\sqrt {\pi} e^{-i\pi k + \frac {\theta +\varphi}{2}}  \sum _{h=-\infty}^{+\infty} a_h (\theta , k) J_{2k+2h}\left (2i e^{\frac {y}{2}} \right ) \, , \ 
U_{0,>}(y)=\sqrt {\frac {2}{\pi}} e^{-\frac {\theta}{2}}  \sum _{h=-\infty}^{+\infty}  b_h (\theta , k) K_{2k+2h}\left (2 e^{\frac {y}{2}} \right ) \label {MST-like} \, , 
\ee
where
\be
a_h (\theta , k)=
\sum _{n=|h|}^{+\infty}e^{4n\theta }  A_h^{(n)}(k) \, , \quad  b_h (\theta , k)=
\sum _{n=|h|}^{+\infty}e^{4n\theta }  B_h^{(n)}(k) \, . 
\label{MME-MST-sols}
\ee
In other words, these are the Floquet and incoming solutions of the `right-kink' MME in the alternative form (\ref  {MST-like}). Following a similar strategy and in particular using again properties (\ref {Bess-prop-1}, \ref {Bess-prop-2}), also the Floquet (\ref {psiform<}) and the outgoing (\ref {psiformV0<}) solutions of the `left-kink' MME can be put in the forms (\ref {pre-MST-like-0}, \ref {pre-MST-like}) or (\ref {MST-like}). 
Although for the CHE, these solution structures are quite common in gravity problems as developed initially in the well-known paper \cite {MST}; in addition here the expressions (\ref{MME-MST-sols}) are already explicit solutions for the unknown MST coefficients. Therefore, contrarily to the previous form with polynomials, this form with the shifted indices does not show any simple relation between $a$'s and $b$'s\footnote{For instance, for the Floquet solution $A_0^{(1)}(k)=\frac {4}{(4k^2-1)^2}$, $A_1^{(1)}(k)=-\frac {1}{2k(2k+1)^2}$, $A_{-1}^{(1)}(k)=A_1^{(1)}(-k)$, while for the outgoing one $B_0^{(1)}(k)=\frac {4}{(4k^2-1)^2}$, $B_1^{(1)}(k)=\frac {1}{2k(2k+1)^2}$, $B_{-1}^{(1)}(k)=B_1^{(1)}(-k)$.}.

Remarkably, also the solutions to the CHE proposed before can be arranged in exactly the same form as in \cite {MST}. For instance, the `left kink' Floquet solution (\ref {paiexp}) 
\be
\psi _<^{(3)}(y';k_3) = \exp \int ^{y'} dy' \Pi _< ^{(3)}(y';k_3) \, , 
\ee
(where,  for future convenience, the dependence on Floquet index was made explicit) is the form ready to be expanded around the radial infinity and then acquires a form analogous to (\ref {psiform<}):
\be
\psi _<^{(3)}(y';k_3)= \left ( \sum \limits_{n=0}^{+\infty} e^{n{\theta _3} } {\cal P}^{(3, n)}_{left}(e^{y'}) \right ) \psi _<^{(3,0)} (y';k_3)+ \left ( \sum \limits_{n=1}^{+\infty} e^{n{\theta _3}} {\cal Q}^{(3, n)}_{left}(e^{y'}) \right ) \frac {d \psi  _<^{(3,0)} (y';k_3)}{dy'} \, , \label {psiform3}
\ee
where ${\cal P}^{(3, n)}_{left}(e^{y'}) , {\cal Q}^{(3, n)}_{left}(e^{y'}) $ are polynomials of degree $n$, which depend polynomially only on the polynomials $P_{left,0}^{(n)}(e^{y'})$ and $P_{left,1}^{(n)}(e^{y'})$ in (\ref {solpin2ter}). The first ones are
\be
{\cal P}^{(3, 0)}_{left}(e^{y'})=1 \, , \  {\cal P}^{(3, 1)}_{left}(e^{y'})=\frac {1}{2}\left (\frac {q_1q_2}{1-4k_3^2}-\frac {1}{4} \right) e^{y'}\, , \
 {\cal Q}^{(3, 1)}_{left}(e^{y'})=-\left (\frac {q_1q_2}{1-4k_3^2}-\frac {1}{4} \right) e^{y'}
\ee
and in general, when $n\geq 1$, 
\be
{\cal P}^{(3, n)}_{left}(e^{y'})=\sum _{m=0}^n {\cal P}^{(3,n)}_{left,m}  e^{my'} \, , \quad {\cal Q}^{(3, n)}_{left}(e^{y'})=\sum _{m=1}^n {\cal Q}^{(3,n)}_{left,m}  e^{my'} \, . \label {Pol-form-3}
\ee
According to the general scheme, the zero-th order wave-function $ \psi _<^{(3,0)} (y';k_3)$ is one of the two Floquet solutions to the zero-th order ODE
\be
\frac {d^2}{d{y'}^2}\psi _<^{(3,0)}(y';k_3)=\left (k_3^2 +2q_3 e^{-y'}+e^{-2y'} \right ) \psi _<^{(3,0)}(y';k_3) \, ,
\ee
namely, recalling (\ref {lead-Flo-1}), $\psi _<^{(3,0)}(y';k_3)={\cal N}\psi _{Coul}(y';k_3)$, with ${\cal N}$ a suitable normalisation factor\footnote {In order to avoid heavy notations here and in the following, we use the same symbol ${\cal N}$ to indicate the different normalisation factor for the various solutions. Hopefully, there will be no confusion.} which implements the boundary condition (\ref {floq3infin}, \ref {symm-choice}) and the Coulomb hypergeometric function defined as
\be
\psi _{Coul}(y';k_3)= i (2i)^{-\frac {1}{2}-k_3}\frac {\Gamma \left(\frac {1}{2}-k_3-q_3 \right )}{\Gamma (1-2k_3)}\exp \left (-e^{-y'}+ k_3y'  \right ) {}_1F_1\left (\frac {1}{2}- k_3+q_3,1 - 2k_3;2 e^{-y'} \right )\, . \label {floq-tv} 
\ee
Relation (\ref {psiform3}) was verified at first instanton orders, using identities between polynomials which give cancellations of unwanted $\left ( \tilde \Pi ^{(0)}(y)\right ) ^n$ terms with the same mechanism as for the MME. Therefore, it seems very reasonable to conjecture that the same cancellations happen at all instanton orders so that this form holds in general: it will be proven rigorously \cite{OFR}. Actually, this is exactly one of the two forms of Floquet solutions to CHE as appeared in \cite{CDFMPP}, the other being, of course, analogously derivable from the expansion (\ref{paiexp-hor}-\ref {Pmrightform}) around the horizon (without need of the kink). We have carefully verified, up to one instanton order, $e^{\theta _3}$, that the coefficients of the polynomials in (\ref {psiform3}) reproduce those ($c$'s) of (69,70) in \cite {CDFMPP} (concerning the Floquet solution).

Furthermore, these relations among the confluent hypergeometric function
\ba
{}_1F_1(a,b;w)&=&w^2 \frac {a(a-b)}{b^2(b^2-1)} \, {}_1F_1(a+1,b+2;w)+w\frac  {b-2a}{b(b-2)} \, {}_1F_1(a,b;w)+{}_1F_1(a-1,b-2;w) \, , \nonumber \\
\frac  {d}{dw}\,  {}_1F_1(a,b;w)&=&w\frac {a(b-a)}{b^2(b+1)}\, {}_1F_1(a+1,b+2;w)+\frac {a}{b} \,  {}_1F_1(a,b;w) \, \label {shift-0}, 
\ea
translate, thanks to (\ref {floq-tv}), into the identities
\be
-i {e^{y'}} \psi _{Coul}(y';k_3)=-\frac {\frac {1}{2}-k_3+q_3}{k_3(1-2k_3)}\psi _{Coul}(y';k_3-1) - \frac {iq_3}{\frac {1}{4}-k_3^2}\psi _{Coul}(y';k_3) - \frac {\frac {1}{2}+k_3+q_3}{k_3(1+2k_3)}\psi _{Coul}(y';k_3+1) \label {shift-1} \, , 
\ee
\be
ie^{y'}\frac {d}{dy'} \psi _{Coul}(y';k_3)=-\frac {\frac {1}{2}-k_3+q_3}{1-2k_3}\psi _{Coul}(y';k_3-1) - \frac {iq_3}{\frac {1}{2}-2k_3^2}\psi _{Coul}(y';k_3) + \frac {\frac {1}{2}+k_3+q_3}{1+2k_3}\psi _{Coul}(y';k_3+1)  \label {shift-2} \, . 
\ee
Actually, the reason for introducing the function (\ref {floq-tv}) is to make contact with the relevant and widespread paper \cite {MST}. Indeed, by using the map $z=i e^{-y'}$, $\nu =-\frac {1}{2}-k_3$, $-2-i\epsilon =q_3$ from parameters of \cite {MST} to ours, it is related in a simple way to the Coulomb wave function $F_\nu (z)$, as defined in \cite {MST}, by 
\be
\psi _{Coul}(y';k_3)=e^{\frac {y'}{2}} F_\nu (z) \label  {maptocoulomb} \, . 
\ee
Of course, the function $F_\nu (z)$ satisfies the properties (3.5, 3.6) of \cite {MST}, which translate into (\ref {shift-1}, \ref {shift-2}) or into the elegant (\ref {shift-0}) in our notation. At this point, properties (\ref {shift-1}, \ref {shift-2}), -- which play for the CHE the same r\^ole as (\ref {Bess-prop-1}) for MME --, allow us to substitute in (\ref {psiform3}) all the powers of $e^{y'}$, coming from the polynomials ${\cal P}^{(3, n)}_{left}$, ${\cal Q}^{(3, n)}_{left}$ (\ref {Pol-form-3}), with integer shifts in the Floquet index of the Coulomb function: eventually, at order $e^{n\theta}$ the function $\psi _<^{(3)}(y';k_3) $  (\ref {psiform3}) becomes a linear combination of $\psi _{Coul}(y';k_3+h)$, with $|h|\leq n$. More explicitly,
\be
\psi _<^{(3)}(y';k_3)= {\cal N} \sum _{n=0}^{+\infty} e^{n{\theta _3}} \left ( \sum _{h=-n}^{n}C_h^{(n)}(k_3)
\psi _{Coul} (y'; k_3+h) \right ) \, , \quad C_0^{(0)}(k_3)=1 \label  {pre-MST-like-CHE} 
\ee
or, equivalently,
\be
\psi _<^{(3)}(y';k_3)={\cal N} \sum _{h=-\infty}^{+\infty} c_h (\theta _3, k_3) \psi _{Coul} (y'; k_3+h)   \label {MST-like-CHE} \, , 
\ee
where
\be
c_h (\theta _3, k_3)=
\sum _{n=|h|}^{+\infty}e^{n\theta _3}  C_h^{(n)}(k_3) \, . \label {c-C-rel}
\ee
Easily, for instance, at the first non trivial order $e^\theta$, 
\be
C_{-1}^{(1)}(k_3)=-i \left (\frac {q_1q_2}{1-4k_3^2}-\frac {1}{4} \right ) \frac {(1+2k_3)\left (\frac {1}{2}-k_3+q_3 \right )}{2k_3(1-2k_3)} \, , \label {Ch-1lp}
\ee
\be
C_{0}^{(1)}(k_3)=-i \left (\frac {q_1q_2}{1-4k_3^2}-\frac {1}{4} \right ) \frac {2iq_3}{\frac  {1}{2}-2k_3^2} \, , 
\label {Ch0lp}
\ee
\be
C_{1}^{(1)}(k_3)=-i \left (\frac {q_1q_2}{1-4k_3^2}-\frac {1}{4} \right ) \frac {(1-2k_3)\left (\frac {1}{2}+k_3+q_3 \right )}{2k_3(1+2k_3)} \, . \label {Ch1lp}
\ee
Impressively, the form (\ref {MST-like-CHE}) coincides with and justify the Ansatz (3.3) of \cite  {MST} for solutions of the 'left-kink' CHE and, in detail, the coefficients (\ref {Ch-1lp}, \ref {Ch0lp}, \ref {Ch1lp}) agree with explicit small frequency computations there (we have explicitly shown this up to the order $e^{\theta _3}\propto M\omega$). 

The same results hold for the other kind of solution around radial infinity, the outgoing, and for the other expansion point, the horizon. The solution outgoing at infinity can be written as  
\be
\psi _{<,up}(y';k_3)= \left ( \sum \limits_{n=0}^{+\infty} e^{n\theta _3} {\cal P}^{(3, n)}_{left}(e^{y'}) \right ) \psi _{<,up}^{(0)} (y';k_3)+ \left ( \sum \limits_{n=1}^{+\infty} e^{n\theta _3} {\cal Q}^{(3, n)}_{left}(e^{y'}) \right ) \frac {d \psi  _{<,up}^{(0)} (y';k_3)}{dy'} \, , \label {psiform3up}
\ee
where the polynomials are the same as in (\ref {psiform3}), while the suitable decaying behaviour (\ref{psi-up-infi}) is insured by 
\be
\psi _{<,up}^{(0)} (y';k_3)= {\cal N}\psi _{U} (y';k_3)\, , 
\ee
with ${\cal N}$ a  ($k_3$-independent) normalisation constant. This is indeed the Tricomi's $U$ up a prefactor as above (or a linear combination of Coulomb functions) 
\ba
\psi _{U} (y';k_3)&=& \left (\frac {i}{2}\right )^{k_3}\exp (-e^{-y'}+k_3 y' ) U\left (\frac {1}{2}-k_3+q_3,1-2k_3; 2e^{-y'}\right) = \\
&=&\left (\frac {2}{i}\right )^{\frac {1}{2}}i^{k_3}\left [ \frac {i^{k_3}\cos \pi (k_3+q_3)}{\sin 2\pi k_3}\psi _{Coul}(y';k_3)- \frac {i^{-k_3}\cos \pi (k_3-q_3)}{\sin 2\pi k_3}\psi _{Coul}(y';-k_3)\right ] \, . 
\nonumber
\ea
The choice of the $k_3$-dependent normalisation is such that $\psi _{U} (y';k_3)$ satisfies the same relations (\ref {shift-1}-\ref {shift-2}) as $\psi _{Coul}(y';k_3)$. Therefore, the outgoing solution enjoys the expansions 
\be
\psi _{<,up}(y';k_3)=  {\cal N} \sum _{n=0}^{+\infty} e^{n\theta _3} \left ( \sum _{h=-n}^{n}C_h^{(n)}(k_3)
\psi _{U} (y'; k_3+h) \right ) \, , \quad C_0^{(0)}(k_3)=1 \label  {pre-MST-like-CHE-up} \, , 
\ee
with the same coefficients $C_h^{(n)}(k_3)$ as for the Floquet solution $\psi _<^{(3)}(y';k_3)$ and then 
\be
\psi _{<,up}(y';k_3)={\cal N} \sum _{h=-\infty}^{+\infty} c_h (\theta _3, k_3) \psi _{U} (y'; k_3+h)   \label {MST-like-CHE-up} \, , 
\ee
with the same definition (\ref{c-C-rel}).

As the form of Floquet solution (\ref{psiform3}) arises around infinity, similarly (\ref {paiexp-hor}-\ref {Pmrightform}) can be rearranged around the horizon in this way
\be
\psi _{+}^{(3)}(y';k_3)= \left ( \sum \limits_{n=0}^{+\infty} e^{n\theta _3} {\cal P}^{(3, n)}_{right}(e^{-y'}) \right )
\psi _{+}^{(3,0)}(y';k_3)+  \left ( \sum \limits_{n=1}^{+\infty} e^{n\theta _3} {\cal Q}^{(3, n)}_{right}(e^{-y'}) \right )
\frac {d \psi _{+}^{(3,0)}(y';k_3)}{dy'}   \label {floq-hor-almost-MST} \, , 
\ee
where ${\cal P}^{(3, n)}_{right}(e^{-y'}) , {\cal Q}^{(3, n)}_{right}(e^{-y'}) $ are polynomials of degree $n$ of the variable $\left (e^{-y'}-\frac {1}{2} \right )$ (with ${\cal P}^{(3, 0)}_{right}=1$ and $ {\cal Q}^{(3, n)}_{right}$ missing the constant term), which depend polynomially only on the polynomials $P_{right,0}^{(n)}(e^{-y'})$ and $P_{right,1}^{(n)}(e^{-y'})$ in (\ref {solpin2bis}) and the zero-th order Floquet solution is
\be
\psi _{+}^{(3,0)}(y';k_3)= {\cal N}  (e^{y'}-2)^{\frac {1-q_1-q_2}{2}}e^{k_3y'} {}_2F_1\left (\frac {1}{2}+k_3-q_1,\frac {1}{2}+k_3-q_2;1+2k_3;\frac {e^{y'}}{2} \right ) \, .
\ee
Anew, this has been checked at first instanton orders, but it should be true at all orders \cite{OFR}. In fact, the following identities of the hypergeometric function 
\ba
&& {}_2F_1 (a,b;c;w)=w^2 \frac {ab(c-a)(c-b)}{c^2(c^2-1)} {}_2F_1 (a+1,b+1;c+2;w)+ w \frac {c(a+b-1)-2ab}{c(c-2)} {}_2F_1 (a,b;c;w) + \nonumber \\
&&+ {}_2F_1 (a-1,b-1;c-2;w) \nonumber \\
&& (1-w)\frac {d}{dw} {}_2F_1 (a,b;c;w)=-w \frac {ab(c-a)(c-b)}{c^2(c+1)} {}_2F_1 (a+1,b+1;c+2;w)+ \frac {ab}{c} {}_2F_1 (a,b;c;w)
\label{funct-right-floq}
\ea
allow us to rewrite (\ref {floq-hor-almost-MST}) as 
\be
\psi _{+}^{(3)}(y';k_3)= {\cal N} \sum _{n=0}^{+\infty} e^{n\theta _3} \left ( \sum _{h=-n}^{n}D_h^{(n)}(k_3)
\psi _{+}^{(3,0)} (y'; k_3+h) \right ) \, , \quad D_0^{(0)}(k_3)=1 \, ,
\ee
with constant coefficients $D_h^{(n)}(k_3)$ and then again a MST form solution. Again around the horizon the incoming solution (\ref {inc-hor}-\ref {solpin2bisin}) can be rewritten as
\be
\psi _{in}(y';k_3)= \left ( \sum \limits_{n=0}^{+\infty} e^{n\theta _3} {\cal P}^{(3, n)}_{right}(e^{-y'}) \right )
\psi _{in}^{(0)}(y';k_3)+  \left ( \sum \limits_{n=1}^{+\infty} e^{n\theta _3} {\cal Q}^{(3, n)}_{right}(e^{-y'}) \right )
\frac {d  \psi _{in}^{(0)}(y';k_3)}{dy'} \label {psi-in-almost-MST} \, , 
\ee
with the same polynomials as in (\ref {floq-hor-almost-MST}) and the zero-th order decaying solution
\be
\psi _{in}^{(0)}(y';k_3)= {\cal N} (e^{y'}-2)^{\frac {1+q_1+q_2}{2}}e^{-y'\left (\frac {1}{2}+q_2 \right )} {}_2F_1\left (\frac {1}{2}+k_3+q_2,\frac {1}{2}-k_3+q_2;1+q_1+q_2;1-\frac {2} {e^{y'}} \right ) \, .
\ee
Now, the following properties of the hypergeometric function 
\ba
&& w \, {}_2F_1 (a,b;c;w)=\frac {a(b-c)} {(a-b)(a-b+1)}{}_2F_1 (a+1,b-1;c;w)+\frac {c(a+b-1)-2ab}{(a-b)^2-1} {}_2F_1 (a,b;c;w) + \nonumber \\
&&+ \frac {b(a-c)} {(a-b)(a-b-1)}{}_2F_1 (a-1,b+1;c;w)  \nonumber \, , \\
&& w(1-w) \frac {d}{dw}\, {}_2F_1 (a,b;c;w)=\frac {ab(b-c)} {(a-b)(a-b+1)}{}_2F_1 (a+1,b-1;c;w)+\frac {(a+b+1-2c)ab}{1-(a-b)^2} {}_2F_1 (a,b;c;w) + \nonumber \\
&&+ \frac {ab(a-c)} {(a-b)(a-b-1)}{}_2F_1 (a-1,b+1;c;w) \, , 
\label{funct-right-in}
\ea
(namely (2.5) and (2.6) of \cite {MST}) are effective for recasting (\ref {psi-in-almost-MST}) in the form
\be
\psi _{in}(y';k_3)= {\cal N} \sum _{n=0}^{+\infty} e^{n\theta _3} \left ( \sum _{h=-n}^{n}E_h^{(n)}(k_3)
\psi _{in}^{(0)} (y'; k_3+h) \right ) \, , \quad E_0^{(0)}(k_3)=1 \, , 
\ee
with constant coefficients $E_h^{(n)}(k_3)$\footnote{Connected with the $D_h^{(n)}(k_3)$.} and then finally a MST form.

\medskip

In conclusion, first our various solutions to the CHE, both around the horizon and radial infinity, can be rearranged into the `MST-Ansatz' form \cite {MST}. Secondly, our alternative derivation gives directly the specific solution for the MST coefficients in each case without mention to their recursive relations \cite {MST}. In fact, the solution provides directly the $\theta _3$ dependence separated from the $k_3$ one. For instance in the relation (\ref {c-C-rel}) for the Floquet around infinity, the dependence on $k_3$ is contained only in $ C_h^{(n)}(k_3)$, which, in our method, is determined polynomially by the coefficients of only the polynomials $P_{left,m}^{(n)}(e^{y'})$, $m=0,1$, -- systematically derived (for details {\it cf.} \cite{OFR}) --, and by the constants appearing upon repeated use of the relations (\ref {shift-1}, \ref {shift-2}).

\subsubsection {Another application to gravity: the dynamical Tidal Love Number}
\label {applgra}

A quantity which can be computed is the dynamical Tidal Love Number (TLN). This number  
captures the so-called tidal response of a body to the external
gravitational field and it is named after A.E.H. Love, who first studied it in the context of 
Earth's response to the tides \cite {LOVE}. The dynamical TLN can be read from the solution $\psi _{in}$, incoming at the horizon, in its form far from the horizon, but not at infinite distance ($a_r\leq  y'<\ln 2$, with $y'$ close to $a_r$ and $-\infty< a_r\ll 0$ or, in gravity variables, $2M\ll r <4Me^{-a_r} <+\infty$, large $r$). Precisely, it can be given by the expression (\ref {00-floq}) for $\psi _{in}$ in terms of Floquet solutions $\psi _\pm ^{(3)}$. In fact, at small instanton coupling $e^{\theta _3}$ or low frequency $M\omega \ll 1$, the Floquet solutions can be read from (\ref {paiexp-hor}, \ref {nf3exp}, \ref  {solpin2bis}):
\ba
&& \psi _\pm ^{(3)}(y')\simeq \tilde B_\pm (e^{y'}-2)^{\frac {1-q_1-q_2}{2}}e^{\pm k_3y'} {}_2F_1\left (\frac {1}{2}\pm k_3-q_1,\frac {1}{2}\pm k_3-q_2;1\pm 2k_3;\frac {e^{y'}}{2} \right ) \cdot \nonumber \\
&& \cdot \exp \left [\frac {4q_3}{1-4k_3^2} \left (\frac {e^{-y'}}{2}+\left (e^{-y'}-\frac {1}{2} \right )
 \tilde \Pi _{+,right}^{(3,0)}(y',\pm k_3) \right ) e^{\theta _3}+O(e^{2\theta _3}) \right ] 
\label{lead-Flo-2-bis} \, .
\ea
With our normalisation of the Floquet functions (\ref {floq3ln2}), 
the constants $\tilde B_\pm$ satisfy the equality 
\ba
&& \tilde B_\pm  2^{\pm k_3}{}_2F_1\left (\frac {1}{2}\pm k_3-q_1,\frac {1}{2}\pm k_3-q_2;1\pm 2k_3;1 \right )\exp \left [\frac {(q_1+q_2)q_3 e^{\theta _3}}{1-4k_3^2}+O(e^{2\theta _3})\right ] = \nonumber \\
&& = \tilde B_\pm  2^{\pm k_3} \frac {\Gamma (1\pm 2k_3) \Gamma (q_1+q_2)}{\Gamma \left (\frac {1}{2}\pm k_3+q_1 \right )
\Gamma \left (\frac {1}{2}\pm k_3+q_2 \right )}\exp \left [\frac {(q_1+q_2)q_3 e^{\theta _3}}{1-4k_3^2}+O(e^{2\theta _3})\right ] =
 \frac {\eta _\pm }{\sqrt {2}} = \frac { e^{\frac {\varphi _\pm^{(3)}} {2}} }{\sqrt {2}}\, . \label {00-norm}
\ea
The exponent in the left hand side, that we have explicitly written only at order $e^{\theta _3}$, is the same for $\psi _+^{(3)}(y')$ and $\psi _-^{(3)}(y')$. Then, this exact computation follows easily
\be
\frac {e^{-\Phi _3} \tilde B_+}{\tilde B_-} = \frac {e^{\frac {\varphi _-^{(3)}-\varphi _+^{(3)}}{2} } \tilde B_+}{\tilde B_-}=\frac {\Gamma (1-2k_3) \Gamma \left (\frac {1}{2}+k_3+q_1 \right ) \Gamma \left (\frac {1}{2}+k_3+q_2 \right ) }{2^{2k_3}\Gamma (1+2k_3) \Gamma \left (\frac {1}{2}-k_3+q_1 \right )\Gamma \left (\frac {1}{2}-k_3+q_2 \right )} \, . 
\label {exactratio}
\ee
Let us come back to the determination of dynamical TLN. In the above stated region and by using (\ref {lead-Flo-2-bis}), the incoming solution (\ref {00-floq}) becomes a superposition of two Floquet 'waves' ($e^{\pm k_3 y' }$)
\be
\psi _{in} (y') \simeq \tilde C  \left [   - e^{-\frac {\varphi _+^{(3)}}{2}}\tilde B_+ (-2)^{\frac {1-q_1-q_2}{2}} e^{k_3 y'} \Bigl (1+e^{\theta _3}f^{resp}(y') \Bigr) + e^{-\frac {\varphi _-^{(3)}}{2}}\tilde B_-  (-2)^{\frac {1-q_1-q_2}{2}} e^{- k_3 y'} \Bigl (1+e^{\theta _3}f^{tidal}(y') \Bigr ) \right ]  \label {r} \, ,
\ee
where $f^{resp}$ and $f^{tidal}$ are two functions which take into account instanton correction in (\ref {lead-Flo-2-bis}). To justify their names, we shall switch to gravity variables. Upon using the parametrisation $k_3=\hat \ell +\frac {1}{2}$, with
\be
\hat \ell =\ell + O\left ((M\omega )^2\right) \, ,
\ee
we can rewrite
\ba
\psi _{in} (r) &\simeq & \tilde C  (-2)^{2iM\omega -\frac {1}{2}}\left [   - e^{-\frac {\varphi _+^{(3)}}{2}}\tilde B_+ 2^{\hat \ell +\frac {1}{2}} \left (\frac {2M}{r} \right )^{\hat \ell +\frac {1}{2}}\Bigl (1-4iM\omega f^{resp}(r)\Bigr ) + \right. \nonumber \\
&+& \left . e^{-\frac {\varphi _-^{(3)}}{2}}\tilde B_- 2^{-\hat \ell - \frac {1}{2}} \left (\frac {2M}{r} \right )^{-\hat \ell -\frac {1}{2}} \Bigl (1-4iM\omega f^{tidal}(r)\Bigr ) \right ]  \label {rbis} \, , 
\ea
where, following for instance \cite {CDGP},  the second term, $ e^{-\frac {\varphi _-^{(3)}}{2}}\tilde B_- 2^{-\hat \ell - \frac {1}{2}} \left (\frac {2M}{r} \right )^{-\hat \ell -\frac {1}{2}}  \Bigl (1-4iM\omega f^{tidal}(r)\Bigr)  $, can be identified with the tidal part of the radial perturbation function and the first term, $-e^{-\frac {\varphi _+^{(3)}}{2}}\tilde B_+ 2^{\hat \ell +\frac {1}{2}} \left (\frac {2M}{r} \right )^{\hat \ell +\frac {1}{2}}\Bigl (1-4iM\omega f^{resp}(r)\Bigr )  $, with the response function. In detail, they expand regularly at large $r$($<4Me^{-a}$) as
\be
f^{tidal}(r)=A^{tidal}(M\omega)+.... \, , \quad f^{resp}(r)=A^{resp}(M\omega)+... \, , 
\ee
where $A^{tidal}(M\omega)$ and $A^{resp}(M\omega)$ are constants, computable as series of non-negative integer powers of $M\omega$, while, the dots represent terms containing powers of $r$ and $\ln r$, which however do not contribute to the Love number. In fact, the dynamical TLN is the ratio between the coefficient of the term of the response function which multiplies $\left (\frac {2M}{r} \right )^{\ell + \frac {1}{2}}$ and the coefficient of the term of the tidal part which multiplies $\left (\frac {2M}{r} \right )^{-\ell -\frac {1}{2}}$:
\be
{\cal L}=-\frac { e^{-\Phi _3}\tilde B_+ 2^{2\hat \ell +1} }{\tilde B_-} \frac {1-4iM\omega A^{resp}(M\omega)}{1-4iM\omega A^{tidal}(M\omega)}   \, .
\ee
Often in literature (see for instance \cite {CDI}) only the real part of ${\cal L}$ is named dynamical TLN. We follow conventions of \cite {BILTKerr} and call dynamical TLN the entire `response' coefficient  ${\cal L}$. It depends on $M\omega$ and its value for $\omega =0$ (static external perturbation) is called static TLN. Within our framework it is easy to compute the dynamical TLN, in particular up to the first order in $M\omega$. In fact, by using (\ref {exactratio}), 
it reads
\be
{\cal L}=\frac {\Gamma (-2\hat \ell -1) \Gamma \left (\hat \ell+3-2iM\omega\right ) \Gamma \left (\hat \ell +1-2iM\omega \right ) }{\Gamma (2\hat \ell +1) \Gamma \left (-\hat \ell +2 -2iM\omega \right )\Gamma \left (-\hat \ell -2iM\omega \right )} [1+O(M\omega )] \, .
\ee
Of course, the static TLN is obtained by setting $\omega =0$ 
\be
{\cal L}=\frac {\Gamma (-2\ell -1) \Gamma \left (\ell+3 \right ) \Gamma \left (\ell +1 \right ) }{\Gamma (2\ell +1) \Gamma \left (-\ell +2 \right )\Gamma \left (-\ell  \right )} \, 
\ee 
and vanishes when computed for physical values integer $\ell \geq 2$ because all the functions $\Gamma (-2\ell -1), \Gamma (-\ell +2), \Gamma (-\ell)$ have simple poles. Due to this fact, the $O(M\omega)$ contribution to the whole dynamical TLN comes only from the $O(M\omega)$ term in 
\be
\frac {\Gamma (-2\hat \ell -1) \Gamma \left (\hat \ell+3-2iM\omega\right ) \Gamma \left (\hat \ell +1-2iM\omega \right ) }{\Gamma (2\hat \ell +1) \Gamma \left (-\hat \ell +2 -2iM\omega \right )\Gamma \left (-\hat \ell -2iM\omega \right )} \, .
\ee
Then, the dynamical TLN reads
\be
{\cal L}=2iM\omega \frac {(\ell +2)! (\ell !)^2 (\ell -2)!}{(2\ell +1)!(2\ell )! } +O((M\omega)^2)
\ee
and agrees with results (28,29) of \cite {CDGP}.

\medskip

We have here built up the foundations for more detailed analysis and calculations on this interesting research field: we will leave this aim for the near future.

\section{Conclusions and outlook}
\label{concl}
\setcounter{equation}{0}

We have studied the change from the ODE/IM (in/up in gravity) to the Floquet basis for the MME, DCHE, CHE and HE. The expressions for the four cases, (\ref {chgbasis}) for the MME, (\ref {nf2psi}) for the DCHE,  (\ref {00-floq}, \ref {0--floq}) for the CHE, (\ref {nf4psi}) for the HE, depend on the acquired phase $\varphi$, the two contributions to it, $\varphi _<$ (from the region $y<0$), $\varphi _>$ (from the region $y>0$) and the quantum momentum at zero, $\Pi _+(y=0)$. Whilst this last quantity can be easily computed as $\theta \rightarrow -\infty$, the computation of $\varphi$ and $\varphi _<$, $\varphi _>$ would naively diverge as suffering from the conflicting $\theta \rightarrow -\infty$ and $y \rightarrow \pm \infty$ limits. A new method has solved the issue by introducing two kink Riccati equations, (\ref {Rictilde}, \ref {Ric<}) for the MME, (\ref {Ricc>}, \ref {Ricc<}) for the DCHE, (\ref {nf4Ricc>}, \ref  {nf4Ricc<}) for the HE and their confluent limits for the CHE, one for region $y>0$ and the other for $y<0$. The solutions of these equations have been written as converging series and found explicitly through a systematic procedure, see (\ref {finding-pol}) recursive order ODEs stemming from MME. Eventually, this provides a converging series for $\varphi _>$, $\varphi _<$ and $\varphi$, with explicit formul{\ae} and, importantly, also for the quantum momentum corresponding to Floquet eigenfunctions: concerning the first orders, (\ref {thetakexp}, \ref {solpin}) for the MME,  (\ref {thetaexpnf2}, \ref {solpin2}) for the DCHE,  (\ref {paiexp-hor}, \ref {solpin2bis}) and  (\ref {paiexp}, \ref {solpin2ter}) for the CHE, (\ref {thetaexpnf4}, \ref {solpin4}) for the HE. Interestingly, the ODE/IM (in/up) solutions can be written as linear combinations of the Floquet ones with coefficients simply depending on $\varphi$. And also more interestingly, their quantum momenta can be expressed as {\it one} converging series: as relevant examples, concerning the MME (\ref {solpindec}) as `incoming' form and regarding the CHE, (\ref {inc-hor}, \ref {solpin2bisin}) as `incoming', while (\ref {up-infin}, \ref {solp-up2ter}) as `outgoing' solutions. Eventually, when the corresponding wave functions are computed (as exponentials), their expansions give those recently found in \cite {CDFMPP} for Floquet solutions and, very surprisingly, more in general all those of the `MST-Anstatz' \cite {MST}. 

The quantity $\varphi $ for the MME (\ref {varphismall4}) and its symmetrisation $\Phi $ for the DCHE (\ref {Phiformulakbis}), the CHE (\ref {nf3Phi1inst}) and the HE (\ref {nf4Phigen}) have been computed at various orders and then conjectured to coincide with $A_D/\hbar$, the corresponding dual gauge periods of ${\cal N}=2$ SYM in the NS background for $N_f=0,2,3,4$, respectively. The rigorous proof of this identification for the MME is given in \cite {FRletter1}, and it will be extended to the other equations in an incoming publication \cite {FR-next1}. As an additional important result, \cite {FRletter1} shows the MME crucial identity   
\be 
\frac  {\partial P ^2 (\theta ,k)} {\partial k }=-\frac  {1} {2} \frac  {\partial  \varphi (\theta ,k)} {\partial \theta } \, ,
\label {cross} 
\ee
which is the condition of irrotationality for the vector $(P^2, -\varphi /2 )$ in the coordinates $\theta, k$ and then implies that the vector $(P^2, -\varphi /2 )$ is the gradient of a `potential'  $-\frac {1}{2}  {\cal F}$
\be
\varphi (\theta, k) =\frac  {\partial  {\cal F} (\theta ,k)} {\partial k} \, , \quad P^2(\theta , k)=-\frac {1}{2} \frac  {\partial  {\cal F} (\theta ,k)} {\partial \theta} \, , \label {crossfinal}
\ee
namely that $\frac {{\cal F} (\theta ,k)}{\hbar ^2}$ coincides with the gauge theory prepotential (with the identifications made just before equation (\ref {PhiAd0}): $\sqrt {2}e^\theta = \frac {\Lambda _0}{\hbar}$, $k=\frac {a}{\hbar}$). In particular, the second equality proves the Zamolodchikov-Matone relation \cite {MAT,FFMP} within the ODE theory (with the /IM counterpart as inspiration, though!). In this paper, the extension of the equation of existence of the prepotential (\ref{cross}) to the DCHE, CHE and HE has been achieved order by order at small $e^\theta$; a functional proof generalising \cite{FRletter1} is not difficult, but left as subject of an incoming paper, where also the derivatives w.r.t. the masses will be considered for reproducing \cite{BILT} without the AGT correspondence \cite{AGT}. Moreover, these identifications and constructions are expected to be valid also in the strong coupling regime of gauge theories, namely at large (adimensional) instanton coupling $\frac{\Lambda}{\hslash}$, and 'strong field' of gravity, namely large (adimensional) frequencies $M\omega$; in both these cases the methods found in \cite{FG} can be easily applied and give results.

As interesting generalisations of the structure (\ref {psiform}) to partial differential equations, in \cite {PP1,PP2} solutions of the BPZ equation with this form are constructed by means of the AGT correspondence, thus emphasising the r\^ole of the Virasoro algebra. Yet, the range of applicability of our extended ODE/IM correspondence (for instance by the introduction of Floquet solutions) seems to be much broader.

\medskip

{\bf Acknowledgements.} 
Special thanks go to D. Gregori, for the effort expended in \cite {DAN}, and to G. Di Russo for many explanations. Discussions with M. Bianchi, D. Bini, G. Bonelli, P. Di Vecchia, F. Fucito, K. Gupta, C. Heissenberg, P. Longhi, F. Morales, A. Nervo, F. Olivi, R. Russo, A. Tanzini, K. Zarembo are kindly acknowledged. This research was supported in part by the grants GAST (INFN), the Cost actions CaLISTA CA21109 and Theory-Challenges CA22113 "Fundamental Challenges in Theoretical Physics", MSCA CaLIGOLA 2021-SE-01-101086123, the EC Network Gatis, the MIUR-PRIN contract 2017CC72MK\textunderscore 003 and the National Science Foundation under Grant No. NSF PHY-1748958.

\appendix

\section{The two Floquet solutions of the CHE}
\label{floqCHE}
\setcounter{equation}{0}

In this Appendix we provide a proof of the equalities (\ref {paiexp-}).
In the right interval ${\cal {A}}<y<\ln 2 -\theta/2$, it holds 
\be
 \Pi _-^{(3)}(y',k_3, q_1,q_2,q_3)=\lim \limits _{H->CH} \Pi _-(y=y'-\theta /2,k, q_+,q_-)=\lim \limits _{H->CH} \Bigl (-  \Pi _+(\theta /2-y',k, q_-,q_+)\Bigr ) \, , \nonumber
\ee
where last step comes from using (\ref {Floq-def}). 
Now, by using property (\ref {SYMM}) and by coming back to the CHE quantum momentum, 
\ba
&& \lim \limits _{H->CH} \Bigl (-  \Pi _+(\theta /2-y',k, q_-,q_+)\Bigr )= \lim \limits _{H->CH}
\Bigl (- \Pi_<(-y',k, q_-,q_+)\Bigr )=\lim \limits _{H->CH} \Pi _>(y',-k, q_+,q_-)= \nonumber \\
&&= \Pi _+^{(3)}(y',-k_3, q_1,q_2,q_3) \, , \nonumber 
\ea
which eventually gives the equality $ \Pi _-^{(3)}(y',k_3, q_1,q_2,q_3)=\Pi _+^{(3)}(y',-k_3, q_1,q_2,q_3)$ for $a<y'<\ln 2$, which is the first of (\ref {paiexp-}).

\medskip

On the other hand, in the left interval $\theta /2-\ln 2<y<{\cal {A}}$, by using again (\ref {Floq-def}) and then (\ref {SYMM}), on finds
\ba
&&  \Pi _-^{(3)}(y',k_3, q_1,q_2,q_3)=\lim \limits _{H->CH} \Pi _-(y=y'-\theta /2,k, q_+,q_-)=\lim \limits _{H->CH} \Bigl (-\Pi _+(\theta /2-y',k, q_-,q_+)\Bigr )=  \nonumber \\
&& \lim \limits _{H->CH} \Bigl (- \Pi _>(\theta-y',k, q_-,q_+)\Bigr )=\lim \limits _{H->CH} \Pi _<(y'-\theta,-k, q_+,q_-)]=\Pi _+^{(3)}(y',-k_3, q_1,q_2,q_3) \, , \nonumber 
\ea
where $\theta-\ln 2<y'<a$, which eventually gives the second of (\ref {paiexp-}).

\end{document}